# Modeling the Multiple Sclerosis Brain Disease Using Agents: What Works and What Doesn't?

Ayesha Muqaddas

Muaz A. Niazi*

*Corresponding author

Department of Computer Science,
COMSATS Institute of IT,
Islamabad, Pakistan

muaz.niazi@ieee.org

**Abstract:**
The human brain is one of the most complex living structures in the known Universe. It consists of billions of neurons and synapses. Due to its intrinsic complexity, it can be a formidable task to accurately depict brain's structure and functionality. In the past, numerous studies have been conducted on modeling brain disease, structure, and functionality. Some of these studies have employed Agent-based approaches including multiagent-based simulation models as well as brain complex networks. While these models have all been developed using agent-based computing, however, to our best knowledge, none of them have employed the use of Agent-Oriented Software Engineering (AOSE) methodologies in developing the brain or disease model. This is a problem because without due process, developed models can miss out on important requirements. AOSE has the unique capability of merging concepts from multiagent systems, agent-based modeling, artificial intelligence, besides concepts from distributed systems. AOSE involves the various tested software engineering principles in various phases of the model development ranging from analysis, design, implementation, and testing phases. In this paper, we employ the use of three different AOSE methodologies for modeling the Multiple Sclerosis brain disease – namely GAIA, TROPOS, and MASE. After developing the models, we further employ the use of Exploratory Agent-based Modeling (EABM) to develop an actual model replicating previous results as a proof of concept. The key objective of this study is to demonstrate and explore the viability and effectiveness of AOSE methodologies in the development of complex brain structure and cognitive process models. Our key finding include demonstration that AOSE methodologies can be considerably helpful in modeling various living complex systems, in general, and the human brain, in particular.

## 1. Introduction:

Cai et al. note that the human brain can be considered as one of the most complex living structures of the known world [1]. Forstmann and Wagenmakers note the complexity of the brain because of its composition of billions of neurons, synapsis, blood vessels, glial cells, neural stem cells, and layered tissues [2]. For many decades, researchers have been attempting to model the human brain. These studies have been conducted primarily to understand the structure, function, connection, dynamics of neurons and overall brain at multiple spatial-temporal scales. Understanding and modeling the brain is extremely important because common brain diseases such as multiple sclerosis, dementia, cancer, Alzheimer, and epilepsy are often caused by a minor distraction of neurons or severe injury.

Due to the its inherent complexity, it is formidable to effectively model and depict brain's structure and functionality at all scales. Numerous studies have previously been conducted on brain disease, structure, and functionality modeling by applying ABM (Agent-Based Modeling), MAS (Multi-Agent systems) and Complex Networks (CN). Few of them have modeled the overall brain structure or functions simply while others have modeled the brain disease. According to researchers' best knowledge, there is no single research that could claim that their findings are complete and absolutely flawless. These studies achieved beneficial results for disease cure and prevention, brain structure and function understanding. However, these marvelous systems have flaws, as these models are developed by using Agent-Oriented technology, without following any AO methodology. To our best knowledge, none of them have used AO methodologies for modeling brain, brain disease, function, and structure.

AOSE (Agent-Oriented Software Engineering) combines MAS, ABM, AI, and distributed systems, and demands the application of AI and software engineering principles in the analysis, design, and implementation phases of a software systems development process. AOSE technology has the broad capability of autonomy, proactivity, reactivity, robustness, and social ability. Due to these capabilities nowadays, AOSE is becoming popular in the development of the distributed and complex application (such as e-commerce, healthcare systems, and social systems). Lucena and Nunes and R. Cunha et al,. state in their study that, AOSE promises to deal with complex and distributed systems[3][4]. For 2000, researchers are making a great effort to produce AOSE techniques for agent-based systems for the guidance of design, development and maintenance process. Now at present, this is a mature technology and has methodologies, architectures, methods, and developmental tools.

As system engineering methodologies focus on technical issues of system development. AOSE methodologies guide step by step development of agent modeling from system requirement to system implementation. When developers jump into system development without any guided methodology, then they ignore the most important aspect of the system to be implemented. Moreover, developers need the high-level expertise of system development, and they face difficulties throughout the development process, and the ongoing project takes more time, which ultimately leads to cost exceeding without any remarkable achievement. To overcome all these mentioned problems we are proposing an idea to follow AOSE methodologies in brain modeling scenarios. Because of neurons in the brain interact in a heterogeneous way, that's why we should use AOSE.

We will model MAS of Multiple Sclerosis disease by following AOSE methodologies GAIA, TROPOS, and MaSE. After applying these methodologies, we will model exploratory agent based model as a proof of concept. The Multiple Sclerosis ( MS) is considered as an inflammatory, autoimmune and demyelination disease of central nervous as stated in articles [5] [6] [7] [8]. In this disease myelin sheath on axonal part of neuron destroys with time. This destruction became an important cause of neurodegeneration which results in impaired muscular performance [9]. It can significant cause of physical and mental disabilities, irreversible neurologic deficits, including paralysis, muscle weakness, ataxia tremor, spasticity, cognitive impairment, body balance disorder, vision loss, double vision, vertigo, pain, fatigue, and depression [10].The dark side of this disease is it is becoming the main cause of neurological disability in young adults[11] that's why intensive research is needed on this disease to save youngsters life. The autoimmune system plays an important role in the disease progression and the body's own immunity assaults the myelin sheath causing injury. However, a satisfactory explanation for the origins and mechanism for MS is lacking in the literature.

In our work, we are taking MS disease as a case study for simulation system development. This case study is presented by Pennisi et al in the article [12]. First of all, we will develop the MS disease model by AOSE methodologies GAIA V.2, TROPOS, MaSE then, we will perform a comparative analysis of these methodologies to find out which methodology is best in a specific scenario. The purpose of AOSE methodologies implementation and comparison is to explore, if agent-based modeling with AOSE methodologies is a suitable paradigm for modeling human brains structure, cognitive process, and disease. And what kind of data would be needed for the sake of validation without spending a considerable amount of time on these models. Moreover, we will evaluate does the AOSE technology have worth in brain modeling. Our findings will prove that the AOSE methodologies are mature enough, that easily can implement the complex real system of any domain without any remarkable background knowledge.

## 2. Related Work:

The ever-growing use of agent-oriented technology demonstrates that it is used in almost every field of science. In this section, we will briefly present the well-known brain models, who have used agent oriented-technology However, the analysis of literature unveils that, the wide use of agent technology in modeling the human brain, leads the idea to use AOSE methodologies, that could aid the development team throughout the system development project. Such as in early requirement analysis phase, design, development and deployment phases. The AOSE methodologies work same like conventional software engineering methodologies, which have proved that methodologies are fundamentally needed for traditional software projects.

Y. Mansury and T. S. Deisboeck have proposed agent-based model for Spatial-temporal progression of tumor cells, according to environmental heterogeneities in mechanical confinement toxic metabolites. Results reveal that tumor cells follow each other along preform pathways. However, this model can be extended to other cancers by incorporating real data[13].

C. A. Athale and T. S. D. Ã, have developed an agent-based system, to simulate progression dynamics of the tumor. This model integrates Transforming Growth Factor α (TGFα) and induces EGFR gene-protein interaction network. Results show the progression rate of tumor cells according to Spatio-temporal and progression speed up when increasing EGFR density per cell [14].

Y.mansuray and .S. Deisboeck have generated a 2D agent-based system, that performs a simulation of two different gene expression, which are Tenascin C and Proliferating-Cell-Nuclear-Antigen for brain tumor progression. Moreover, this model investigates the effect of an environmental factor in gene expression changes. The results reveal that Tenascin C plays a crucial role in the migration of Glioma cell phenotype, and the expression of Proliferating-Cell-Nuclear-Antigen is responsible for proliferating behavior of tumor cell [15].

Y.Mansury and T.S.Deisboeck have proposed a 2D agent-based system, to examine the ongoing progressive performance of multiple tumor cells in human brain. The results demonstrate that these dangerous tumor cells proliferate and migrate on an adaptive grid lattice. And there is acorrelation among the progression dimension of the tumor and the velocity of tumor expansion[16].

Zhang et al. have developed a 3D multi-scale agent-based system, to perform a simulation of the cancer cell decision process. The results reveal that tumor cells do not only oscillate between migration and proliferation area, besides this these dangerous cells directly disturb the entire SpatioTemporal expansion patterns of brain area which is affected by cancer[17].

Zhang et al. have proposed a hybrid agent-based model that predict severe cancer areas in the brain. This model simulates cancer cells, either active or inactive cluster [18].

Germond et al. have proposed a multi-agent model, with remarkable features of deformation and edge detector. This model performs segmentation on MRI data and clearly shows each dynamic dimension of a scanned brain. When the results compared with real brain image then a real brain phantom is reported[19].

Richard et al have advocated multi-agent framework for brain MR image segmentation. That focus on radiometry tissue interpretation. This framework performs segmentation on a complete volume in less than 5 min with about 0.84% accuracy[20].

Vital-Lopez et al. have developed a 3D mathematical agent-based system, that performs a simulation of tumor progression as a collective behavior of individual tumor cells. The simulation results conclude that vascular network damage is one of the main reasons for tumor growth and invasiveness[21].

Soc,R has proposed an agent-based model for action selection in autonomous systems. He incorporated biological details in this Agent-Based system, in order to generate a set of predictions for decision selection. This research concludes that agent-based technology has the broader implication of action selection mechanism for both natural and artificial sciences related systems[22].

Signaling et al. have expanded their already proposed 3D Agent-Based and multi-scale system to perform simulation on the brain tumor. Which examine the simplified progression of Glioma pathway. Moreover, intracranial pressure is introduced to examine the influence of clonal heterogeneity on the human tumor growth. This research concludes that the brain regions that are near to blood vessels and are affected by tumor cells, these regions are the best nutrient source for tumor[23].

The scientist of Health Agent project González-Vélez et al have proposed an Agent-Based decision support model, for the detection and the diagnosis of brain tumor classifications. This study determines that now prediction of tumor classification is more accurate and optimize because reasoned argument is performed among intelligent agents[24].

Zhang et al., have developed ABM for cancer simulation according to multiple dynamic scales in space-time. This model proposes an idea of incorporating macroscopic expression patterns and microscopic cell behavior with molecular pathway dynamics. The results show that at the same time this 3D model efficiently model five different cancer cell clones[25].

Zhang et al., have proposed a Multiscale Agent-Based Model system to perform the simulation of Glioblastoma Multiforme cancer. This simulation model considers the progression and proliferation of cancer in the real time. The results conclude that this model is 30 times faster than the previous models. Because this model's extracellular matrix have large fine grids[26].

Haroun et al., have proposed a simulation model, based on multi-agents for brain MRI segmentation. The results show a global and local view of te image and these images are more clear than the other competitive models result. This research also demonstrates that an appropriate quantity of agents in the simulation is important to improve the segmentation quality of MRI images[27].

Pennisi et al., have proposed an ABM to formulate medication for MS disease and to reveal the mechanism of distraction and new potential formulation of the blood-brain barrier (BBB). The results suggest that vitamin D is effective for blood blockage [12].

Koutkias and Jaulent have developed ABM for Pharmacovigilance to detect complete potential signals associated with drugs and adverse effects. This model enables clinicians to detect timely and accurate drug signals effect[28].

Leistritz et al. have proposed an agent-based framework for agents, that have Belief, Desire, and Intention capabilities. As a result, this model satisfies a key integration challenge, of coupling time stepped ABM with event-based BDI systems[29].

Barrah has developed a MAS for magnetic resonance imaging (MRI) with a median filter to decrease computational time and save data details. The proposed FRFCM with MAS is fastidious, accurate and take less computational time[30].

Pennisi et al., have developed ABM for RRMS (Relapsing Remitting Multiple Sclerosis) disease. The outcome results show that the occurrence of genetic disposition in neurons is one cause of MS disease. However, the main cause is a breakdown of peripheral tolerance mechanism.[31].

| Ref | Title | Author | Year | Journal | MAS | ABM | S.E | AOSE |
|---|---|---|---|---|---|---|---|---|
| [13] | "The impact of ''search precision'' in an agent-based tumor model" | Mansury and Deisboeck | 2003 | Journal of Theoretical biology | ✗ | ✓ | ✗ | ✗ |
| [14] | "The effects of EGF-receptor density on multiscale tumor growth patterns" | Athale and Ã | 2006 | Journal of Theoretical Biology | ✗ | ✓ | ✗ | ✗ |
| [15] | "Simulating the time series of selected gene expression profile in an agent – based tumor model" | Mansury and Deisboeck | 2004 | Physica D | ✗ | ✓ | ✗ | ✗ |
| [16] | "Simulating 'structure-function' patterns of malignant brain tumors" | Mansury and Deisboeck | 2004 | Physica A | ✗ | ✓ | ✗ | ✗ |
| [17] | "Development of a three-dimensional multiscale agent-based tumor model: Simulating gene-protein interaction profiles, cell phenotypes and multicellular patterns in brain cancer" | Zhang, Athale and Deisboeck | 2007 | Journal of Theoretical Biology | ✗ | ✓ | ✗ | ✗ |
| [18] | "Multi-scale, multi-resolution brain cancer modeling" | Zhang, Chen and Deisboeck | 2009 | MATHEMATICS AND COMPUTERS IN SIMULATION | ✗ | ✓ | ✗ | ✗ |
| [19] | "A cooperative framework or segmentation of MRI brain scans" | Germond et al., | 2000 | Artificial Intelligence in Medicine | ✓ | ✗ | ✗ | ✗ |
| [20] | "Automated segmentation of human brain MR images using a multi-agent approach" | Richard, Dojat and Garbay, | 2004 | Artificial Intelligence in Medicine | ✓ | ✗ | ✗ | ✗ |
| [21] | "Modeling the Effect of Chemotaxis on Glioblastoma Tumor Progression" | Vital-lopez, Armaou and Hutnik, | 2011 | American Institute of Chemical Engineers | ✗ | ✓ | ✗ | ✗ |
| [22] | "Introduction. Modeling natural action selection" | Soc,R | 2007 | The royal society | ✗ | ✓ | ✗ | ✗ |
| [23] | "Simulating Brain Tumor Heterogeneity with a Multiscale Agent-Based Model : Linking molecular signatures, phenotypes and expansion rate" | Signaling, Bias and Rate | 2009 | Mathematical and Computer Modeling | ✗ | ✓ | ✗ | ✗ |
| [24] | "HealthAgents: distributed multi-agent brain tumor diagnosis and prognosis" | González-Vélez et al | 2009 | Applied Intelligence | ✗ | ✓ | ✗ | ✗ |
| [25] | "Multiscale agent-based cancer modeling" | Zhang et al., | 2009 | Journal of Mathematical Biology | ✗ | ✓ | ✗ | ✗ |
| [26] | "Developing a multiscale, multi-resolution agent-based brain tumor model by graphic processing" | Zhang et al., | 2011 | Theoretical Biology and Medical Modelling | ✗ | ✓ | ✗ | ✗ |
| [27] | "A Massive Multi-Agent System for Brain MRI Segmentation" | Haroun et al., | 2005 | Massively Multi-Agent Systems | ✓ | ✗ | ✗ | ✗ |
| [12] | "Agent-based modeling of the effects of potential treatments over the blood–brain barrier in multiple sclerosis" | Pennisi, Marzio, et al | 2015 | Journal of immunological methods | ✗ | ✓ | ✗ | ✗ |
| [28] | "A multi-agent system for integrated detection of pharmacovigilance signals" | Koutkias, V, and Jaulent, M. C | 2016 | Journal of medical systems | ✗ | ✓ | ✗ | ✗ |
| [29] | "Time-variant modeling of brain processes" | Leistritz, L., Schiecke, K., Astolfi, L., and Witte, H. | 2016 | Proceedings of the IEEE | ✗ | ✓ | ✗ | ✗ |
| [30] | "MAS based on a Fast and Robust FCM Algorithm for MR Brain Image Segmentation" | Barrah, H. et al. | 2016 | . International Journal of Advanced Computer Science & Applications | ✓ | ✗ | ✗ | ✗ |
| [31] | "Agent based modeling of Treg-Teff cross regulation in relapsing-remitting multiple sclerosis" | M. Pennisi et al. | 2013 | BMC Bioinformatics | ✗ | ✓ | ✗ | ✗ |

## 3. Material and Methods

This section describes in detail the case study of multiple sclerosis (MS) disease and the implementation of the MS model by using AOSE methodologies like Gaia V.2, Tropos, and MASE. According to our best knowledge that AOSE methodology has never been used in any biological model. First time we tried to model biological problem, according to well defined agent methodologies. To do this, here we are copying same rules and regulations that are used to model other systems in engineering field. We implemented MS model in Net-Logo environment. This model is already implemented by Pennisi et al. in [12].

### 3.1. Case study:

MS is one of the central nervous system diseases. In which myelin sheath from the axonal part of the neurons in the brain and spinal cord is removed and the communication between neurons is disconnected.In this disease, different factors and cells are involved. The overall mechanism of MS disease is given below.

The thymus gland is a lymphoid organ that produces T-cells (T-reg and T-eff). The T-reg and T-eff both cells have two states, "active and resting". EBV (virus) is an external factor that latent infection. EBV cause activation of both T-eff and T-reg. The active T-eff (A.T-eff) attacks the myelin of axons and cause the neural communication damage and in return duplicate itself. A.T-eff produce one cytokine against one attack. Active T-reg (A.T-reg) try to catch A.T-eff and suppress it. In return, A.t-reg receive a positive feedback and will duplicate. After the attack of A.T-eff the amount of myelin in BWM will be lowered. If the damaged portion still has any amount of myelin then it is recoverable. If myelin amount reaches zero, then it is unrecoverable. A.T-eff produce cytokines after damage myelin. Cytokine attacks BBB and damages it. The damaged BBB allows other T-cells and virus to enter into the brain. The main purpose of the BBB is blocked and bounce back to all T-reg, A.T-reg, T-eff, A.T-eff, Virus, Cytokines.

At the start of the simulation, all agents are introduced in the model in the resting state. The agents are considered as brain cells. All agents have life counter that decrement by one at every step of the agent. If the life counter of agent reaches to zero, then that agent will be removed from the simulation automatically. In running simulation, all agents will move randomly in the environment (actually brain) and new agents will be introduced at random time.

### 3.2. GAIA V.2:

The original GAIA methodology was presented by Wooldridge in 2000, It was the first complete AOSE methodology developed for large-scale real-world applications [32]. GAIA V.2 focuses on two main development phases of a system, analysis phase, and development phase. Moreover, these models are considered as a guideline for developing the real models of the to-be-developed complex system.

This methodology is applied to ABM and MAS modeling after requirements are gathered and specified. After adoption of Gaia methodology for complex systems development, researchers realize that it is easy to use and implement. However, it is not much suitable for complex systems design [33]. Because modeling notations are poor for expressing complex problems, such as complex and multiphase interaction protocols. Moreover, Gaia has a deficiency of requirement phase, environment model, and does not provide an appropriate domain knowledge. To overcome these limitations Zambonelli, Jennings and Wooldridge proposed Gaia V.2 agent methodology in [34]. In our MS model we are using Gaia V.2, same like Gaia it has two development phases 1) analysis and 2) design as shown in the following picture.

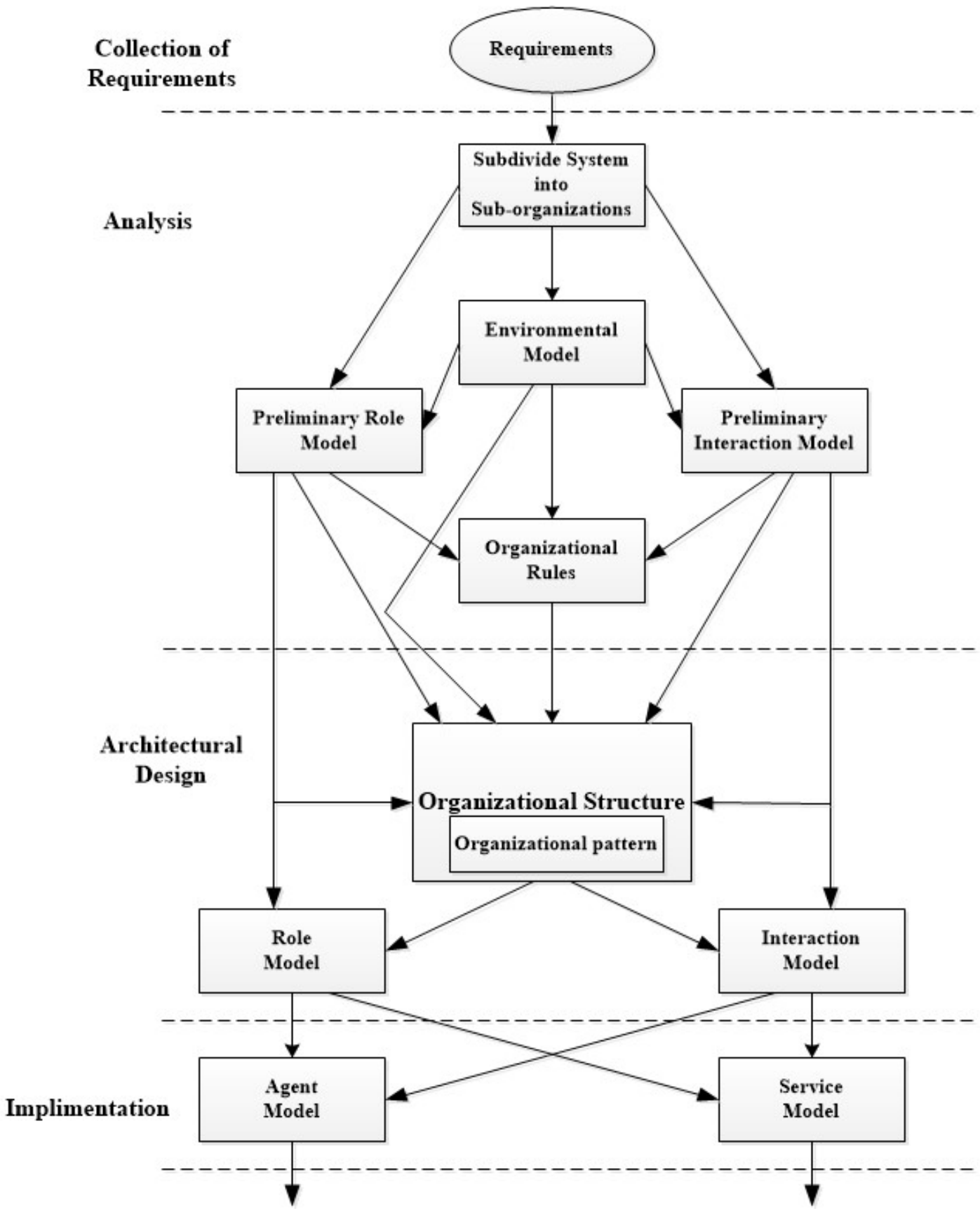

Figure 1: GAIA V.2 methodology [35]

### 3.2.1 Analysis Phase:

Generally, the analysis phase emphasizes on understanding what the MAS and ABM system will have to be as their properties or specifications (without reference to any implementation detail). The output of the analysis phase became the base input for the design phase. In actual, in this phase analyst decide about the functional and non-functional characteristics of the to-be-system. The output of this phase is five basic models such as 1) sub-organizations, 2) environmental model, 3) preliminary role model, 4) preliminary interaction model, and 5) organizational rules model.

#### 1) System Sub-Organizations

This model tries to identify all the involved organizations, and used a fruitful way to categorize the whole system into loosely coupled sub-organizations. The identification of sub-organization is easy if

1) The Sub-organizations are already identified through system specifications.

2) The system itself mimics the overall structure of the real world, in which multiple organizations interact.
3) Apply the modularity technique. This technique splits the overall complexity of the to-be-system into a set of smaller and more manageable components.

According to our experience, the sub-organizations of the developing system can be found easily, when we divide the overall system into portions. And each portion exhibit a specific behavior and they interact with each other to achieve a subgoal. Any portion of the system which performs any task or interacts with other to perform a specific task, then we can take it as a sub-organization.

| **Sub-organizations** | **Description** |
|---|---|
| **T-eff** | The goal to achieve is to “become active” by catching the virus. It will interact with the virus for activation and BBB to enter into the brain. |
| **T-reg** | The goal to achieve is to “become active” by catching the virus. It will interact with the virus for activation and BBB to enter into the brain. |
| **Virus** | The goal to achieve is to “enter into the brain” and make “ T-reg and T-eff active”. It interacts with T-reg, T-eff to make them active and BBB to enter into the brain. |
| **A.T-eff** | The goal to achieve is to “attack on myelin” to damage brain, “duplicate itself” and “produce cytokines”. It interacts with BBB to enter into the brain. |
| **A.T-reg** | The goal to achieve is to “attack on ” A.T-eff to kill them and “duplicate itself”. It interacts with A.T-eff to kill them and interact with BBB to enter into the brain. |
| **Cytokines** | The goal to achieve is “attack on BBB” to damage it. It interacts with only BBB. |
| **BWM** | The goal to achieve is “recover damaged myelin” in the brain to maintain neural communication. |
| **BBB** | The goal to achieve is “stop the entrance” of T-eff, T-reg, A.T-eff. A.T-reg,virus, Cytokine into the Brain. |

Table 1: Sub-Organizations and their goals

In our’s MS disease model, we identified eight sub-organizations T-reg, T-eff, virus, A.T-eff, A.T-reg, Cytokines, BBB and BWM as showed in table 1. After identification of sub-organizations, we divided them into two main groups by arranging roles into groups with logical or physical similarities as Silva has done in his work [35]. The first group is brain organs and the second one is neuron cells. The Brain organ group consist of two types of agents (BBB, and BWM). On the other hand cell group contain EBV ,T-reg, T-eff, A.T-reg, A.T-eff and Cytokines. According to AOSE term, in general, we would call “agents” to all neuron cells and brain organs. The agents in the second group (neuron cells) are tightly connected and they also interact with the second group (brain organ) agents. Although GAIA is lenient with complex systems development, however, due to the lack of requirement analysis phase, the accurate identification of sub-organizations is not feasible from GAIA, as it also lacks of established hierarchy model and organizational structure[35] [36].

**2) Environment Model**

The environmental model describes the real world in which the system operates with all its variables, resources, and uncertainties. There is not a single easy way, to provide a general modeling abstraction and the general modeling techniques for the development of the system environment. Since the environments of the applications are divergent in nature because in most cases systems suffers from compatibility issues with the technology. As Passos stated in his study that the development of the environmental model is a separate one type of agent-oriented methodology [37].  That's why we can not model complete environmental model of any system.

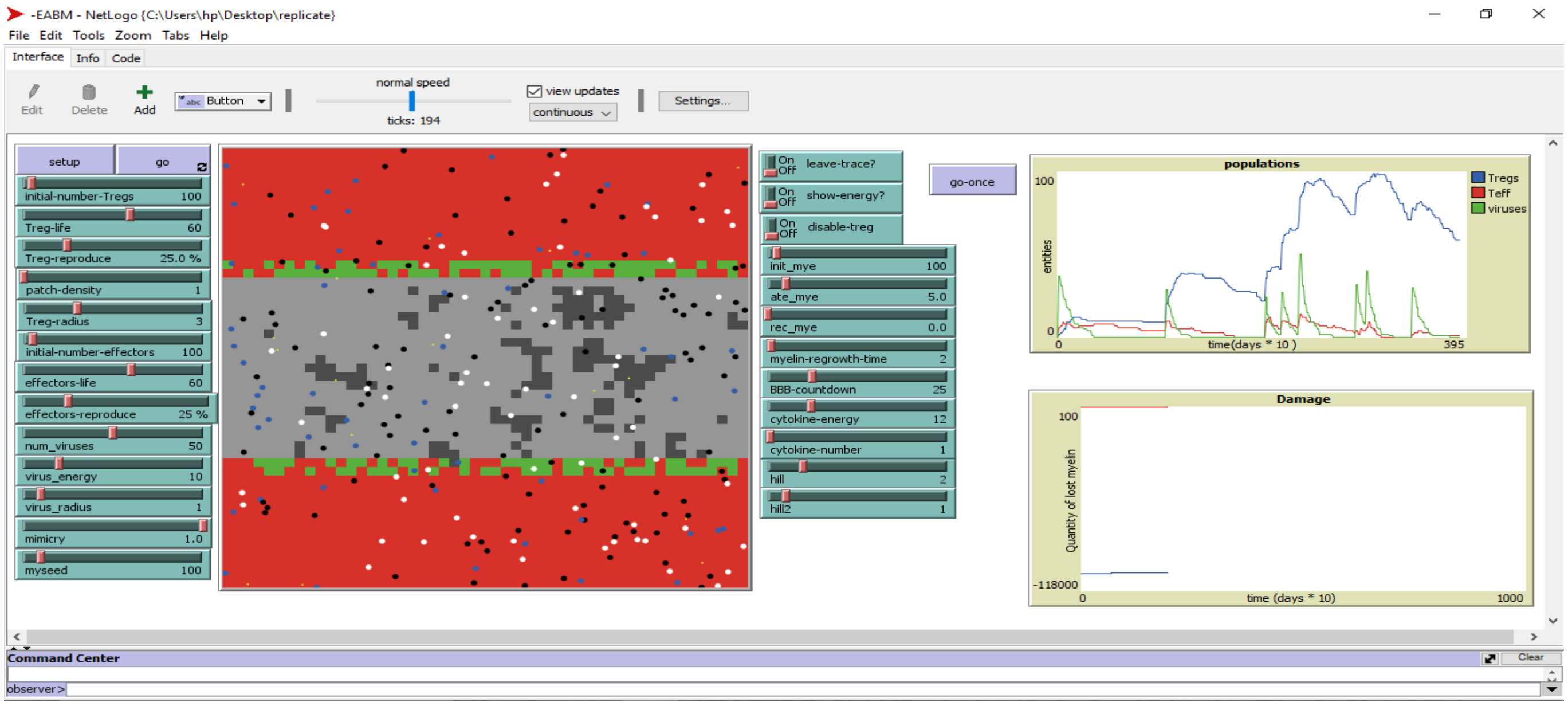

Figure 2 Environmental Model of MS disease simulation

The environment of our model is Netlogo development tool and resources are Netlogo's buttons, sliders, plots, switches eg. We divide this model into two environments, static and dynamic according to Gaia methodology. The static part of our model is buttons, sliders, switches that do not change frequently. And the dynamic part is disply monitor in which agents move randomly in environment to perform complex operation and to access resources. The discription of the environmental model of MS disease is given in below table.

| Environment Resources | Description |
|---|---|
| **Display Screen** | The screen shows all the time movement of agents and changing in the model. It also displays what is going on among agents. |
| **Buttons** | The buttons stop or start the simulation and initialize special characteristics of agents, as creation or deletion agents etc. |
| **Sliders** | Sliders assign special values to agents in the start or during the simulation to observe effects of changes. |
| **Switches** | Switches on/off characteristics of agents in simulation. |
| **Plots** | The Plot shows the behavior of agent interaction. |
| **Agents** | Agents representative  of characters of one type that are involved in the |
| | simulation to perform a specific task. |
| **Patches** | The patches are also representative of agents of another type that are involved in the simulation to perform a specific task. |

Table 2: Description of Environmental Model

**3) Preliminary Role Model**

A role model provides an abstraction of the agent with a set of its expected behavior[34]. The final goal of the analysis phase is not to identify the all involving entities and model them into the real system organization. The main purpose is to identify the active roles and their interaction with others roles at a very abstract level. Particularly, the role abstraction leads to the identification of basic specifications for the developing system, to achieve its main goal. As well as the important interactions that are required for the completion of system's specification. However, this type of identification can be performed without knowing that, what the actual structure of the developing system will be. There is a chance to identify some characteristics that will remain same independently throughout the organizational structure. This identification can be more beneficial if the analysis phase carefully models the system specifications in terms of involved characters and their responsibilities at a very beginning stage. In our model, we identified 8 roles T-reg, T-eff, A.T-reg, A.T-eff, Virus, BBB, Cytokines and BWM (Brain White Matter) which are shown in figure 2.

**4) Preliminary Interaction or protocol Model**

The interaction model identifies all dependencies which show the relationship among roles by defining the protocol[37]. Protocols are a request for resources or to complete a task, that a role does to interact with other roles. In protocol development, more focus is on the purpose of interaction and the nature of interaction, than to the type of message exchange and sequence of execution steps. A standard protocol model of GAIA V.2 consists of these points:

1) Protocol title/name: The protocol name is a description of the roles interaction nature. For example, it specifies that a request is for resource share or to assign a task.
2) Initiator: The role who starts the conversation.
3) Partner: that role will be partner which responds to initiator during conversation.
4) Input: The information used by the initiator role as a reason to initiate the protocol.
5) The final action that will take the responder role. It can be an information, resource or a request for anything.
6) Description: The description explains the whole scenario of protocol processing in detail. e.g which roles will involve and what will be their intentions and what will be the consequences.

In MS disease system five protocols are identified. That are Attack, Catch, Produce, MakeActive, and stop. At the end, the general model of preliminary interaction and role express the overall scenario of protocols among the roles of MS disease. The interaction model is also called a protocol model. In the below tables, all protocols are described in detail.

| |
|---|
| **Protocol Name:** catch |

| | | |
|---|---|---|
| **Initiator:**<br>T-reg | **Partner:**<br>Virus | **Input:**<br>Information about virus existence |
| **Description:**<br>When any T-reg knows, that there is virus around it then, it tries to catch the virus to become active | | **Output:**<br>A.T-reg |

Table 3: T-reg's Protocol

| **Protocol Name:** catch | | |
|---|---|---|
| **Initiator:**<br>T-eff | **Partner:**<br>Virus | **Input:**<br>Information about virus existence |
| **Description:**<br>When any T-eff knows, that there is virus around it then, it tries to catch the virus to become active | | **Output:**<br>A.T-eff |

Table 4: T-eff's Protocol

| **Protocol Name**: makeActive | | |
|---|---|---|
| **Initiator:**<br>Virus | **Partner:**<br>T-reg, T-eff | **Input:**<br>Information about T-reg, T-eff existence |
| **Description:**<br>When any virus knows, that there is T-reg, T-eff around it. Then, virus tries to catch the T-reg, T-eff to make it active | | **Output:**<br>A.T-reg |

Table 5: Virus's Protocol

| **Protocol Name:** Attack | | |
|---|---|---|
| **Initiator:**<br>A.T-eff | **Partner:**<br>BWM | **Input:**<br>Information about myelin existence |
| **Description:**<br>When any A.T-eff finds, that there is myelin in the brain, then it attacks on myelin and damage it. | | **Output:**<br>Damaged myelin |

Table 6:A.T-eff's Protocol

| | | |
|---|---|---|
| **Protocol Name:** Attack | | |
| **Initiator:** A.T-reg | **Partner:** A.T-eff | **Input:** Information about A.T-eff existence |
| **Description:** When any A.T-reg finds, that there is A.T-eff in brain then, it attacks on A.T-eff and kill it. | | **Output:** A.T-eff killed |

Table 7: A.T-reg's Protocol

| | | |
|---|---|---|
| **Protocol Name:** Attack | | |
| **Initiator:** Cytokine | **Partner:** BBB | **Input:** Information about BBB existence |
| **Description:** When any Cytokine finds, that there BBB, then Cytokine attacks on BBB and damage it. | | **Output:** Damaged BBB |

Table 8: Cytokine's Protocol

| | | |
|---|---|---|
| **Protocol Name:** Produce | | |
| **Initiator:** A.T-reg | **Partner:** A.T-eff | **Input:** Information about A.T-eff existence |
| **Description:** When any A.T-reg finds, that there is A.T-eff in brain then, it kills A.T-eff and produce itself duplicate. | | **Output:** A.T-reg |

Table 9: A.T-reg's Protocol

| | | |
|---|---|---|
| **Protocol Name:** Produce | | |
| **Initiator:** A.T-eff | **Partner:** Cytokine | **Input:** Information about myelin existence |
| **Description:** When any A.T-eff finds, that there is myelin in the brain, then it attacks on myelin and duplicate itself. | | **Output:** A.T-reg |

Table 10A.T-eff's Protocol

| Protocol Name: Stop | | |
|---|---|---|
| **Initiator:** BBB | **Partner:** T-reg, T-eff, Virus, A.T-reg, A.T-eff, Cytokines, | **Input:** Information about enterance of cells into brain |
| **Description:** When ever BBB knows that any cell is trying to enter into brain then BBB stop its entrance. | | **Output:** A.T-reg |

Table 11:BBB's Protocol

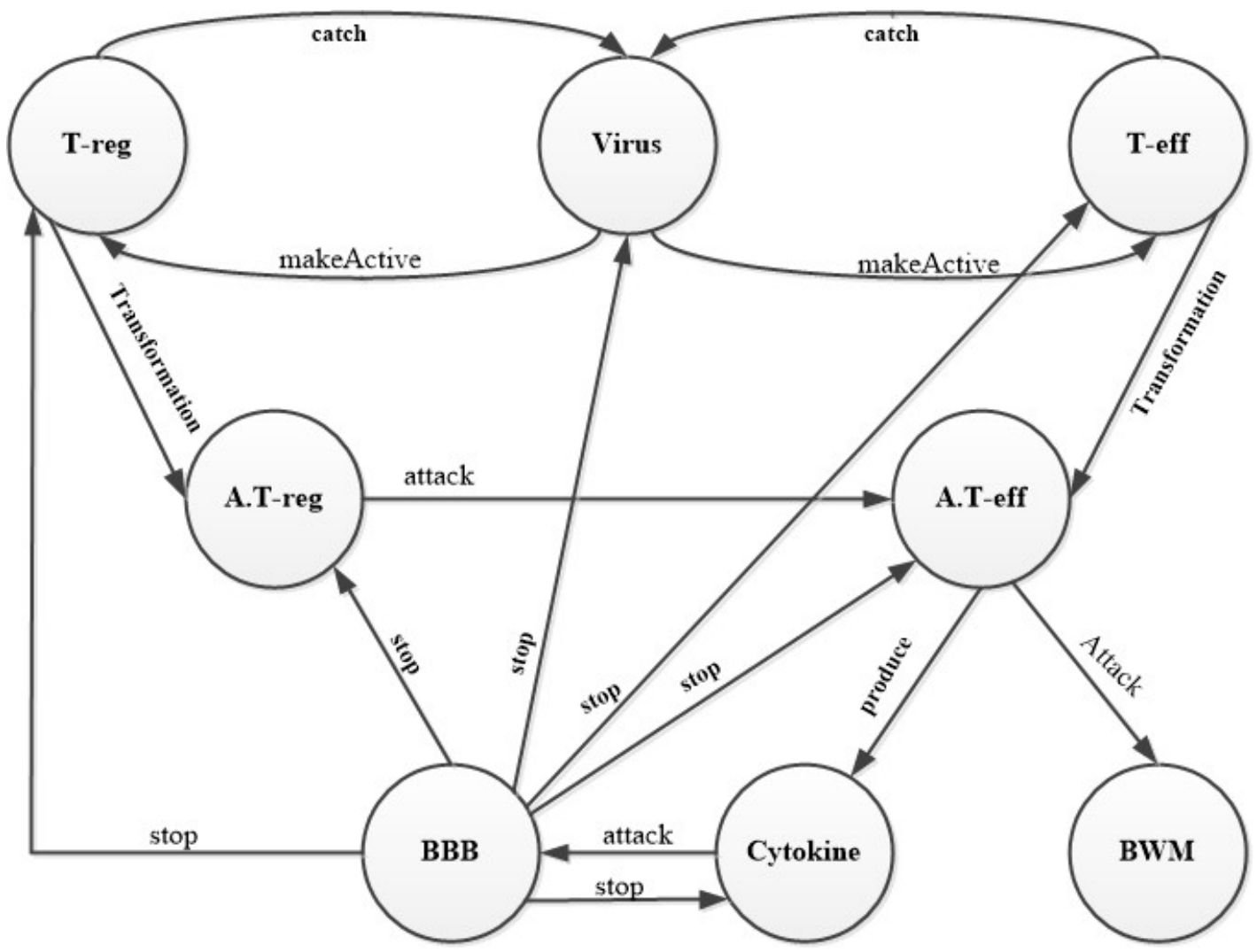

Figure 3: Preliminary Role and Interaction Model of MS

This is a combined figure of the preliminary role and their interaction model of MS disease.
In this figure, there are all identified role and protocols. T-reg and T-eff roles interact with the virus by using catch protocol. These both roles aim is to become active by catching the virus. In general, the protocol defines the purpose of communication among agents. In the same way, the virus role interacts with T-reg and T-eff by using "makeActive" protocol to make them active. A.T-reg (Active T-reg) and A.T-eff (Active T-eff) are active agents of T-reg and T-eff community. A.T-eff interacts with BWM by using "attack" protocol. The "attack" protocol is used to damage the BWM. A.T-eff also interact with Cytokine by using "produce" protocol. A.T-reg role interacts with A.T-eff by using "attack" protocol. The purpose of this protocol or communication is to kill the A.T-eff role to stop its dangerous activities in the brain. In the same way, the Cytokine interacts with BBB by using "attack" protocol to damage BBB. The BBB role interacts with almost all other roles except BWM by using "stop" protocol. The purpose of this protocol is to stop all other roles, to entering the brain.

### 5) Organizational Rules

The organizational rules define whether a new agent can be added into the organization. If new agents added then what would be their position in the organization and which type of behavior would be expected from the added agents. In simple words, the organizational rule defines the overall responsibility of the concerned organization in an abstract way. There are two types of organizational rule:

(1) Liveness: The liveness rules take concern about the evolution of system dynamics according to time. It also takes concern about that a specific agent will play a specific role and allow the agent to play next task if it has played the previous role. Same like roles, a specific protocol may execute only after the execution of the other specific protocol. Moreover, liveness organizational rule decides that which role would be played by which agent.

(2) Safety: The safety rules consider all unexpected events that can occur during processing of a specific task. These events are considered as time-independent events. To overcome all the unexpected faults, the organization rule force to apply the concept that a single role must play by a distinct entity or agent and two concurrent tasks should not be played by a single entity.

By following these conditions, in our MS disease system sub-organizations define their rules separately.

| **Organization Name: T-eff** |
|---|
| **Liveness: This role would become active if only it catches a virus, or it would transform from T-eff to A.T-eff if only virus attacks on it.** |
| **Safety: Only this agent can transform from T-eff to A.T-eff agent.** |

Table 12: T-eff Organizational Rules

| **Organization Name: T-reg** |
|---|
| **Liveness: This role would become active if only it catches a virus, or it would transform from T-reg to A.T-reg if only virus attacks on it.** |
| **Safety: Only this agent can transform from T-reg to A.T-reg agent.** |

Table 13: T-reg Organizational Rules

| **Organization Name: Virus** |
|---|
| **Liveness: This agent can enter into the brain if BBB is broken.** |
| **Safety: Only this agent trigger the MS disease.** |

Table 14: Virus's Organizational Rules

| **Organization Name: A.T-eff** |
|---|
| **Liveness: This agent can damage myelin if the only virus makes it active.** |
| **Safety: Only this agent can damage the myelin.** |

Table 15: A.T-eff Organizational Rules

| |
|---|
| **Organization Name: A.T-reg** |
| **Liveness: This agent can kill the A.T-eff if only it is active.** |
| **Safety: Only this agent can kill the dangerous agent A.T-eff.** |

Table 16:A.T-reg Organizational Rules

| |
|---|
| **Organization Name: Cytokine** |
| **Liveness: These agents born after myelin damage.** |
| **Safety: Only these agents can damage BBB.** |

Table 17: Cytokine's Organizational Rules

| |
|---|
| **Organization Name: BBB** |
| **Liveness: This agent stops the other agent who tries to enter into the brain.** |
| **Safety: This agent stops all agents to enter into the brain and recover damaged BBB.** |

Table 18: BBB Organizational Rules

| |
|---|
| **Organization Name: BWM** |
| **Liveness: This agent recovers myelin after myelin damage occurs.** |
| **Safety: This agent recovers myelin for neural communication.** |

Table 19: BWM's Organizational Rules

### 3.2.2. Design Phase

The design phase is a much important phase in system engineering, where the final decisions take about what will be the specifications of the developed system, what will be the operational environment. The main purpose of this phase is to transform the analysis models into low-level design models. Low-level design models are abstract level models which can be easily implemented in the development phase.

This phase is also important in this way that, it identifies missing and conflicting requirements and helps developers when system specifications should be considered mature to develop a complex system. The design phase of Gaia V.2 is divided into two sub design phases, first one is architectural design and the second one is detailed design.

#### 1. Architectural Design

#### 1.1 Organizational structure

The selection of organizational structure of a system is a very crucial decision in ABM and MAS development since, it affects all subsequent phases. The one important benefit of an organizational structure is, it organizes roles as a topology, which is easily understandable by inexperienced persons.

And the other main objective of this structure is to explicate the inter-role relationship type among roles/agents. In general, there are three types of relationships:

I. In the control relationship, one role has authority over the other role, in this relationship a role can partially or fully control the action of the other role.
II. The peer relationship defines that, in the organization, the involved roles have equal status.
III. In the dependency relationship, one role depends on the other role for resources or knowledge, which is compulsory for its accomplishment.

However, the organizational structure same like sub-organizations model does not accurately be implemented, because there is no fixed hierarchical structure exists. The MS disease system is implementing the generic architectural model by considering whole MS circumstances as a hierarchy, and dependency between agents as a control structure. In our model the organizational structure manifest roles, sub-organizations and the relations and association between them.

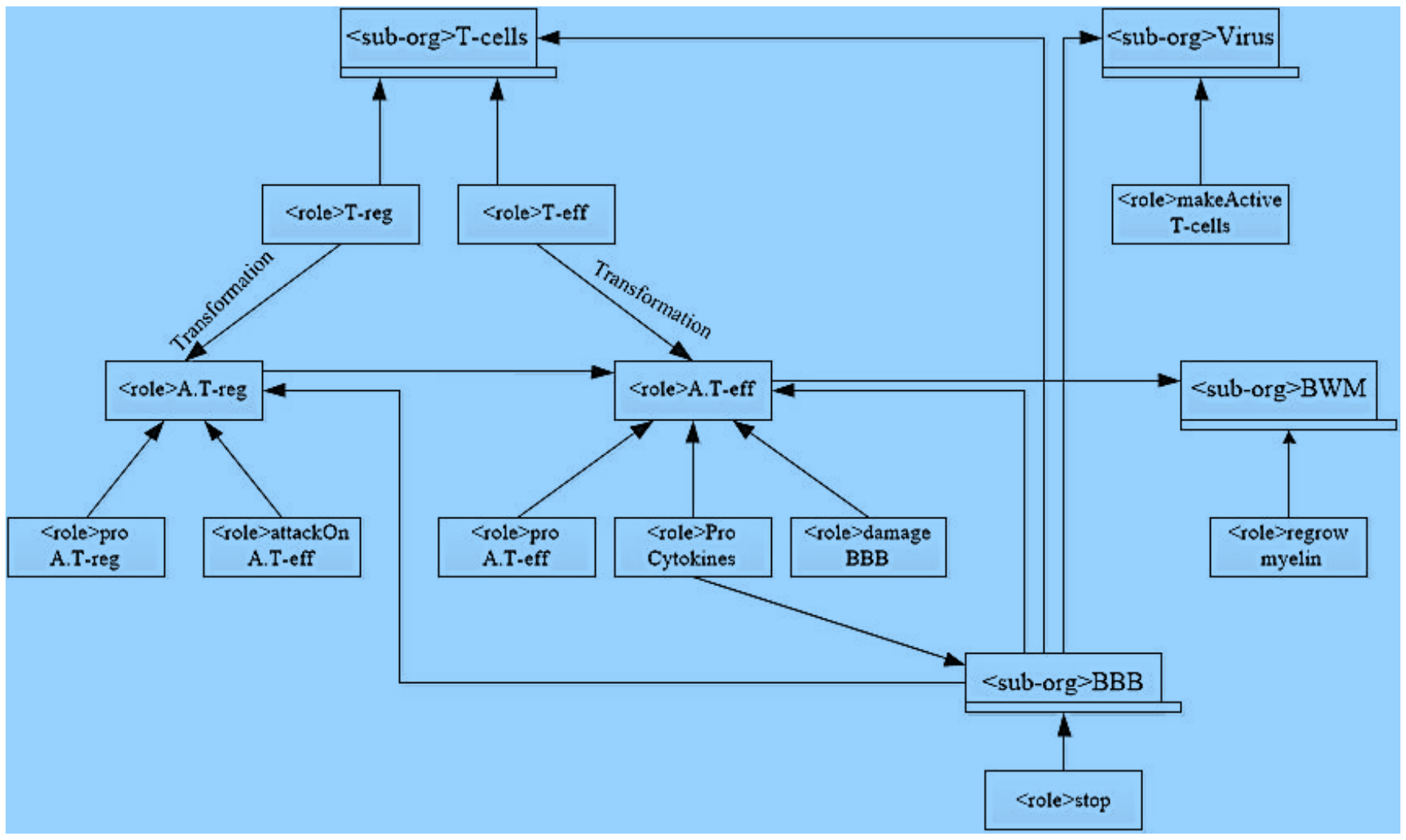

Figure 4: GAIA's Organizational Structure

The MS model defines brain organs (BBB, BWM), T-cells, Cytokines, and virus as sub-organizations. The arrows show organization's communication with other organizations. This communication can be direct organizations' communication or with the role of other organizations. One organization can communicate with the roles of other organizations.
IN MS model T-cells, Virus, BWM, and BBB are sub-organizations because they all are different in structure and nature. T-reg and T-eff cells belong to the same cell category because they are produced by the same body organ. That's why they belong to the same sub-organization. However, after activation, they behave differently according to their builtin nature. In the same way, BBB organization communicates with all others organizations' agents to stop their entrance into the brain. Further, A.T-reg develops hierarchy structure. In this structure, A.T-reg is organized according to its goals. A.T-eff also develops a hierarchy structure. Which, is organized according to roles.

### 1.2 Role and interaction model

The design phase transforms the preliminary role model and the role interaction model into absolute detailed role and interaction model to form an organizational structure. Which clearly defines the roles of each agent and interaction type among them. To construct detailed role and interaction model a developer should consider these mentioned instructions:

I. Identify the new roles and organizations which was not identified by analysis phase.
II. Developers should identify all the possible activities in which a role can be involved, as well as classify role's safety and liveness responsibilities contribute to overall organizational rules.
III. Identify all the possible protocols, needed by the developing system. As well as classify which roles will be involved in the relationship to complete the protocol execution.
IV. Moreover, developers should identify the protocols for the organizational level relationship.

In short, for absolute role model, we identify the complete role's activities and services. And for the absolute interaction model, we identify and model interaction of all roles involved.

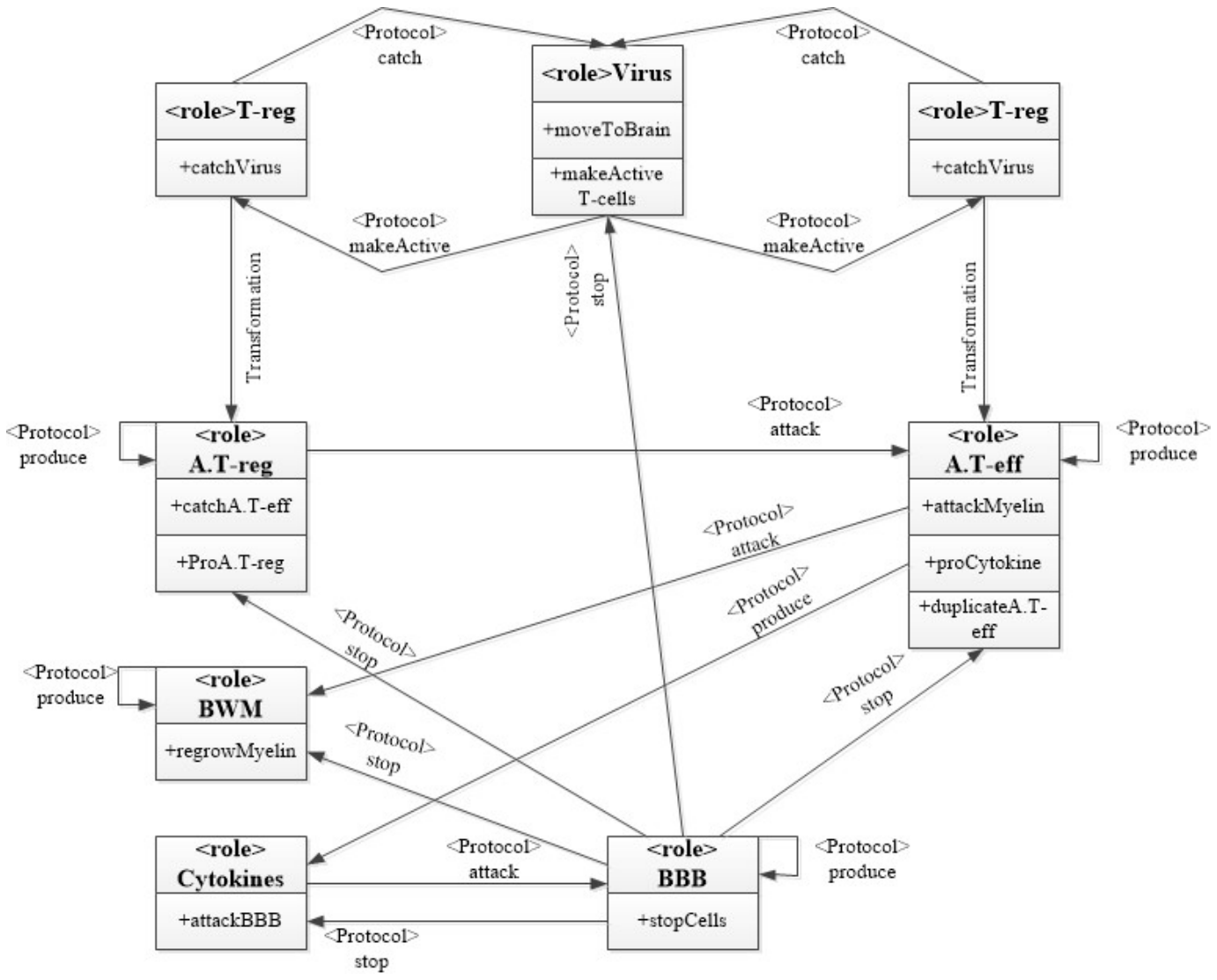

Figure 5: Role and Iteraction Model

In this model, we defined roles with their responsibilities, services, and protocols. The difference between preliminary phase models and detailed design models is, in the design phase, we identify all possible and complete information about the roles characteristics and protocols.

As already defined that 8 MS disease roles and two types of agents are identified. BBB and myelin belong to brain organ type and virus, T-cells belong to cells agent type. T-cells (T-reg and T-eff) belong to the same community of cells, but they react to brain according to their built-in nature. In other words, some of T-cells became dangerous as T-eff and some became part of immune system T-reg. The T-reg role is responsible to catch viruses to get active. The active T-reg (A.T-reg) support immune system by killing dangerous Active T-eff (A.Teff) cells and it produce duplicate A.Treg cell. The duplicate T-cell perform functionality same like original T-cell. Virus role is responsible for T-reg and T-eff cells activation. The virus searches T-cells and try to attack them after that it died. A.T-eff role is responsible to attack myelin and damage it. It also duplicates A.T-eff and produces cytokines. Cytokine roles, responsibility is to attack the BBB to damage it and allow viruses and all other agents to move toward the brain. BBB role belongs to brain organ agent type and is responsible to block unwanted molecules and cells to enter into the brain. It also manages to allow wanted molecules (water, glucose, and water) and cells, T-reg to enter into the brain to protect the brain. BWM or myelin is responsible to maintain communication between neuron cells and if any damage occurs, then regrow myelin to maintain communication.

In this MS model, there are five types of protocols “stop, catch, attack, produce, and makeActive”. BBB initiates stop protocol to stop the virus, T-reg, T-eff, A.T-reg, A.T-eff, Cytokine, to enter the brain. Virus initiates “makeActive” protocol with T-reg, T-eff to make them active. T-reg and T-eff agents initiate catch protocol with a virus to get active. A.T-reg agent initiates “attack” protocol with A.T-eff agent to stop their unhealthy activities. On the other side, A.T-eff initiate “attack” protocol for demyelinate BWM. Cytokines initiate “attack” protocol with BBB to destroy the BBB. Moreover, “produce” protocol is initiated by A.T-reg, A.T-eff to produce their duplicates. The “produce” protocol is also initiated by BBB and BWM to recover themselves.

## 2. Detailed Design Phase

### 2.1 Agent Model

In the detailed design phase, developers are more experienced about the developing system. This phase helps developers to identify the actual role model and the actual interaction model which in return will assist the developers in the implementation phase. The Agent model captures all agent types and agent’s roles that will be implemented in the system. This phase implements agent model same like class model in UML. As Castro and Oliveira state that GAIA does not provide modeling notations for agent modeling. It simply suggests to adopting UML class diagram in article[38]. Zambonelli, Jennings and Wooldridge states in article [34], according to GAIA “An agent is a software that plays a set of roles of a certain type”. Thus the agent model carefully identifies, what type of agent classes should be involved to play the specific roles and how many instances should be in each class of the actual system.

### 2.2 Service Model

The service model identifies all the possible services associated with each agent class and consistently with the roles. The service model can be applied in both cases: 1) the static assignment of roles to agent classes, 2) the assignment of dynamic roles to agent classes[34]. The service is a function of the considered agent, and it is derived from the agent’s protocols, liveness and responsibilities, and the activities of the role that each agent implements. This fig is representing service and agent model. Services express what agents are contributing for the beneficial of the overall models working.

In our MS disease model, we are combining agent and role model in a single model.

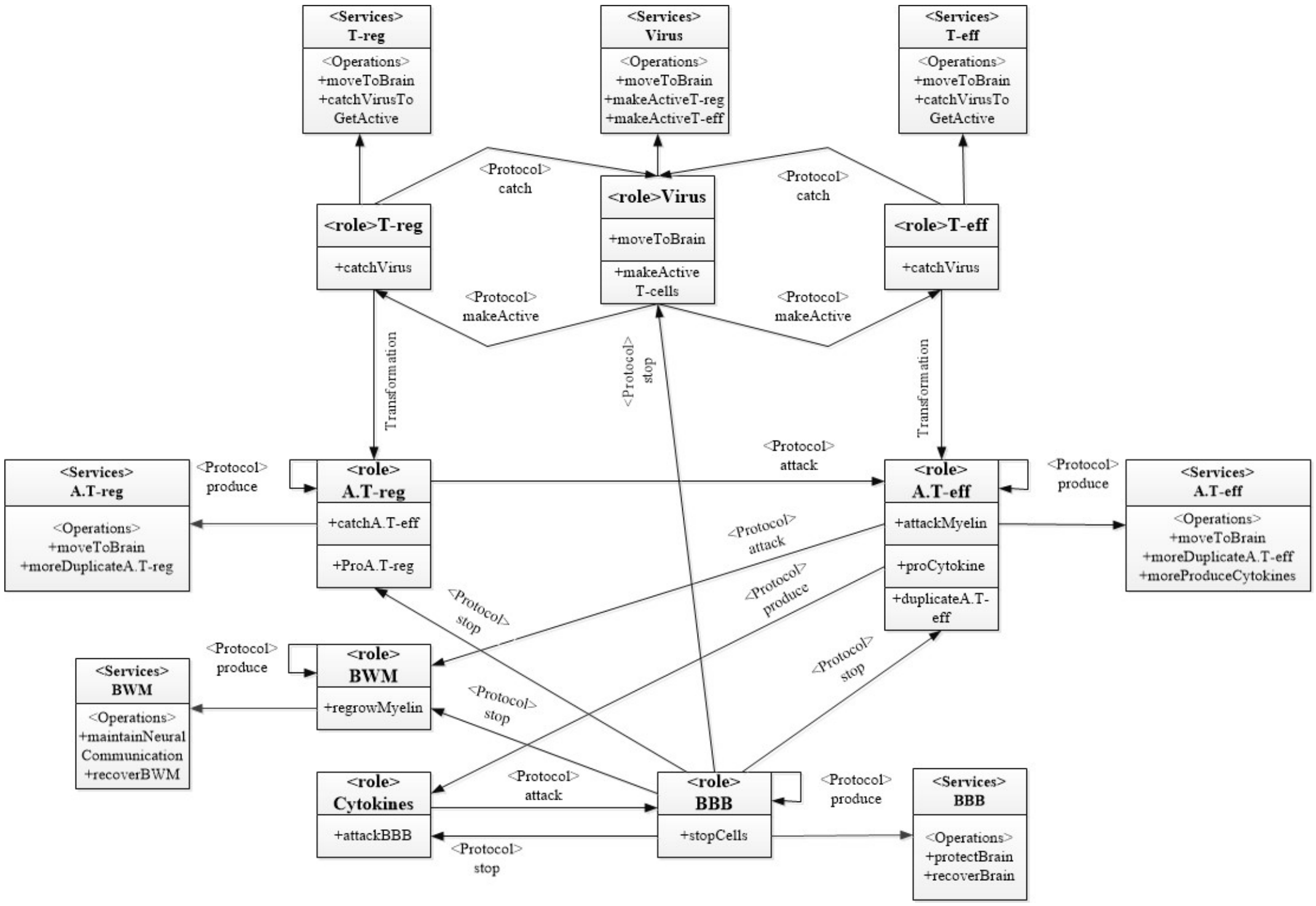

Figure 6: Agent and Service Model

In the MS disease model, there are two types of agents patches and turtles. The patches represent to brain organ (BBB, BWM) and turtles represent to T-Cells (T-eff, T-reg, A.T-reg, A.T-eff), Virus, Cytokines. In figure 6, all agents are defined with precise services.

### 3.3 TROPOS

The TROPOS is an agent-oriented software engineering methodology, which was proposed by Castro et al. in 2002. The name was derived from a Greek word "trope" which means that it is easily adaptable and modifiable[39]. The complete methodology and its implementation proposed for Media shop in the article [40]. The principle aims to propose this methodology was to overcome the semantic gap between the operational environment and the developed system. Since the structured and object-oriented software development methodologies only have the programming concepts, not organizational ones [40].

Basically, this methodology was founded on two features [41]:

1. This methodology was developed according to the agent's concept and the mentalistic notions of the agent such as goal, plan, and resource. Then these notions help the developers throughout the development process of the system.
2. This methodology introduces legitimate requirement specifications to give a crucial role to requirement analysis phase.

This methodology adopts *i** model, which was proposed by Yu in [42]. The i* model explicitly describes to the actor(can be role, position, and agent), actor dependencies and goals as primitive concepts for all models which will be developed in different phases of software development. Tropos more specifically covers four phases of software development 1) Early Requirement Analysis, 2) LateRequirement Analysis, 3) Architectural Design and 4) Detailed design. And in somehow way it also supports implementation phase [43][44][39].

### 3.3.1 Early Requirements

In the early requirement analysis phase, the developers focus on the identification of the problem domain. After the identification of problem, they study the existing organizational setting where the system to-be operated. The Study reveals that the core intention of requirement analysis phase is, derive a set of functional and nonfunctional requirements for the system to-be[40][43]. Both the early and late requirement phases share the same methodological and conceptual approach. The early requirement analysis phase analyzes and identifies the involved stakeholders and their intentions The stakeholders of the system to-be is modeled as social actors or agents, that depend on one another for plans to be performed, goals to be achieved, and resources to be shared. In MS disease model the stakeholders are cells who participate in disease cause. And the agent's intentions are modeled as the goals to be achieved.

In this phase, we will implement i* model, that consist of 1) strategic dependency model as actor diagram and 2) strategic rationale model. The rational model supports and describes the reason to fulfill its goals and to make a relationship with the other agents. The outcome of this phase is an organizational model. The organizational model includes all the involving actors, their dependency relationship with other actors and their particular goals[40]. However, the ultimate goal of this phase is to clearly understand the environment and context of the organization, where the system to-be will perform[41].

In the depth of modeling, Tropos depend on *i** notations as described in below.

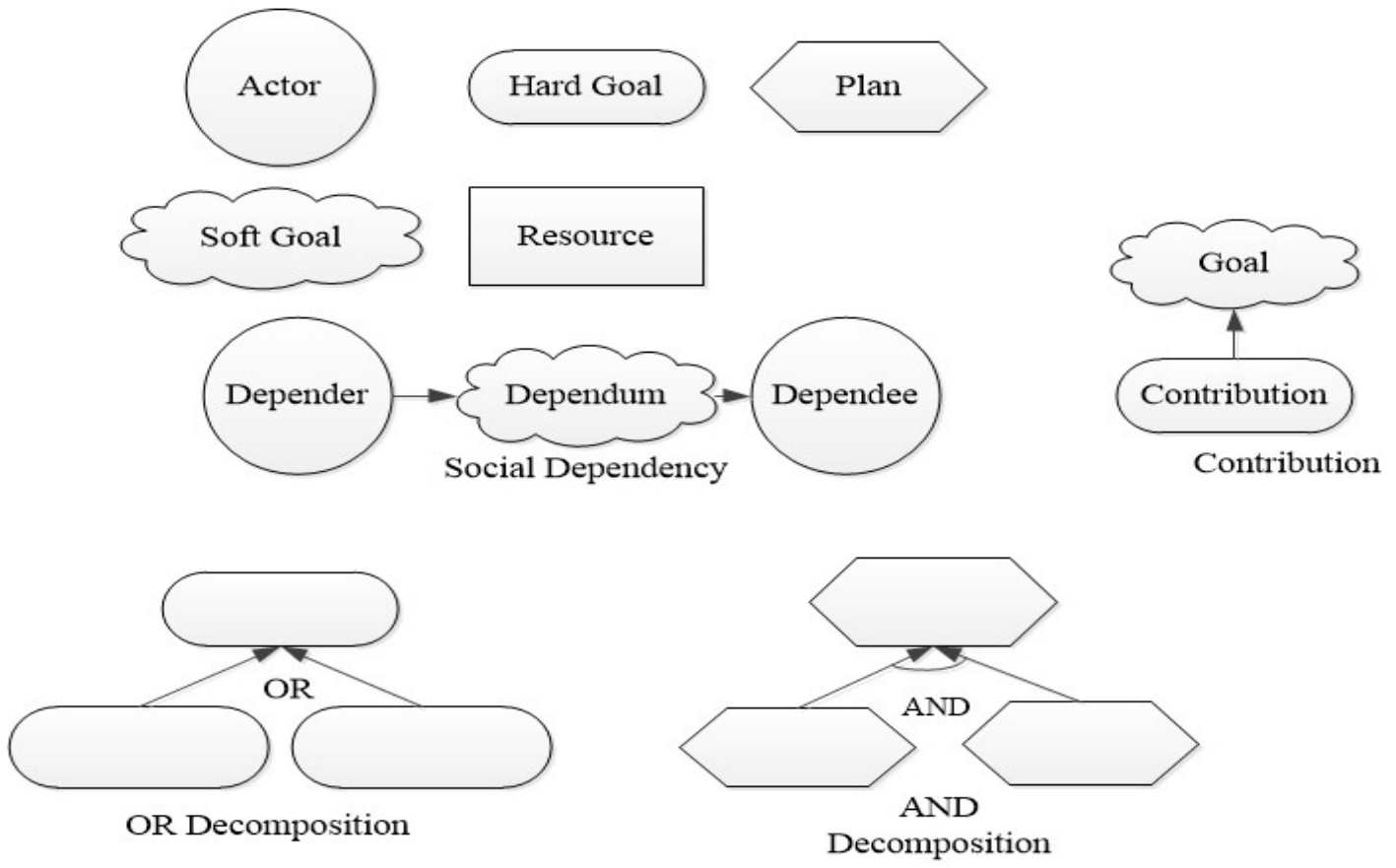

Figure 7: Notation Diagram

**Actor:** An actor is a representative of physical entity or position in the organization, a software entity, a social role. Who is responsible for goals to be achieved, resources to be shared, and tasks to be performed.

**Goal:** Dam and Winikoff states that goals represent actor's strategic interest in [45]. There are two types of goal: hard goal and soft goal. The hard goal represents functional requirement or specification of the system. However, soft goal have no clear definition for deciding whether they are achieved or not.

**Plan:** A strategy that is adopted to fulfill a specific goal or soft goal.

**Resource:** Resources are something such as physical or informational entity, which required for agent to fulfill specific tasks.

**AND/OR Decomposition:** The AND decomposition divides the main goal or task into more than one sub goal. In this case, all sub goals must be achieved to fulfill the root goal. In the OR decomposition, the main goal is divided into other alternative ways. In this case, the root goal can be achieved by any sub goal's completion.

**Mean-end Analysis:** In mean-end relationship a mean (in term of goal, resource, and plan) completely satisfy the root goal.

**Social Dependency:** The social dependency is an agreement between two actors: the depender and the dependee, to achieve their common goal. Their common goal is called the dependum.

In TROPOS implementation, we are considering each neural cell and brain organ as an agent.

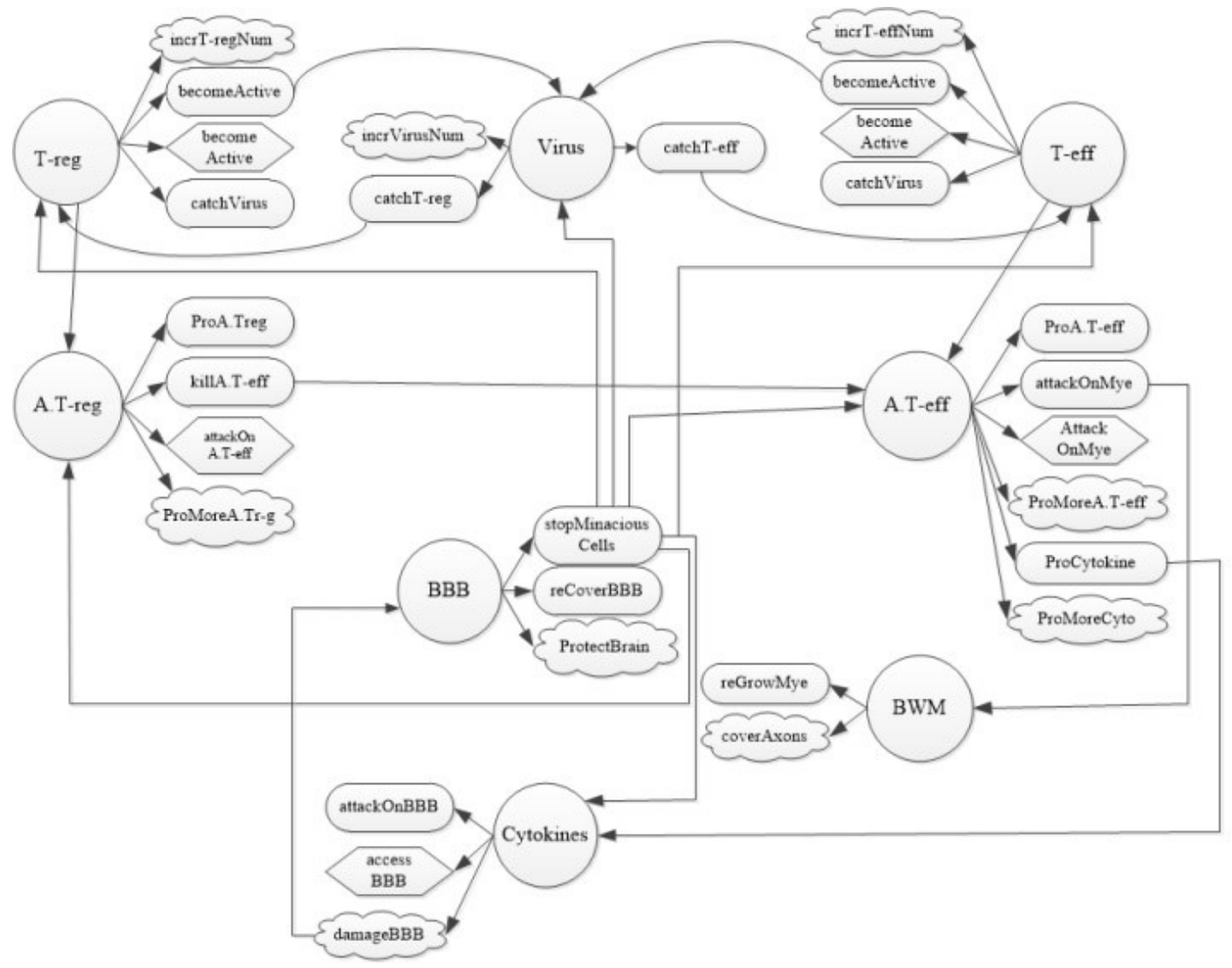

Figure 8: Actor Diagram

In the early requirement analysis phase of TROPOS, we identified 8 actors T-reg,T-eff, Virus, A.T-reg (active T-reg), A.T-eff (active T-eff), BBB (Blood Brain Barrier), BWM (Brain White Matter or myelin) and cytokine. Each role has its own goals, soft goals, resources, and plans. Actor **T-reg** has a goal "catch virus" and "get active" to perform its responsibility such as protect myelin by killing A.T-eff cells. To achieve goals T-reg depends on the virus. If it detects any virus in the brain, then it became active. As actors can depend on other actors by soft goals, resources and plans. The **T-eff** also depend on virus to complete its goal which is "catch the virus" and "become active". **The Virus** agent depends on T-eff and T-reg agents to make them active without knowledge that T-ref will slow down damage rate. **A.T-reg** has goals attack injurious agents as A.T-eff and kill them and in return produce A.T-reg. To fulfill its goals, it depends on A.T-eff agent. And the soft goal is to increase the amount of T-reg agents to slow down damage rate. **A.T-eff**'s goals are "attack on myelin" to damage brain as well as "produce A.T-eff and cytokines". To achieve its all goals it depends on BWM. After the attack on myelin then, it is able to produce A.T-eff and Cytokines. Its soft goal is increase brain damage rate by producing more A.T-eff and Cytokines. In fact, these agents play injurious activities in the brain. The **BBB's** goals are "stop agents entering into brain" and "repair damage part of the BBB". The intention and goal of BBB is to protect the brain by blocking all minacious agents.To fulfill goal this agent depends on all agents except BWM. The **BWM** agents' goal is to repair damaged brain by regrowing myelin. Its soft goal is cover axons to maintain neural communication between neurons. The **Cytokine** agents have a threatening behavior for BBB. It depends on BBB to achieve goal, its goals are "damage BBB" and "allow all minacious agents to enter into brain".

**Rational models**

The strategic rational model is a balloon like circle contain on four types of nodes task, goal, resource, soft goal, and two types of links task decomposition link and mean-end link. The rational model captures how an agent makes plan to fulfill its root goal and how it makes relation with other agents to fulfill the system's goal.

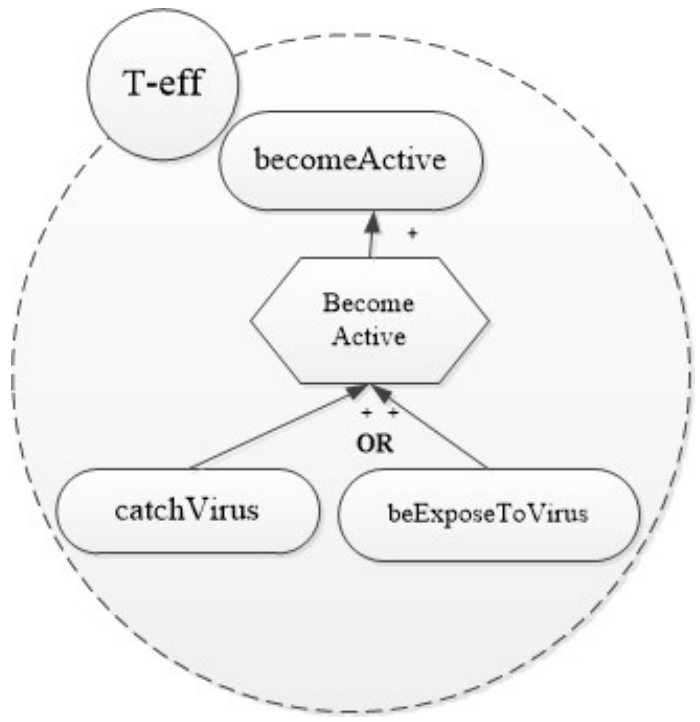

Figure 9: T-eff Rationale model

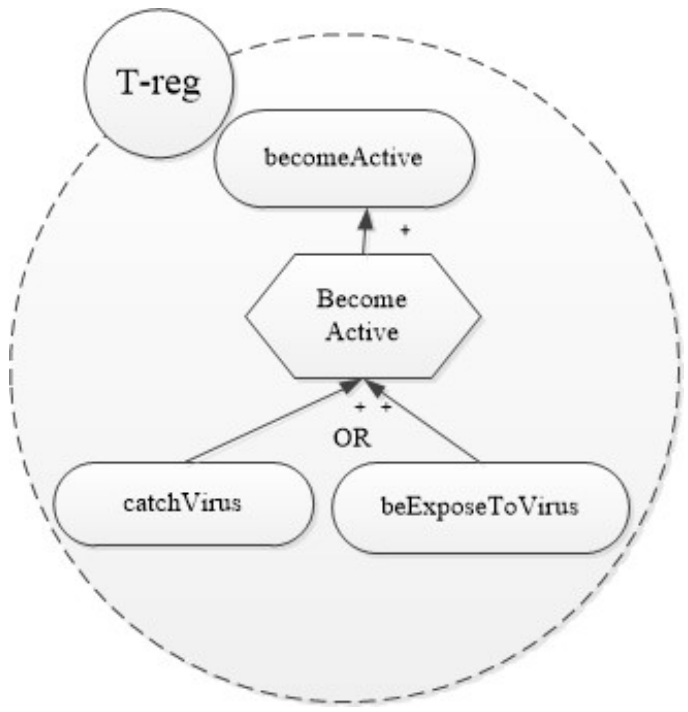

Figure 10:T-eff Rationale model

The goal of T-eff agent is to become active to perform its fundamental activities. For this, it plans either active by catching a virus or be exposed to the virus. OR decomposition represent to an option to fulfill a single aim and these options positively effect the plan. As a consequence plan positively supports the main goal. This is the same scenario in the case of T-reg agent.

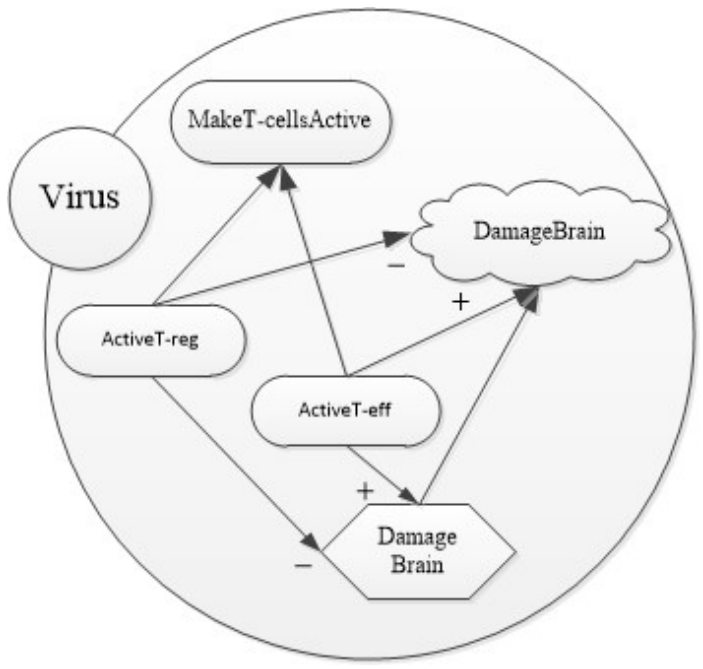

Figure 11 Virus Rationale Model

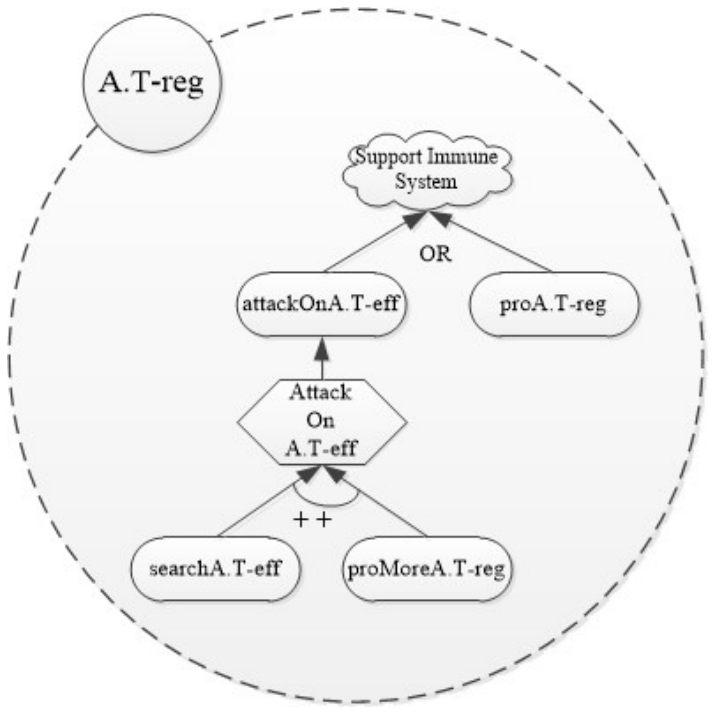

Figure 12: A.T-reg Rationale Model

In Rationale model virus have goals made T-cells active and damage brain. To achieve damage the brain goal, it has two options, either make more T-eff active or make less T-reg active. These options provide alternative ways to fulfill a single goal. The Active T-Reg goal has a negative effect on soft goal and plan.

In A.T-Reg Rationale Model the soft goal of the agent is to support the immune system. This goal is achieved by OR decomposition of killing A.T-eff or producing A.T-reg agents. The plan to kill A.T-eff can succeed by producing more A.T-reg agents or by searching more A.T-eff agents.

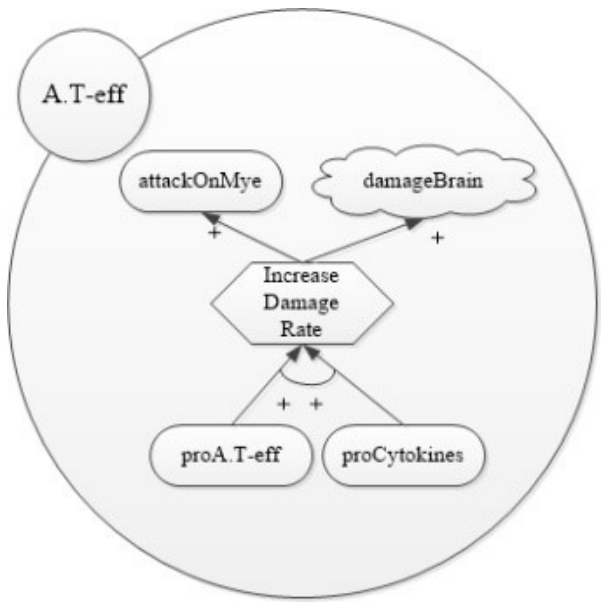

Figure 13: T-eff Rationale Model

The overall objective of A.T-eff is to damage the brain. To achieve this goal A.T-eff agent set a hard goal attack on myelin (BWM). For this goal this agent plans for increase damage rate by producing more A.T-eff and cytokine agents. The Plan is decomposed into two goals. These goals have a positive effect on the plan.

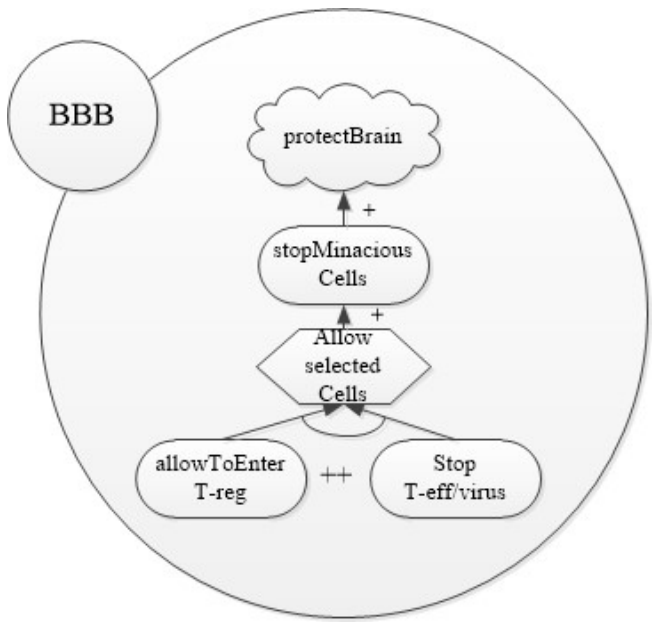

Figure 14: BBB Rationale Model

The main objective of the BBB is to protect brain from minacious molecules and cells. To achieve this goal this agent set a hard goal to stop entrance of minacious cells into the brain. To fulfill this goal, it plans to enter only T-helper cells T-reg and block T-eff, virus and dangerous molecules.

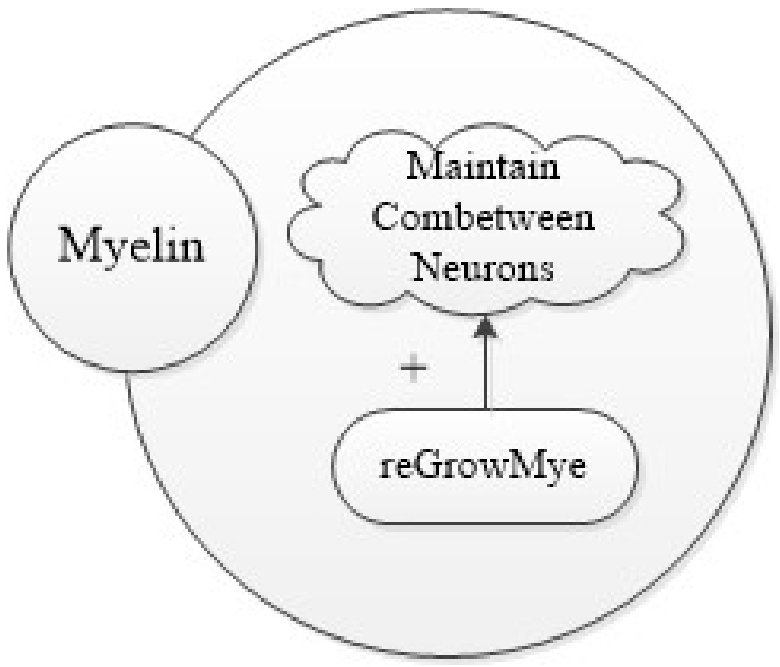

Figure 15: BWM (Myelin) Rationale Model

BWM or myelin maintains neural communication between neurons that would be its soft goal. To fulfill its duty It regrows myelin if any damage occurs.

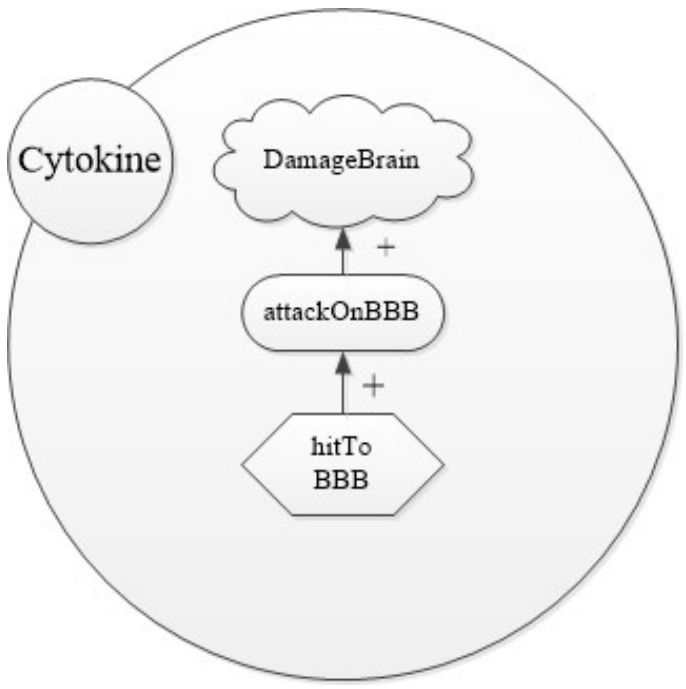

Figure 16: Cytokine Rationale Model

Cytokines are produced from A.T-eff agents. So they have the same goal as T-eff have "damage brain". But they achieve this goal from another way by damage BBB. To achieve this goal cytokine try to reach at BBB and hit it.

### 3.3.2 Late Requirements

In this phase, the developers identify all possible specifications of the system to-be and its operational environment. The output of this phase is the final specifications of the system, in the form of functional and nonfunctional requirement. This phase introduces the operational view of the system as an actor model. To iterate, the conceptual model of the system to-be is extended by introducing system actor in the overall system model and shows the dependency relationship between system actor and the other involved actors. These dependencies show the main reason of interaction of system agent with the other involved agents[43].

As Bertolini et al. state that, the system actor has the same features, which the other social actors have, in terms of social dependency and goals to analyze[46]. The addition of system actor helps the developers to recheck and conform that all involved actors and their dependencies are identified and the system actor can be more than one actor. To summarize that, the actual system actor comes into a picture of one or more actors, who helps the other actors to fulfill their goals.

In MS disease case MS is the system actor which describes the overall phenomena, how this disease appear and how other actors contribute to proliferate this disease.

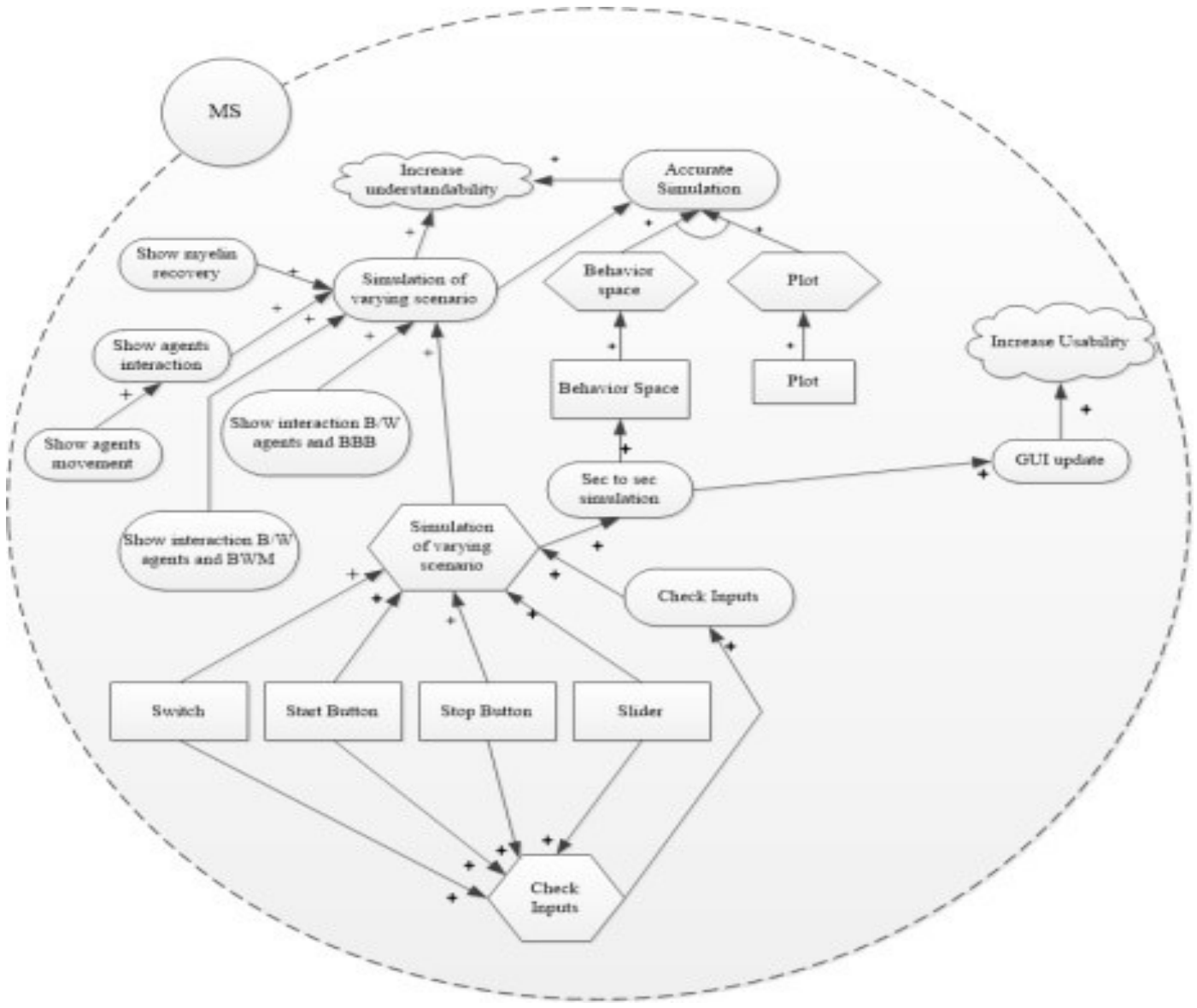

Figure 17: System Rationale Diagram

The system model controls overall system's environment and the involvement of actors. To iterate, system actor is responsible for the entire behavior of the system. The behavior of system actor is, how it responds and fulfill the requirements of other actors and the users of the system. In MS model the user actor will be, who will run the simulation against particular factors or agents attributes. The hard goal of the MS system actor is to show an accurate simulation of MS disease. This simulation shows, the varying behavior of agents, for example, it shows at which point, which agents interact, how they interact, the reason of agent's interaction and what was the effect of their interaction. To achieve this goal MS actor plan for behavior space and plot. This plan uses behavior space as a resource. Actually, system actor shows how and why other actors depend on it. The other hard goal of this system actor is to show a simulation of varying scenario. Many other sub-goals (like show myelin recovery, show agent's interaction, show agent's movement, show interaction b/w agents and BBB, show interaction b/w agents and BWM) participate to complete this goal. System actor also plans for this goal and plan succeed by Buttons, sliders, and switches. The soft goals are increasing understandability and usability. Understandability support users to operate the whole system easily understand the output results of the simulation.

### 3.3.3 Architectural Design

The architectural design phase expresses the overall architecture of the developing system. In architecture sub-system or agents are connected by data flow, control flow, or other dependencies[40]. In actual, this phase is an extended version of the actor diagram. In which each actor is introduced in detail according to its dependencies with the other actors.

Moreover, the architectural design phase performs a mapping on the system actors, and the outcome is a set of software agents. And each agent is characterized according to its specific abilities[43]. The system architecture helps the designers to understand how system components work together and how to constitute a relatively small, intellectually manageable model of system structure. The main objective of this phase is, to make a check that either all involved agents are identified? If new agent is identify then it should be incorporated in the system and also identify its control interconnections in the form of dependencies.

In MS disease model, new actors are not identified in architectural design phase; however the overall organizational structure is presented. This model includes rational diagram of each actor and the dependency between actors through goal, soft goal, plans, and resources.

Create rationale diagram of each agent and link them to show architectural design.

### 3.3.4 Detailed design

The detailed design phase further elaborates the architectural design and the behavior of its interconnected components[40]. The main concern of this phase is, actively deals the requirement specification document, and all the involved agents at micro and macro level. Bresciani and Perini suggest that mostly we use AUML activity diagram during detailed design for representing capabilities, plans and interaction of agents in the article [43].

Activity diagram captures the dynamic process of the overall system and each involving agent. The agent dynamics are the behavior of participating agents in the protocol and internal plans to achieve a specific goal. For capability modeling, the UML's activity diagram assist to model the capability from the stand point of a specific agent. For this purpose, the external events stars up the starting state of the activity diagram, action state models plan, transition arcs model finishing condition of action state and interaction, and the beliefs are models as object. Furthermore, for the plan and interaction modeling, the AUML's sequence diagram can be exploited.

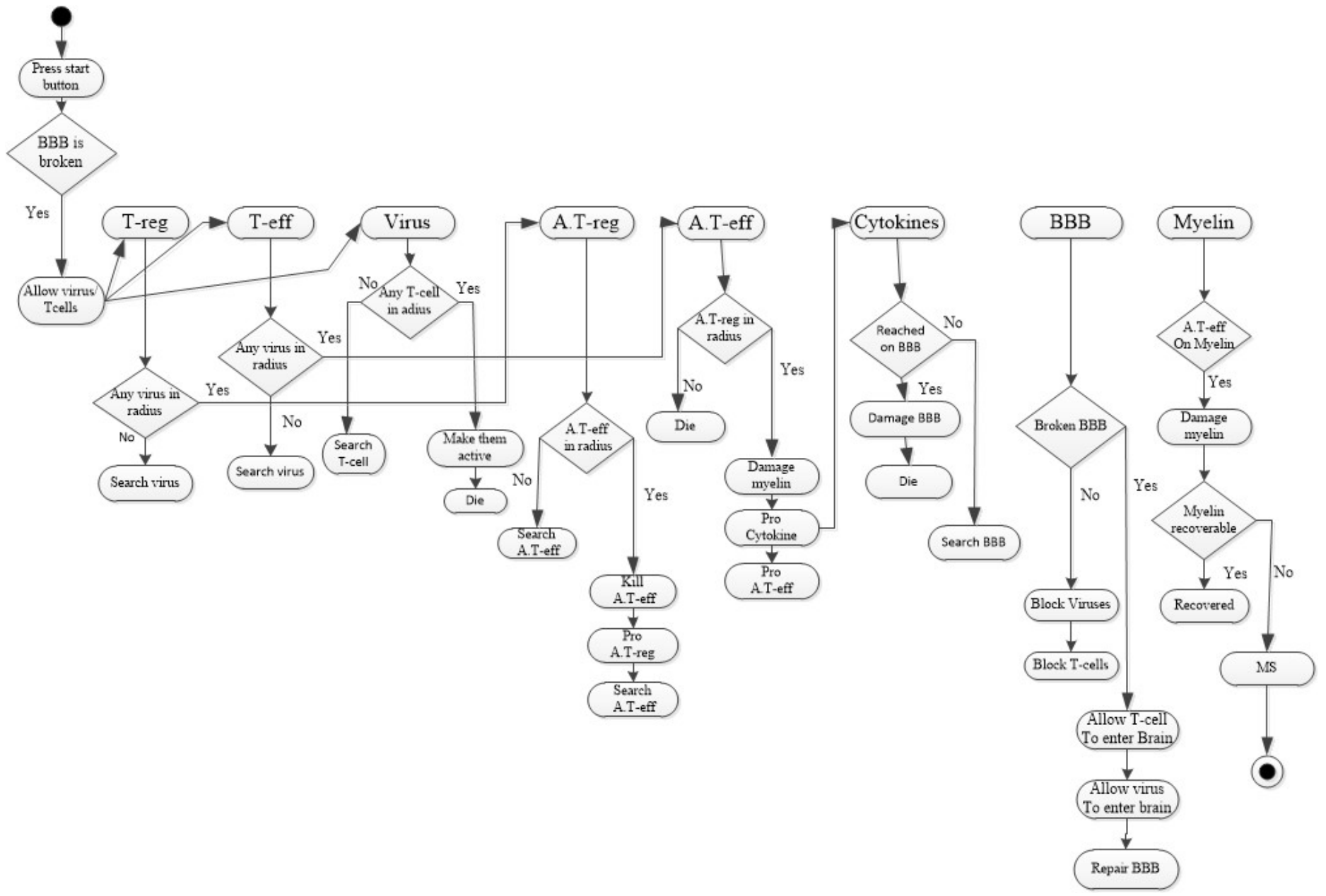

Figure 18: Activity Diagram

The activity diagram shows the general scenario of the MS disease model. In our developed model the start button is “go” button which act as external event to start the simulation. When users, press start button, then simulation starts. Go button checks the condition either BBB is broken, then agents are allowed to go into the brain. In brain T-cells (T-reg, and T-eff) search for viruses. If they find a virus in their radius, then they change their state from inactive to active else they search for viruses. In the same way, virus search for T-cells to make them active and die. In the case of A.T-reg, it searches for A.T-eff. If it finds any A.T-eff agents in its radius, then it kills them and produces a duplicate of A.T-reg agent. The A.T-eff agents search for BWM, if they find BWM then, they damage BWM and produce duplicate of A.T-eff, Cytokines and at the end they die . The cytokine agent moves randomly and if it finds BBB then it damages BBB and die. The BBB agent checks a condition that if it is damaged, then allow agents to move towards brain otherwise block them. And with time BBB try to recover from damage. The BWM checks a condition if any A.T-eff attacks on myelin then it became damaged and checks another condition if the damaged portion is recoverable then recovered otherwise remain damaged. The damaged portion shows the severity of MS in the model.

### 3.4 MaSE

At the very first time, MaSE methodology was proposed by Deloach in 1999 in the article [47] and after that, the improved version of MaSE methodology was proposed by Deloach, Matson and Li in 2003 for the development of a team of rescue robots which are autonomous, heterogeneous searcher, and rescuer [48]. Several attempts have been made at creating methodologies and their tools for building autonomous, heterogeneous, distributed and complex dynamic systems. However, mostly the proposed methodologies and their tools have focused on either the agent architecture, or the methodology lack of sufficient details to adequately support the designing of complex systems[49].

The significant purpose of the development of this methodology was, MaSE should be independent of any particular agent-based system architecture, specific agent architecture, programming language, and precise agent communication framework. Then the developed methodology proved that, the MaSE seemed a good fit for the cooperative robotic systems[48].

The main objective of this methodology is, guide a system developer throughout the system development process, by following the set of interrelated system model[48][50]. The previous research on intelligent agent has focused, on the structure and the capabilities development of an individual agent. Now, researchers have realized that to solve the complex system's problems,  agents coordination is mandatory for the heterogeneous environment.

The one another constructive aspect of this methodology is, it has its own developmental tool "agentTool". Same as the MaSE methodology, the agentTool is also independet of the particular agent architecture, the agent's programming language, and the agent communication language. The analysis and design phases perform a transformation that shows how to derive new models from the existing models. The MaSE is implemented after gathering requirements, which further divided into sub-goals as defined in the article[49][51][50].

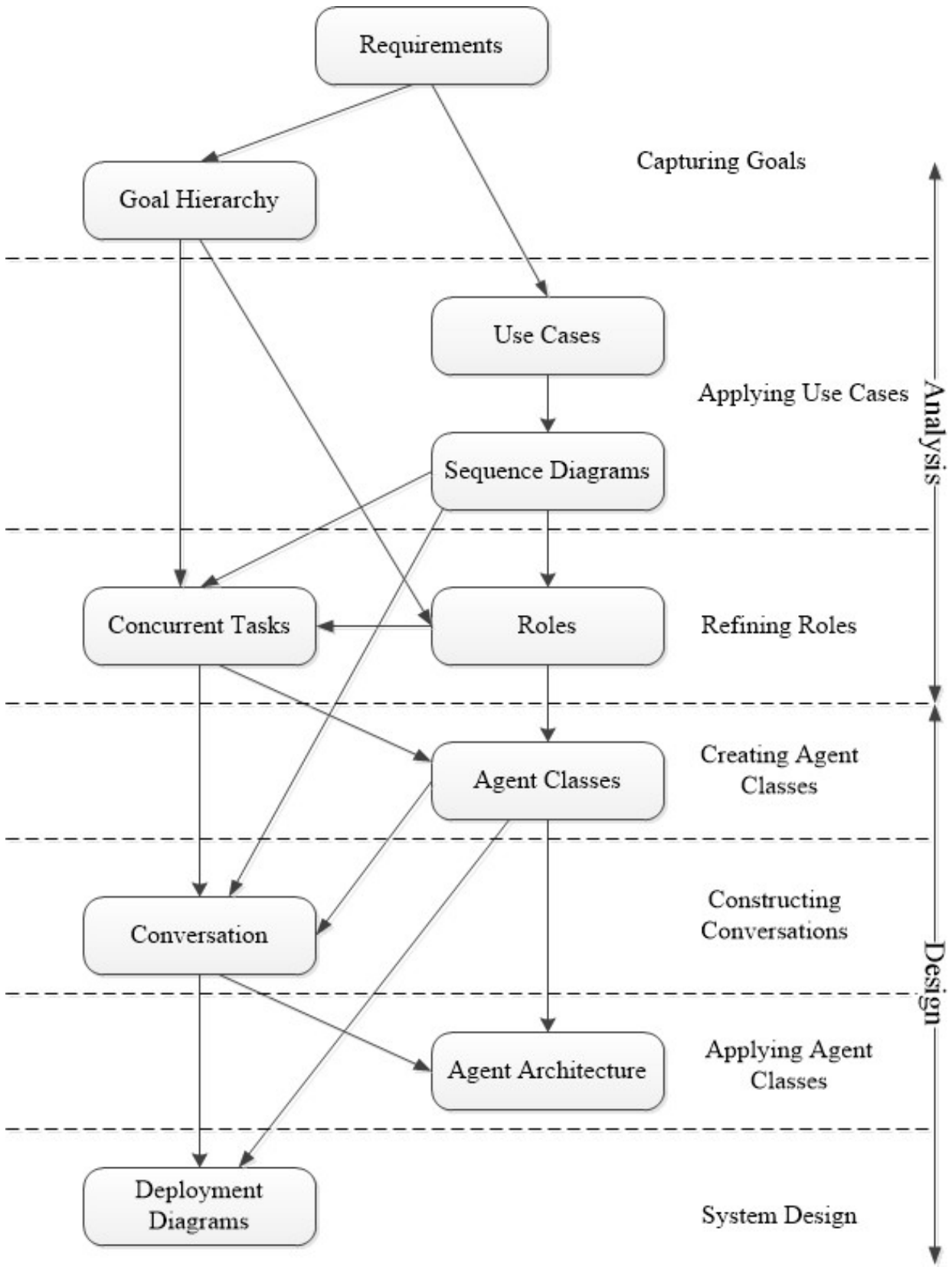

Figure 19: MaSE Methodology Models [49].

### 3.4.1 Analysis Phase:

The analysis phase is further divided into three precise phases, such as

1. Capturing Goals:

This phase develops a goal hierarchy of the system to-be, in which each goal has system level objectives. The developers take system requirements and organize them into a sequential set of system goals. Simply, The developers collect all the requirement and set them into hierarchy of basic goals of the system to-be. We developed the goal hierarchy model of MS disease case study, as researchers implemented MaSE in different scenarios. [48][49]. Goals must be identified through initial system context, which gives a starting point to the analyst for system analysis.

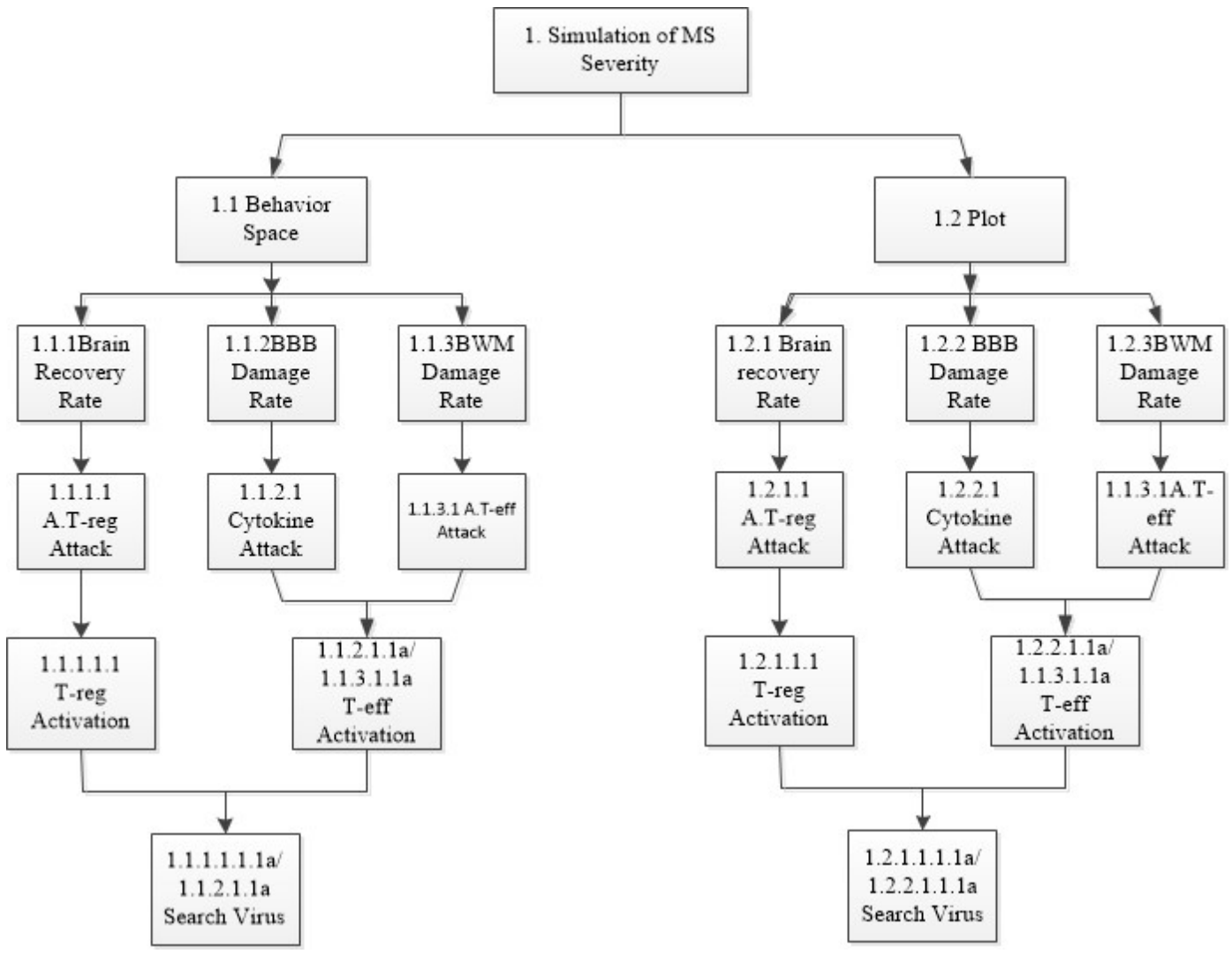

Figure 20: Goal Diagram

The figure 20 shows that, the main goal of the MS disease model is to simulate the severity of disease. This cental goal is further decomposed into sub-goals. However, each goal should always be defined as a system level goal.

2. Applying Use Cases:

This phase develops Use Cases and sequence diagram. The use cases of a developing system explain the complete scenario that a system would perform in the real operating environment. To put it simply, a use case of a system explains the sequence of all working events that must be performed by the developed system, these events may be its failure and hanging events. For the development of a flawless system, the analyst should develop enough use case which covers every possible event that can occur in the system by using different data and event scenario.

The sequence diagram helps the developer to understand the overall system scenario and find out all involved agents with their communication paths. Moreover, the sequence diagrams provide assistance in capturing the use cases of the system. Furthermore, these use cases would use later in the analysis phase, in which a particular role would be assigned to a specific goal. In general, a single sequence diagram is considered as a representative of each use case.

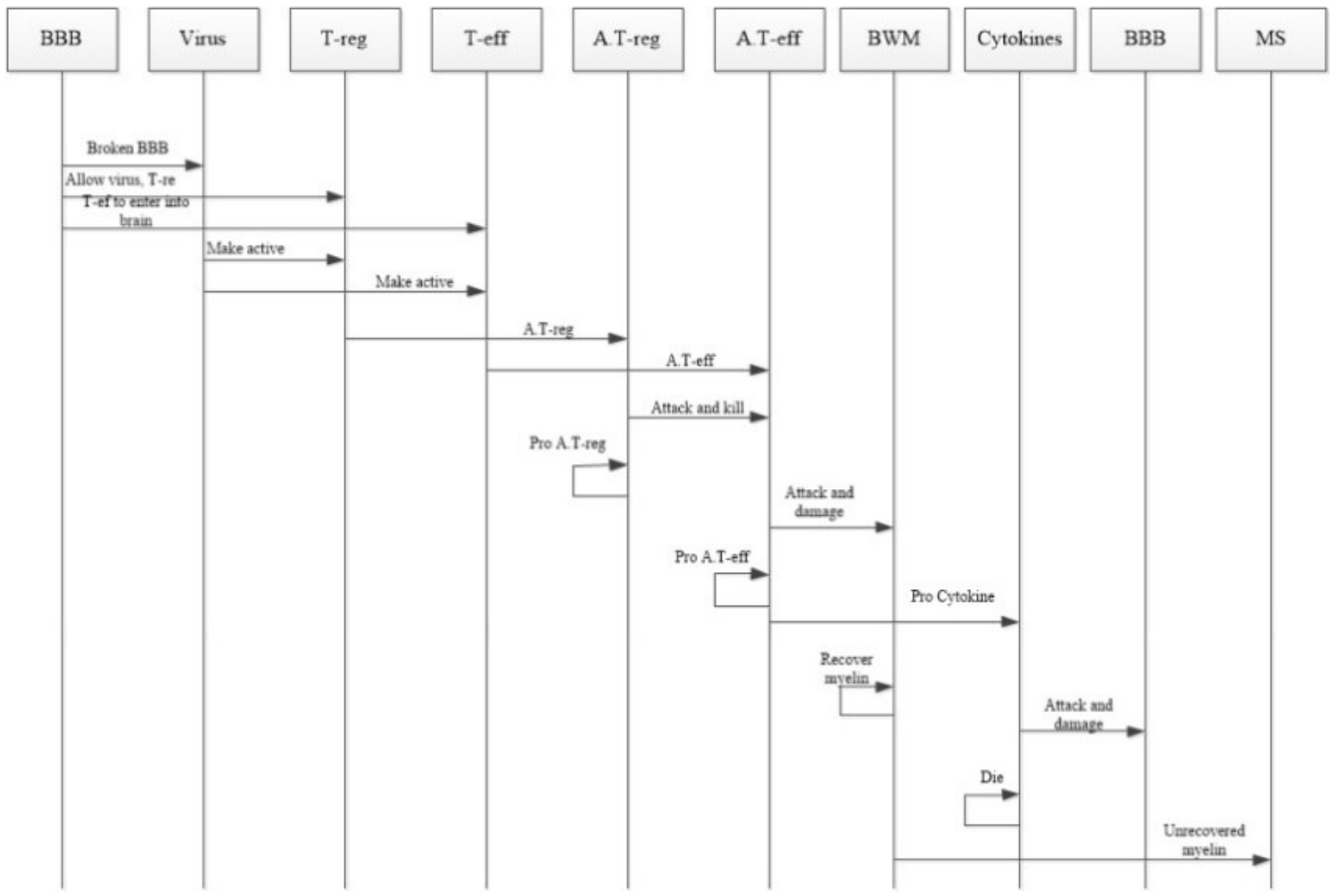

Figure 21: Sequence Diagram

In the sequence diagram, normally roles are represented by rectangular boxes which are placed at the top place of the diagram and the connecting arrows of the roles represent to ongoing events among the roles. And the time is assumed as a sequence of events from the top of the sequence diagram to the bottom.

In MS disease' sequence diagram we identified nine agents. To precisely represent the sequence, we used BBB two time as an agent. If the BBB is broken then, it informs to T-Cells and virus. When virus finds T-cells around then it makes them active. An A.T-eff damage the BWM and in return produce Cytokine and duplicate A.T-eff. When A.T-reg finds an A.T-eff in around then it attacks and kills. In return A.T-reg duplicate itself. In the case of damage, BWM and BBB recover themselves. Cytokine search BBB and damage it.

3. Refining Roles:

The main objective of refining role model is, transform the structured goals and sequence diagram into the final rules of the system to-be. Furthermore, these roles provide a foundation the agent class model and became system goals during the design phase. Refining role is one of the important steps of the MaE methodology because, the system goals would be satisfied with the only one way, if every single goal is considered as an identical role, and every role is played by an identical agent class. This phase develops concurrent tasks and goal hierarchy models. In general, the transformation of goals to roles is a one-to-one mapping. However, a single role may have multiple goals.

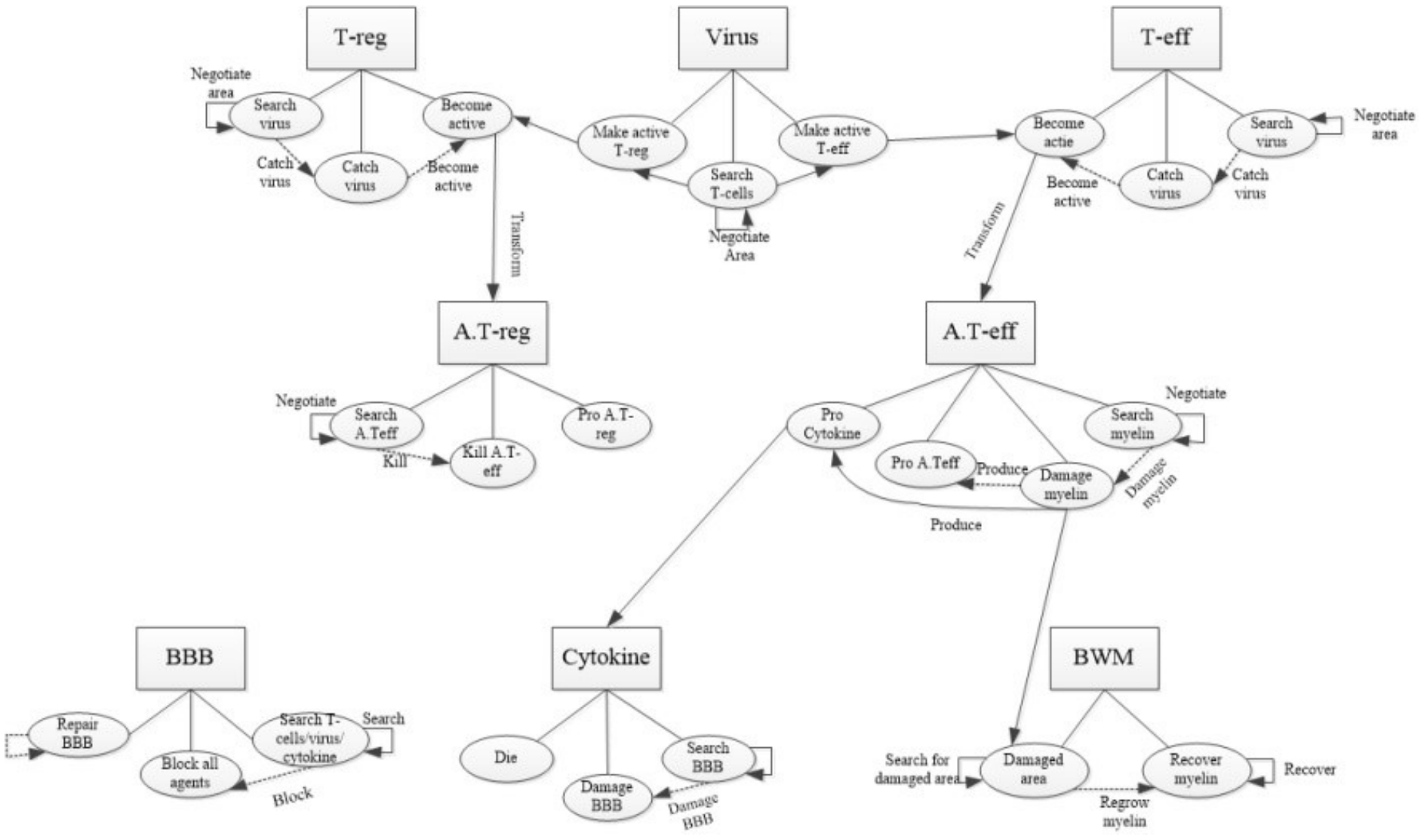

Figure 22: Refined Role Model

In the MS disease model we identified eight roles, which are shown in rectangle boxes. According to Agent definition we select those entities as an agent who have some responsibilities or activities to fulfill the system requirement. All selected roles have multiple goals as shown in figure 22.

### 3.4.2 Design Phase:

1. Creating Agent Classes:

This phase develops agent class diagram, which consists of agent classes and the conversation between them. The agent class model clearly depicts the overall organization of agents. Additionally, the agent is an actual instance of an agent class, and class is a template for a single type of agent in the system. Additionally, the agent is an actual instance of the agent class, and class is just a template for a single type of active role in the system to-be. In this phase, the agent classes are developed in terms of the roles, that must be played. And the conversation is developed as protocols, in which they must participate.

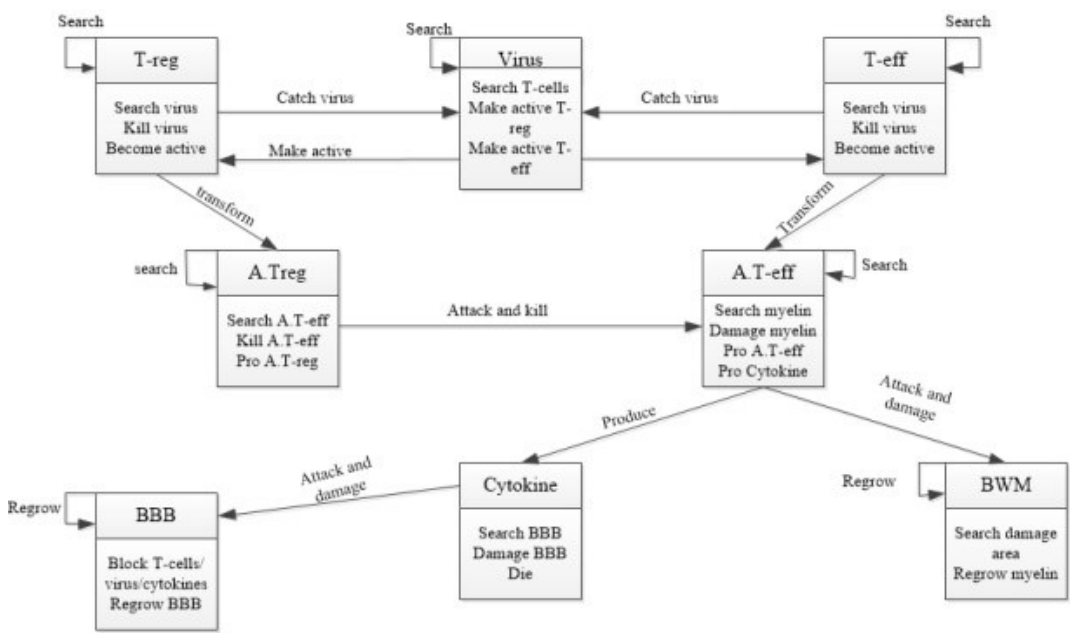

Figure 23: Agent classes Interaction

2. Constructing Conversation:

This phase develops conversation diagram of each software agent. The MaSE conversation model defines the conversation of two software agents as a coordination protocol[1][2]. In a conversational event, two agent classes participate, one is an initiator and the second is the responder. The actual purpose of the conversational model is to define the purpose and the detail of the conversation. When a software agent wants to communicate with the other agent then it sends the initiator message to the partner agent to start the communication.

When the responder agent received a conversation request, then it compares it with all its permitted conversations. If it finds a match, then it performs the required task that can be a request of resource, data share, or a coordination request. Otherwise, the responder agent assumes that the conversation message is a request to start a new conversation. In this case, the responder compares the request with all its possible conversations in which this agent can participate and all the agents' with whom it can make the conversation. If it finds a match, it begins a new conversation.

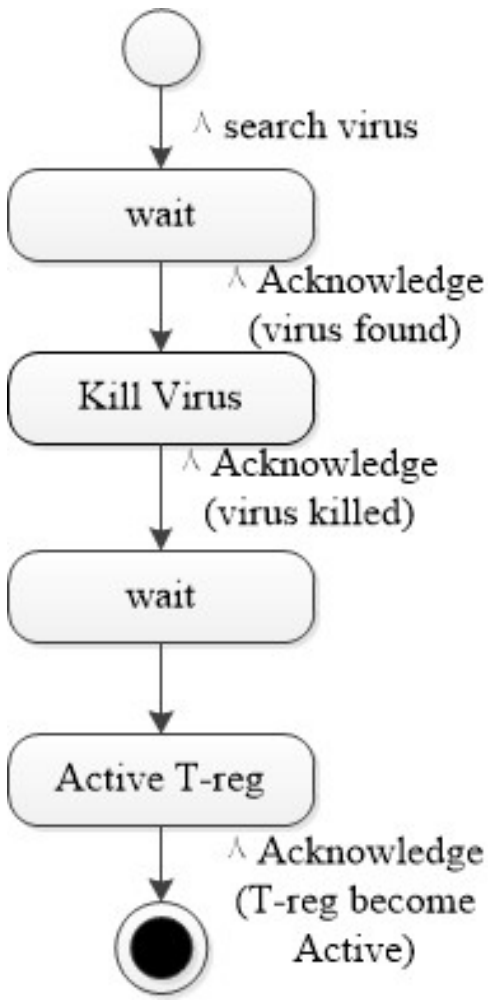

Figure 24: T-reg Conversation model. This conversation take place between T-reg and Virus.

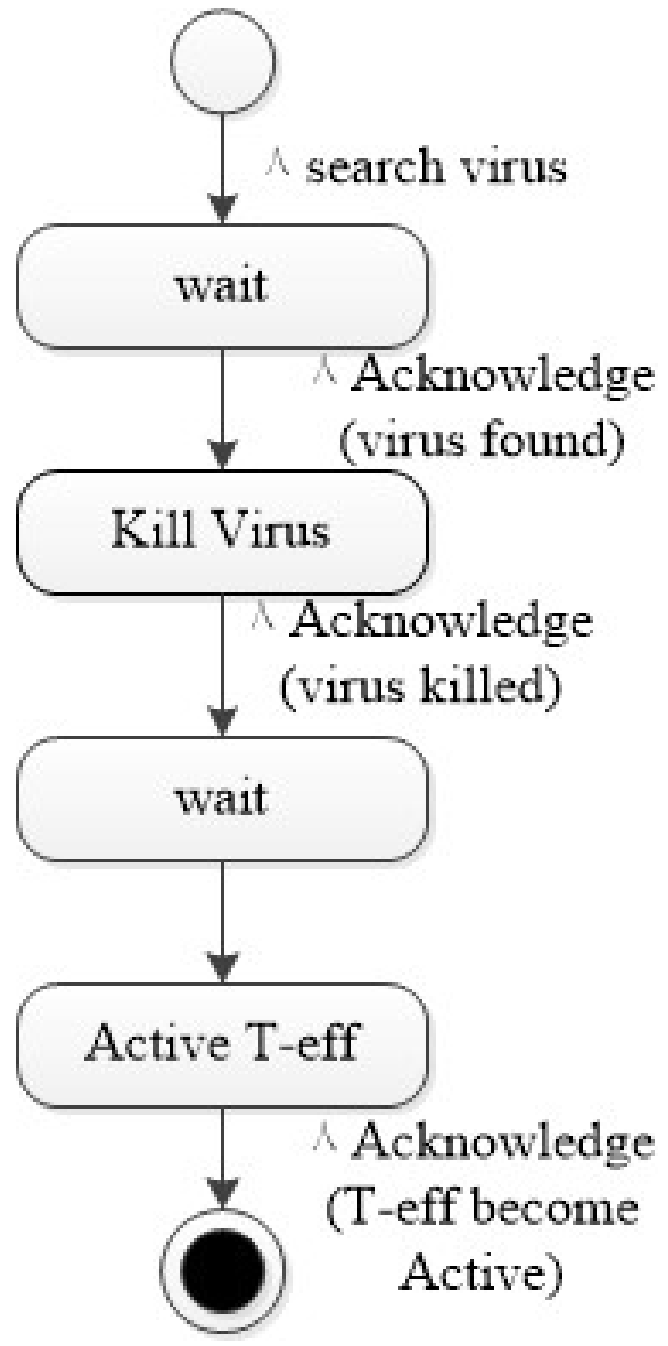

Figure 25: T-eff conversational model with Virus.

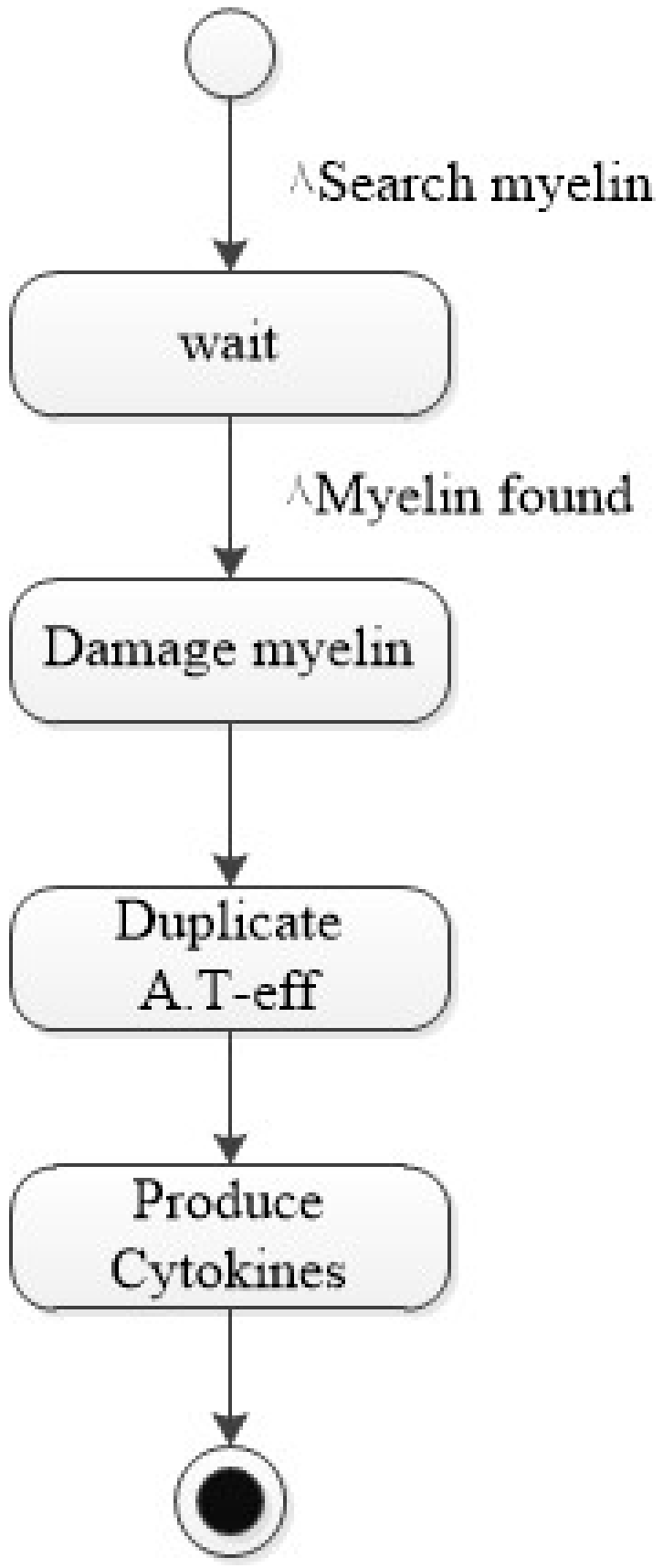

Figure 26: A-T-eff Conversational Model

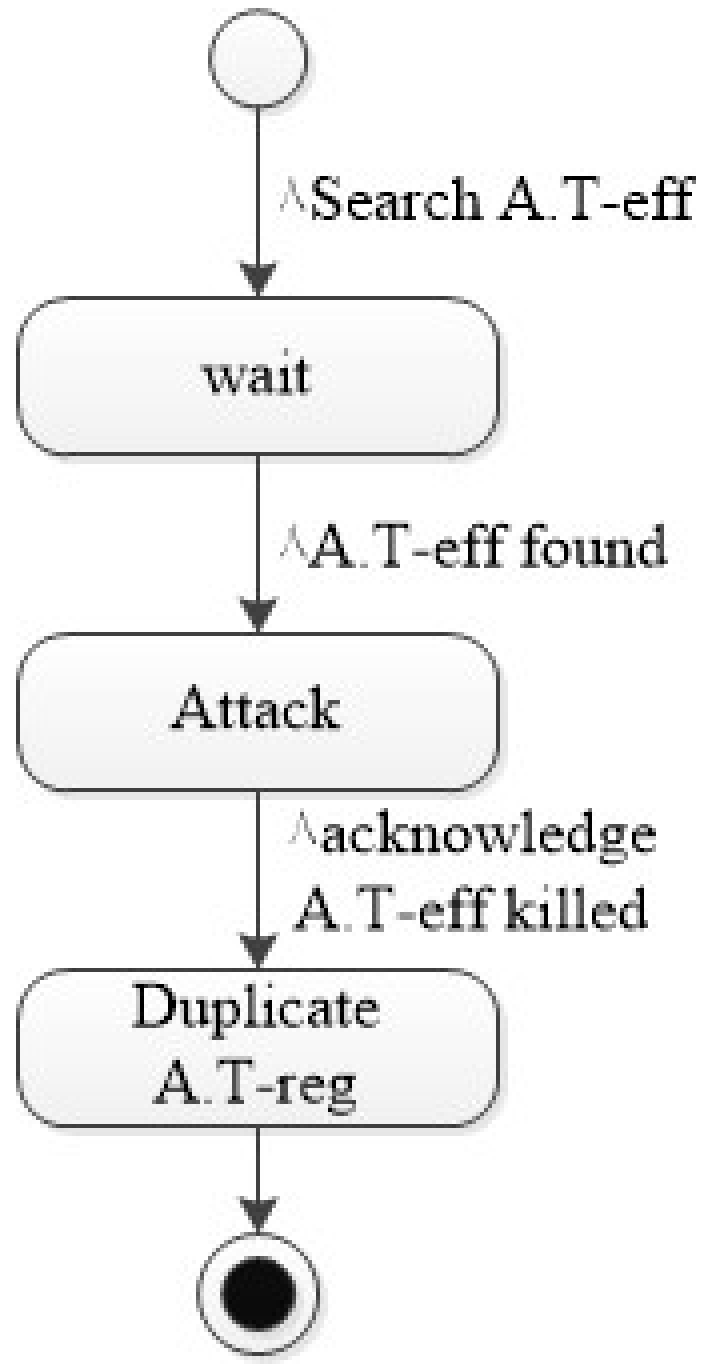

Figure 27:A.T-reg Conersational Model

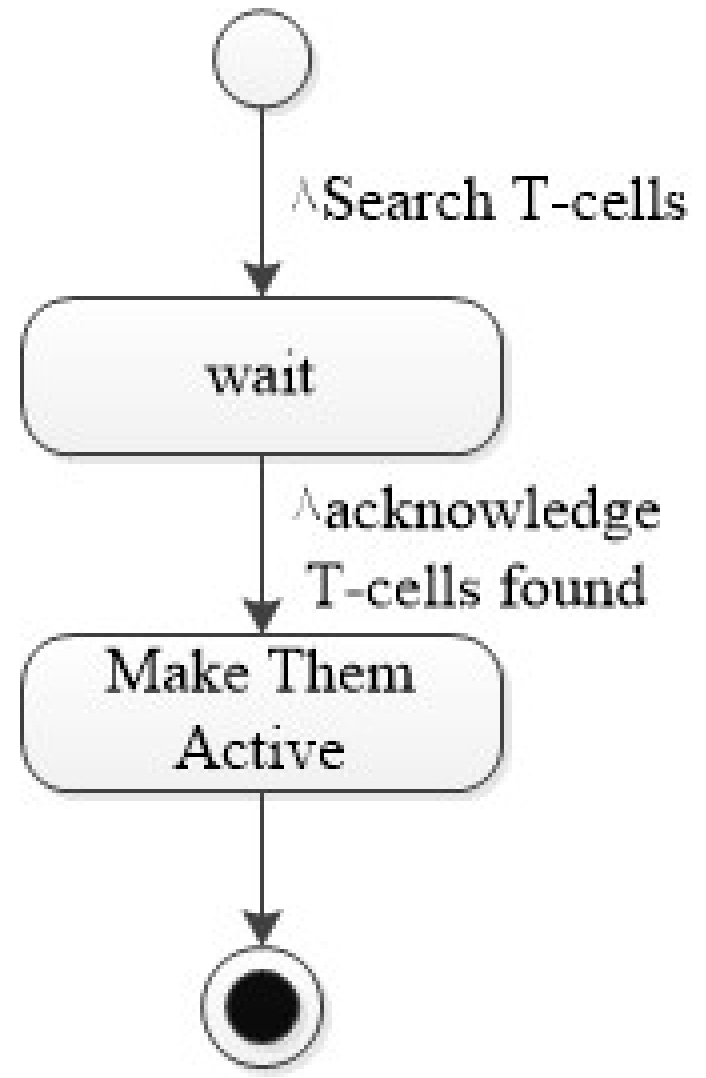

Figure 28: Virus makes conversation with T-cells to make them active.

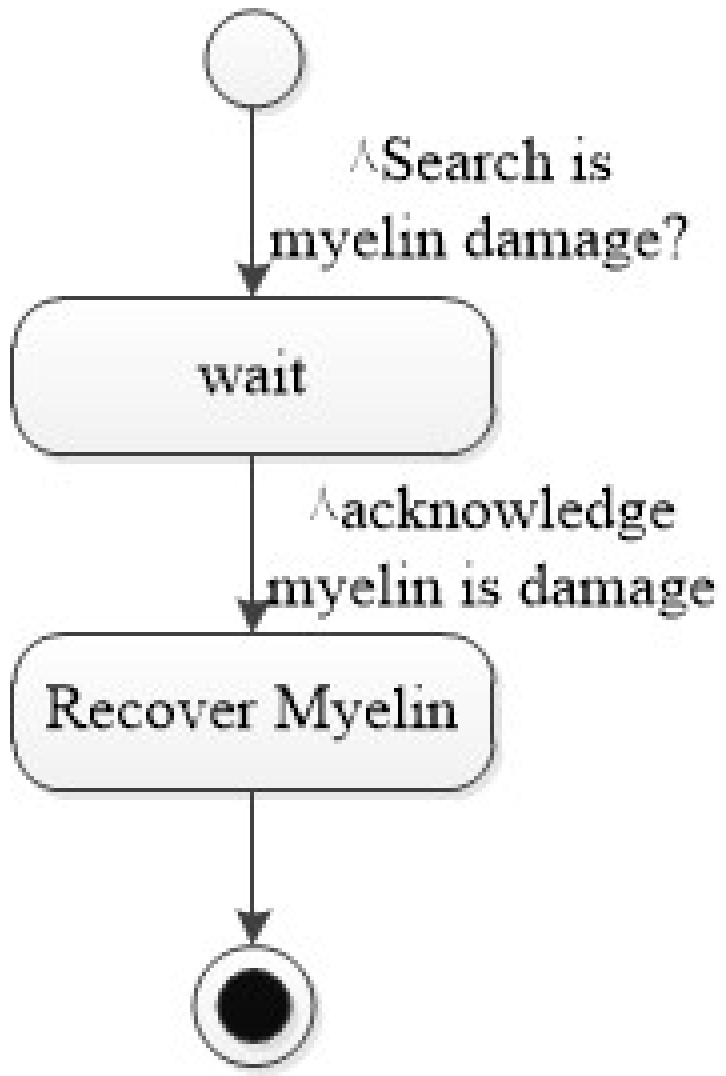

Figure 29: BWM makes conversation with itself.

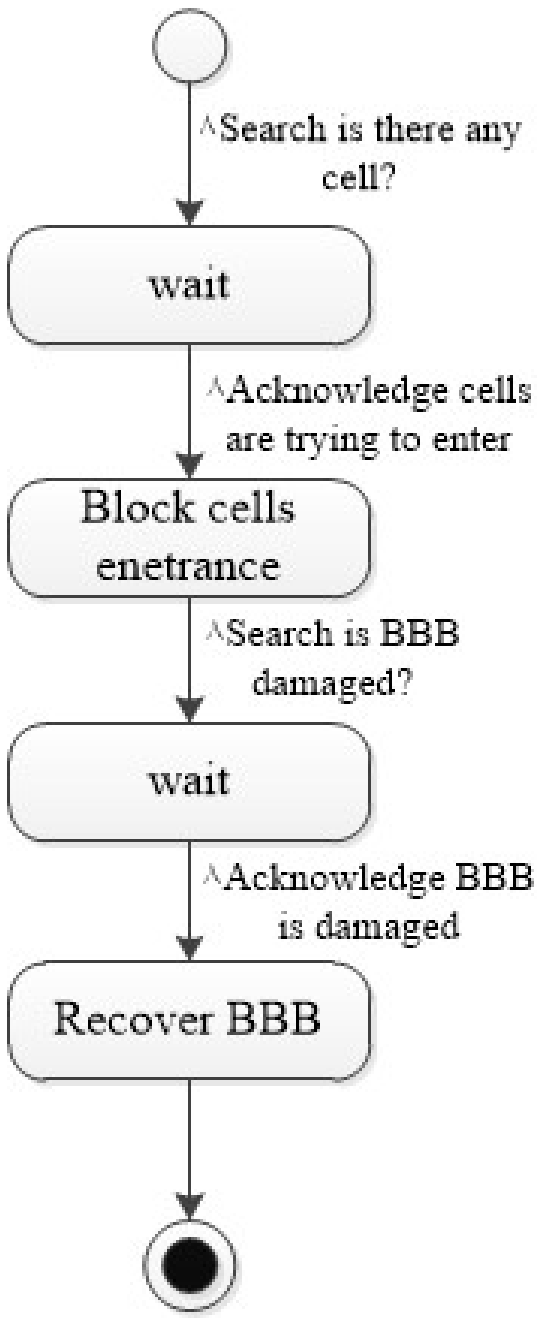

Figure 30: BBB's conversational Model.

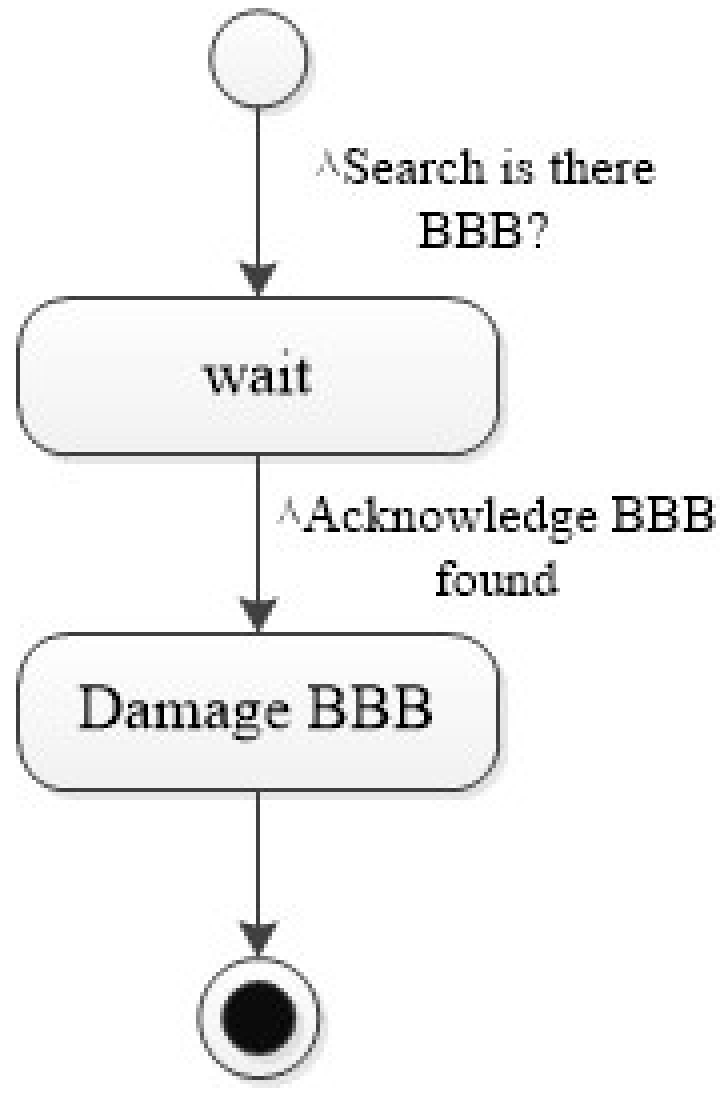

Figure 31: Cytokines Conversational Model.

3. Assembling Agent Classes:

This phase develops the architecture diagram of agent software by integrating different agent classes into a single model. This development phase is accomplished by accomplishing two sub-steps: 1) by defining the agent architecture, 2) by defining the all components that makeup the agent architecture

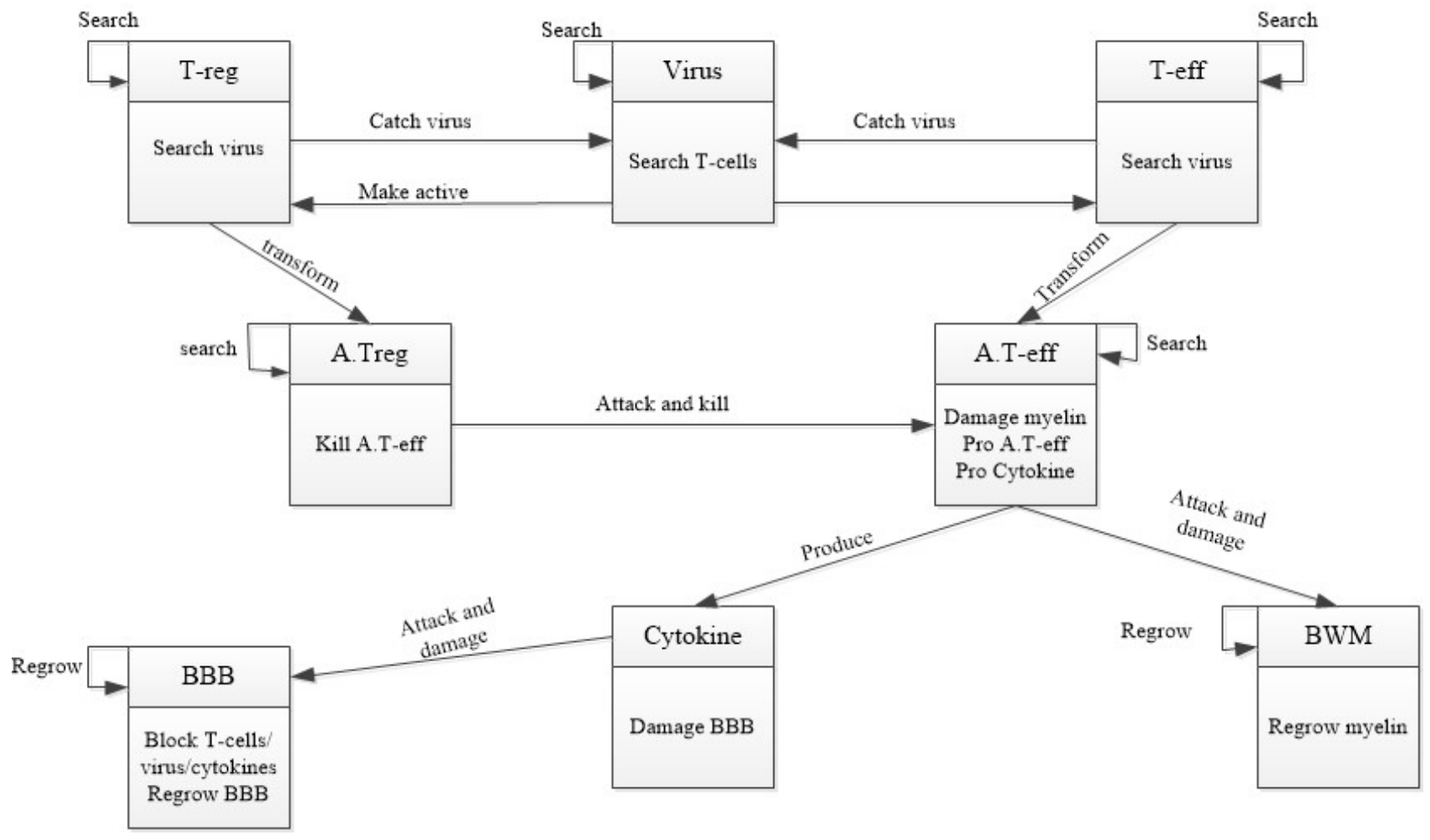

Figure 32: Assembling agent classes.

4. System Design:

Generally, this phase develops the deployment diagrams. This phase receives the agent classes of each involved agents and introduces them as actual agents of the system to-be. And go through a data flow diagram to demonstrate the location and control flow of each agent.

Moreover, this final step of MaSE performs the configuration on the involved agents and their responsibilities by developing deployment diagrams. Which shows, the total numbers of agents, their location and agent's type within the system.

## 4. Comparative Analysis

As biological researchers have no clear idea about engineering methodologies. And the other problem is, at present, more than two dozen AOSE methodologies exist. Therefore, the process of choosing the right methodology for their specific problem agonize them. To escape from this tension they just model their problem without any guided methodology. So this hastens development create some serious problems such as missing essential information about the problem, the wrong decision of choosing development tool, the doubt on systems authenticity. To overcome all these mentioned nervousness, in this work, we applied three well known AOSE methodologies: GIA, TROPOS, and MASE on MS brain model. In this work we performed a comparative analysis of these methodologies to analyze that, which methodology is more suitable for biological models development.

For comparative analysis, we used a framework that focuses on four major facets of methodologies: 1) notations and modeling techniques, 2) concept and properties, 3) development process and 4) pragmatics. To be very clear, that our work would not attempt to state that which the right methodology is. Rather, it examines the existing methodologies to advise researchers to choose right methodology according to their specific scenario. A lot of comparative analysis has been done in literature as shown in the below table.

| Ref | Paper Title | Author | Journal | Year | I.F 2017 | Compared methodologies |
|---|---|---|---|---|---|---|
| **[52]** | "Multi-agent approach for cancer automated registration" | Sanislav et al | Control Engineering and Applied Informatics | 2010 | 0.695 | 1. GAIA<br>2. PASSI<br>3. INGENIAS<br>4. MASE |
| **[53]** | "A Methodology to Evaluate Agent Oriented Software Engineering Techniques" | Lin et al. | Proceedings of the 40th Hawaii International Conference on System Sciences | 2007 | | 1. Tropos<br>2. GAIA<br>3. MASE |
| **[54]** | "A framework for the evaluation of agent-oriented methodologies" | Abdelaziz, Elammari and Unland | Innovations'07: 4th International Conference on Innovations in Information Technology, IIT | 2008 | | 1. GAIA<br>2. MASE<br>3. HLIM |
| **[55]** | "A comparative analysis of i* agent-oriented modelling techniques" | Grau et al. | In Proceedings of The Eighteenth International Conference on Software Engineering and Knowledge Engineering (SEKE'06) | 2006 | | 1. Tropos<br>2. GBM<br>3. ATM<br>4. BPD<br>5. RiSD<br>6. PriM |
| **[56]** | "Agent-Oriented Methodologies - Towards a Challenge Exemplar" | Yu and Cysneiros | Proceedings of the International Bi-Conference Workshop on Agent-Oriented Information Systems | 2002 | | They just proposed question to evaluate methodology |
| **[57]** | "On the evaluation of agent oriented modeling methods" | [56] | Proceedings of Agent Oriented Methodology Workshop | 2002 | | They just proposed terms for methodology evaluation |
| **[58]** | "ASPECS: An agent-oriented software process for engineering complex systems" | Cossentino et al | Autonomous Agents and Multi-Agent Systems | 2010 | 2.103 | 1. PASSI<br>2. INGENIAS<br>3. ANEMONA<br>4. GAIA<br>5. ROADMAP<br>6. TROPOS<br>7. PROMETHEUS<br>8. ADELFE<br>9. ASPECS |
| **[33]** | "A COMPARISON OF THREE AGENT-ORIENTED SOFTWARE DEVELOPMENT METHODOLOGIES : MASE , GAIA , AND TROPOS" | Jia et al. | In Information, Computing, and Telecommunication, 2009. YC-ICT'09. IEEE Youth Conference on 2009. | 2009 | | 1. GAIA<br>2. TROPOS<br>3. MASE |
| **[59]** | "Evaluating how agent methodologies support the specification of the normative environment through the development process" | Garcia et al. | Autonomous Agents and Multi-Agent Systems | 2015 | 2.103 | 1. OMASE<br>2. OPERA<br>3. TROPOS<br>4. GORMAS |
| **[60]** | "METHODOLOGIES AND | F. | Springer Science \& | 2009 | | 1. GAIA |
| | SOFTWARE ENGINEERING FOR AGENT SYSTEMS” | Bergenti, Federico and Gleizes, Marie-Pierre and Zambonelli | Business Media | | | 2. TROPOS<br>3. MaSE |

### 4.1 The Proposed Evaluation framework

The structure of the proposed analysis framework is taken from the study [60] however, the evaluation attributes are taken from different literature. The framework is divided into four subparts. 1) concept and properties, 2) Notations and Modeling Techniques, 3) Development Process, 4) Pragmatics. The proposed evaluation framework is based on feature analysis attributes. These attributes evaluate feature of each examined methodology from different aspects. Before implementing the evaluation framework on methodologies, first of all, we will discuss in brief the main objective of each section of the framework.

#### 4.1.1 Concepts and properties

The “Concepts and Properties” evaluation criteria are important for agent-oriented methodology evaluation. “A concept is an abstraction or a notion derived from a specific instance within a problem domain”. And the property represents a special characteristic or capability of an agent. This facet is concerned with the question whether a methodology addresses the basic notions such as concepts and properties of agents in MAS or ABM. In the following, there are concepts and properties according to which methodologies should be evaluated.

| [52] | [53] | [54] | [55] | [56] | [57] | [58] | [33] | [59] | [60] | Gaia V.2 | TROPOS | MaSE | Concepts and properties |
|---|---|---|---|---|---|---|---|---|---|---|---|---|---|
| ✓ | ✓ | ✓ | × | ✓ | ✓ | × | ✓ | × | ✓ | ✓ | ✓ | ✓ | Autonomy |
| ✓ | ✓ | ✓ | × | × | × | × | × | × | × | ✓ | ✓ | × | Adaptability |
| × | ✓ | × | × | × | × | × | × | × | × | ✓ | ✓ | ✓ | Agent Abstraction |
| × | ✓ | ✓ | × | × | ✓ | × | × | × | ✓ | × | ✓ | ✓ | Belief |
| ✓ | ✓ | ✓ | × | × | ✓ | × | × | × | × | ✓ | ✓ | ✓ | Communication |
| × | ✓ | × | × | × | × | × | × | × | × | ✓ | ✓ | ✓ | Concurrency |
| × | ✓ | × | × | × | × | × | × | × | × | ✓ | ✓ | ✓ | Collaboration |
| ✓ | × | × | × | ✓ | × | × | × | × | × | ✓ | ✓ | ✓ | Cooperation |
| × | ✓ | ✓ | × | × | ✓ | × | ✓ | × | ✓ | × | ✓ | ✓ | Desire |
| × | × | × | ✓ | × | ✓ | × | × | × | × | ✓ | ✓ | ✓ | Problem Decomposition |
| × | × | ✓ | × | × | × | × | × | × | × | ✓ | × | ✓ | Events |
| × | × | × | × | × | ✓ | × | × | × | × | × | ✓ | × | Multiple Interest |
| × | ✓ | ✓ | × | × | ✓ | × | ✓ | × | ✓ | × | ✓ | ✓ | Intention |
| × | × | × | × | ✓ | ✓ | ✓ | ✓ | × | × | × | × | × | System Interface Guidance |
| ✓ | × | ✓ | × | × | × | × | ✓ | × | ✓ | ✓ | ✓ | ✓ | Pro-activity |
| ✓ | ✓ | ✓ | × | × | × | × | ✓ | ✓ | ✓ | ✓ | ✓ | ✓ | Protocol |
| ✓ | × | × | × | × | × | × | ✓ | × | ✓ | ✓ | ✓ | ✓ | Organization |
| × | × | × | × | × | ✓ | × | ✓ | × | ✓ | ✓ | × | × | Message |
| × | × | ✓ | × | × | × | × | ✓ | × | ✓ | ✓ | ✓ | ✓ | Reactivity |
| × | × | ✓ | × | × | × | × | ✓ | × | ✓ | ✓ | × | ✓ | Role |
| × | × | × | × | × | × | × | ✓ | × | ✓ | ✓ | ✓ | ✓ | Tasks |
| ✓ | × | × | × | × | × | × | ✓ | × | ✓ | ✓ | ✓ | ✓ | Norms |
| × | × | × | × | × | × | × | ✓ | × | ✓ | ✓ | × | × | Society |
| × | × | ✓ | × | × | × | × | ✓ | × | ✓ | ✓ | ✓ | ✓ | Soci-ality |
| × | × | × | × | × | × | × | ✓ | × | × | ✓ | × | × | Service |
| × | ✓ | × | × | × | × | × | × | × | × | ✓ | ✓ | ✓ | Agent- Oriented |

## 5. Results and Discussion

In this section we present ……

1. Autonomy: The ability of an agent to perform its tasks without any supervision.

In GAIA analysis phase, each role has some responsibilities, and they are independent to perform their responsibilities. In the design phase roles are replaced with agents. So autonomy is expressed in the way that roles encapsulate its functionality.

The requirement analysis phase of TROPOS defines the autonomy through actor's individual goals, the association between agents, and their plans. In the requirement analysis phase, the agent's autonomy is expressed by its goal and plan to achieve goals. In the architectural design, the autonomy is stated by elaborating actors agenda for achieving the system goal. In the detail design phase, the activity diagram, which expresses autonomy.

In the analysis phase of MaSE, autonomy is expressed by tasks. And agents are responsible to execute their tasks on its own responsibility. In the design phase autonomy is expressed through agent classes, in which roles encapsulate their functionality. And this functionality is internal states of agents that not affected by the environment. The autonomy is followed by each phase of each methodology, therefore; we can say that autonomy is high in these three methodologies.

2. Adaptability: The ability of an agent or methodology to deal with the variety of computing environments, and adjust itself according to changing settings.

In GAIA the adaptability is expressed by environmental model. That deal with its internal states and expresses how other agents and environment effect on it. In TROPOS, activity diagram express the adaptability and deals with the variety of computing environments, and changing settings. The MaSE methodology does not support adaptability. The environmental model of GAIA does not explicitly define the negative or alternate responses to change, however, the activity diagram in TROPOS explicitly explains both worse and better responses to change. In that way, adaptation is stronger in TROPOS as compare to GAIA.

3. Agent Abstraction: The ability of a methodology to describe agents using high-level abstraction.

In GAIA, the preliminary role model support agent abstraction by extraction roles from requirement specification document. And In the design phase the agent model expresses the agent abstraction property. In TROPOS the actor model in the early requirement analysis phase and the architectural design phase define the actor abstraction property. In MaSE the role model in the analysis phase and, agent classes in the design phase expresses the agent abstraction property. Almost all phases of each methodology support to agent abstraction, therefore agent abstraction is strong in these three methodologies.

4. Belief: The belief is a faith of an agent, which believes that it is always true about the world.

GAIA does not have a belief. The goals of the actor in TROPOS express the belief concept. In MaSE methodology, belief is expressed by goals, tasks, and states. MaSE strongly support belief property as compare to TROPOS and GAIA.

5. Communication:

The interaction model in the analysis and design phase of GAIA defines communication. The activity diagram explicitly expresses the communication in TROPOS. In MaSE methodology conversation model defines communication. Due to activity diagram and conversation model the communication is strong in TROPOS and MASE as compared to GAIA.

6. Concurrency:

In GAIA the service model, in the detailed design phase support concurrency. The activity diagram in the detailed design phase of TROPOS defines concurrency. In MaSE the role model in the refining role model explicitly explains concurrency.

7. Collaboration: An agent has methods to cooperate with other agents to achieve goals.

In GAIA's analysis and design phase, the interaction model expresses collaboration between agents. In the detailed design phase of the TROPOS sequence diagram defines collaboration between agents. The sequence diagram in the analysis phase and a conversation model in the design phase of MaSE express collaboration between agents. Due to sequence diagram, collaboration is more precise and strong in TROPOS and MaSE as compare to GAIA.

8. Cooperation: The cooperation is a collaborative activity with one objective, but it is distributed among several actors. In cooperation, each agent performs actions according to the shared objectives.

The organizational structure of GAIA expresses the cooperation of agents. The architectural design in TROPOS expresses cooperation of agents to perform a specific task. The agent class model in the design phase of MaSE expresses the cooperation of agents. Due to the class diagram, cooperation is strongest in MaSE as compared to GAIA and TROPOS.

9. Desire: A goal of an agent to be achieved.

GAIA does not support Desire property of the agents. The goal of the agent in TROPOS expresses the desire of the agent. The goal, task models and state in the conversation model of MaSE express the desire of agents. Desire is much stronger in the MaSE methodology after that TROPOS have strong desire property. And GAIA does not have desire property.

10. Problem Decomposition: The ability of a methodology to divide the large problem into smaller and more manageable parts. Basically, this property tackle complexity.

The analysis phase, architectural design, and detailed design phases of GAIA support problem decomposition property. The earlier requirement, late requirement, architectural design and detailed design phases of TROPOS support problem decomposition property. The analysis and design phases of MaSE support problem decomposition property. The problem decomposition is stronger in these three methodologies since each phase of these methodologies is divided into subparts. The division of phases helps to decompose and understand complex problems.

11. Events: The ability of a methodology to control event triggering. The events trigger the interaction and agents become responsible for a new goal.

In GAIA the analysis phase, the sub-organization model interacts with the environmental model, role model, and interaction model. And these models interact with the organizational rules model. The interaction generates events in models to interact with other models to perform a required task. The same event phenomena happen in the design phase. In this way, GAIA expresses event generation property. In TROPOS early requirement or any other phase does not interact with late requirement phase and with others phases. Since TROPOS phases do not generate events or interaction, that's why this methodology does not express an event concept. MaSE methodology's phases interact with other phases and models in a phase interact with each other, this interaction expresses an event concept. Only TROPOS does not support to event concept.

12. Multiple Interests: At a time, an agent may have multiple tasks such as co-operate with other agents, be independent, or help the other agents to achieve a goal in the environment.

GAIA does not support this concept. TROPOS express this concept through goals, plans, and resources. Since, TROPOS agents interact with other agents through goals, plans, and resources. Since they have multiple interests to achieve their goals or to interact with other agents. The MaSE methodology does not support multiple interests concept. Only TROPOS have a strong concept of multiple interests.

13. Intention: A fact that represents the way of realizing a desire, sometimes referred to as a plan.

The GAIA does not support intention concept. The agents of TROPOS express intention through its plans. The goals, task, and states in a conversation model of MaSE express the intention concept. The Intention is stronger in MaSE since three different models of MaSE express this concept. After the MaSE intention is strong in TROPOS.

14. System Interface Guidance:

The GAIA does not support interface to the external world. The TROPOS methodology does not offer an interface for the external world. The MaSE methodology does not support or offer an interface for the external world. These three methodologies lack an interface to the external world. This flaw should improve in all these methodologies.

15. Pro-activity: The ability of an agent and methodology to pursue new goals.

The service model in the detailed design phase of GAIA expresses pro-activeness. The plans of an agent to achieve a goal express the pro-activeness in TROPOS. Any phase of MaSE methodology does not support pro-activeness. The pro-activity is high in GAIA and TROPOS and MaSE lack of this property.

16. Protocol: A set of messages that defines the purpose and detail of a particular interaction among the agents.

The role and interaction model in the architectural design phase and agent and service model in the detail design phase of GAIA explicitly define protocol concept. The sequence diagram in the detailed design phase of TROPOS somehow defines protocols, however; this methodology does not clearly define this concept. The conversation model in the analysis phase and a conversation model in the design phase of MaSE explicitly expresses protocol concept. GAIA and MaSE methodologies strongly support protocol concept. On the other side, TROPOS does not express this concept explicitly. TROPOS lack conversation model.

17. Organization: A group of agents working together to achieve a common purpose. An organization consists of roles that characterize the agents, which are members of the organization.

The Sub-organization model in the analysis phase and the organizational structure model in the design phase of GAIA express organization concept. The architectural design phase of TROPOS expresses organization concept. The agent architecture model in the detailed design phase of MaSE expresses organization concept. GAIA strongly expresses an organization model as compared to TROPOS and MaSE.

18. MESSAGE: The message is a request for making conversation between agents for resources and task completion.

The protocols in the interaction model, agent model and services model of GAIA express this concept. The activity diagram in TROPOS expresses message concept explicitly. The conversation model in the MaSE defines message concept. The message concept is defined by all these methodologies.

19. Reactivity: The ability of an agent and methodology to respond to changes in environment on time.

The responsibilities of a role in the role model and agent model show liveness. This liveness expresses reactivity in GAIA. The state change in activity diagram and sequence diagram of TROPOS methodology expresses reactivity. The conversation model and sequence diagram in MaSE express reactivity. All these methodologies somehow meet the requirement of reactivity. There is no explicit or clear model to support reactivity.

20. Role: An abstract level description of the agent's function, its services or its specific identification within a group.

The role models in the analysis and design phases of GAIA define the role concept explicitly. TROPOS does not express role concept. The role model in the analysis phase of MaSE expresses the role concept. The role concept is strong in the GAIA methodology as compared to MASE. In TROPOS role abstraction concept is weak.

21. Task: A precise piece of work that is assigned to the agent of the system to be in the form of its function.

Task represents to agent responsibilities. In GAIA the role model and service model express task concept. In TROPOS methodology the task concept represents to a capability of the agent. And capability is expressed by activity diagram. The concurrent task model in MaSE expresses a task concept explicitly. In MaSE the task concept is stronger as compared to TROPOS and GAIA.

22. Norms: A set of rules that characterize a society and the agents of this society are bound to follow all the mentioned norms.

GAIA: The norms defined by the organizational rules model in the analysis phase of GAIA. The norms are defined by the organizational structure in TROPOS. The agent architecture in the design phase expresses norm concept in MaSE. In GAIA norm concept is stronger as compared to TROPOSE and MaSE methodology.

23. Society: A collection of agents and organizations that collaborates to promote their individual goals.

The organizational rules and organizational structure models define society in GAIA. The organizational structure in TROPOS somehow way defines society. The agent architecture in MaSE somehow defines society. GAIA methodology has a strong society's concept as compared to TROPOS and MaSE.

24. Sociality: The capability of an agent to communicate with the other agents of the system by sending and receiving messages and cooperate with them to perform a specific task.

In GAIA the sociality is expressed within the interaction model that defines the communication links among agent types. In TROPOS the sociality is expressed by the system model in the late requirement phase. The sociality is somehow expressed by agent architecture in MaSE. All methodologies weakly support sociality concept.

25. Service: The service is a "knowledge level analogue" of an agent's operation to achieve a specific goal.

The service is expressed by service model in the detailed design phase of GAIA. The TROPOS methodology does not support service concept. The MaSE methodology does not support service concept. Only GAIA methodology explicitly expresses service feature.

26. Agent-Oriented: The agent-oriented features focus on whether the methodology addresses Agent-based features during the analysis and design.

GAIA methodology somehow defines all attributes of an agent such as pro-activity, autonomy, reactivity, and sociality so, it is an agent-oriented methodology. The TROPOS methodology somehow defines all attributes of an agent such as pro-activity, autonomy, reactivity, and sociality so, it is an agent-oriented methodology. The MaSE methodology somehow defines all attributes of an agent such as pro-activity, autonomy, reactivity, and sociality so, it is an agent oriented methodology. All these methodologies are agent-oriented.

#### 4.1.2 Notations and modeling techniques

Notations are a set of symbols that technically represent agents and their functional goal in a system to-be. These modeling techniques collectively build a precise model that represents developing system at different levels of abstraction and express their different facets such as structural and behavioral sides. This section deals with the properties of notions which a modeling methodology should have. The list of properties is given below:

| **[52]** | **[53]** | **[54]** | **[55]** | **[56]** | **[57]** | **[58]** | **[33]** | **[59]** | **[60]** | **GAIA V.2** | **TROPOS** | **MaSE** | **Notation and Mdeling Technique** |
|---|---|---|---|---|---|---|---|---|---|---|---|---|---|
| ✗ | ✗ | ✓ | ✗ | ✗ | ✗ | ✗ | ✗ | ✗ | ✗ | ✓ | ✓ | ✓ | Agent Attributes |
| ✗ | ✗ | ✗ | ✗ | ✗ | ✗ | ✗ | ✓ | ✗ | ✓ | ✓ | ✓ | ✓ | Accessibility |
| ✗ | ✗ | ✗ | ✗ | ✗ | ✗ | ✗ | ✓ | ✗ | ✓ | ✗ | ✓ | ✓ | Analyze-ability |
| ✓ | ✓ | ✗ | ✗ | ✗ | ✗ | ✗ | ✓ | ✗ | ✓ | ✓ | ✓ | ✓ | Complexity Management |
| ✗ | ✗ | ✗ | ✗ | ✗ | ✗ | ✓ | ✗ | ✗ | ✗ | ✗ | ✓ | ✗ | Dynamic Structure |
| ✗ | ✓ | ✗ | ✗ | ✗ | ✗ | ✗ | ✓ | ✗ | ✓ | ✓ | ✓ | ✓ | Expressiveness |
| ✗ | ✗ | ✗ | ✗ | ✗ | ✗ | ✗ | ✓ | ✗ | ✗ | ✓ | ✗ | ✓ | Consistency |
| ✓ | ✗ | ✗ | ✗ | ✗ | ✗ | ✓ | ✗ | ✗ | ✗ | ✗ | ✓ | ✗ | Open system |
| ✗ | ✗ | ✗ | ✗ | ✗ | ✗ | ✗ | ✗ | ✗ | ✓ | ✗ | ✓ | ✓ | Execute-ability |
| ✗ | ✗ | ✗ | ✗ | ✗ | ✗ | ✗ | ✗ | ✗ | ✗ | ✓ | ✓ | ✓ | Unambiguity |
| ✗ | ✗ | ✗ | ✗ | ✗ | ✗ | ✗ | ✓ | ✗ | ✗ | ✓ | ✗ | ✓ | Consistency |
| ✗ | ✓ | ✗ | ✗ | ✗ | ✗ | ✗ | ✓ | ✓ | ✗ | ✓ | ✓ | ✓ | Trace-ability |
| ✗ | ✗ | ✗ | ✗ | ✗ | ✗ | ✗ | ✓ | ✗ | ✓ | ✓ | ✓ | ✓ | Preciseness |
| ✗ | ✗ | ✗ | ✗ | ✗ | ✓ | ✗ | ✓ | ✗ | ✗ | ✗ | ✓ | ✗ | System View |
| ✗ | ✗ | ✗ | ✗ | ✗ | ✓ | ✗ | ✓ | ✗ | ✓ | ✓ | ✓ | ✓ | Modularity |

1. Agent Attributes: The ability of a methodology that concern about the description of the agent's parts that make up the internal structure.

The internal structure of the agent is strong in the GAIA as the role model of the analysis phase transforms to agent model in the design phase. This strength is due to that every agent plays a specific role and is independent in making decisions. In TROPOS the actor diagram of the requirement phase transforms to the activity diagram in the design phase. So in this sense, the internal structure of an agent is strong in TROPOS. In MaSE the role model transforms to agent model in the design phase. All these methodologies explicitly manifest agent attribute for system development.

2. Accessibility: The ability of a methodology that assist the developers to easily adapt it, understand it and implement it.

All GAIA models and the phase distribution are simple and understandable. Engineers can use easily these models and can develop complex systems. Gaia does not support accessibility when developers transform its models to the overall system model. The notation understandability and modeling in TROPOS is easy. However, the transformation of models from rational models to activity diagram and sequence diagram is difficult. Same like GAIA, MaSE methodology has simple models that can be understood by developers easily, but the transformation of these models into the overall system model is a difficult task. All these methodologies in some way are accessible; however, some improvement is required in this area.

3. Analyze-ability: The capability of a methodology to check the internal consistency of each agent. And also point out the unclear aspects of models and agents.

GAIA does not deal with the analyze-ability. TROPOS check analyze-ability of models through the agent plan, resource, and goals. However, the analyze-ability compromise between rational models and architectural structure**.** The MaSE methodology provides analyze-ability within models through the transformation of models from the analysis phase to design phase. However, analyze-ability compromise when transforms activity diagram into a sequence diagram. All these methodologies need a development in this area.

4. Complexity Management: An ability of the agent-oriented methodology to deal with the different levels of complexity. For example, sometimes only high-level requirements are needed while in the other situation more details are required. In this case, the agent oriented methodology should provide all level information.

GAIA gives somehow favor to manage complexity by giving different levels of abstraction by different models. However, it does not have a hierarchical structure to manage system's complexity. The TROPOS controls complexity by goal model in the early requirement phase, but it cannot control the details within it. The hierarchical structure somehow controls complexity. The MaSE methodology provides an abstraction of almost every concept of the goal, agent, conversation model, in this way it controls a limited extent of complexity. However, it does not have any hierarchy model to control complexity and detail of complex tasks. Generally, GAIA and MaSE are suffering from this deficiency. TROPOS is somehow better on this property.

5. Dynamic Structure: The ability of a methodology to provide support for the dynamic structural reconfiguration of the system.

The GAIA does not deal with a dynamic structure. The TROPOS somehow deal with the dynamic structure of activity diagram. The MaSE does not deal with the dynamic structure. The future work should be done on the dynamic structure feature of all methodologies.

6. Expressiveness: The ability of a methodology of presenting system concepts such as system structure, encapsulated knowledge of models, ontology structure, data flow, control flow, and concurrent activities of the involving agents.

Due to the generic structure, GAIA can handle and model a large variety of systems. However, the system structure of the system to-be is not presented explicitly. Usually, TROPOS helps to develop BDI (Belief, Desire, and Interaction) systems. However, the system structure of the system to-be is not presented explicitly. The MaSE methodology explicitly describes the system structure by using an agent architecture and system design diagrams. However, the encapsulated knowledge of the system is not presented explicitly. All methodologies express expressiveness in a limited way.

7. Consistency: The ability of methodology to maintain the quality of a system throughout the development process.

Each model in the GAIA explicitly defines each concept clearly, means that agent model, role model, environmental model, service model. There is no chance of ambiguity between models. In TROPOS the rational model and the architectural design confuse the developers due to ambiguity. However, in detail design phase it controls this lack through activity diagram. The MaSE explicitly express consistency within each model through control flow and data flow. Moreover, each model clearly defines a single concept such as goal model, role model, and agent model, etc. Each methodology tries to be consistent in some way. However, MaSE strongly expresses the consistency as compared to GAIA and TROPOS. On the second number, GAIA is stronger in consistency than TROPOS.

8. Open System: The ability of a methodology to add or remove new agents in the development system.

The GAIA methodology completely supports open system property because this methodology has separate models for each concept. New agents, roles, and protocols, etc easily can be removed or add into the system. TROPOS methodology does not support to open system property due to activity and sequence diagrams. Same like GAIA, MaSE explicitly supports to open system property. GAIA and MaSE are strong, according to this aspect.

9. Execute-ability: The ability of the agent-oriented methodology, to provide the facility of performing a simulation to validate the system specification. Or at least generate a prototype of the overall system to-be.

GAIA does not deal with the execute-ability issue. TROPOS dealt with the execute-ability by using JACK. The MaSE methodology has a developmental tool "agentTool" which support partially to code generation. MaSE methodology is stronger as compared to the GAIA, and TROPOS, according to this aspect.

10. Unambiguity:

As GAIA has clear models for each concept, so in this way, it tries to overcome this issue. In TROPOS the architectural design and rationale models lead to ambiguity and create confusion for developer's understanding. Same like GAIA, MaSE has clear models for each concept. MaSE and GAIA methodologies have less ambiguity as compared to TROPOS.

11. Trace-ability: The ability of a methodology to handle the main concept of the system to-be throughout the system development.

The models in GAIA's analysis phase become input for models in the design phase. In this way we can say it maintain trace-ability throughout the development phases. Same like GAIA, the TRPOS methodology also maintains trace-ability. The models in the analysis phase of MaSE become the base for the design phase models. Each mode can be traced by its base model. All these methodologies in somehow way fulfill the trace-ability aspect.

12. Preciseness: The ability of the methodology to handle the ambiguity throughout the system development. It assists the developers in avoiding the misinterpretation in system development.

Each model has a clear meaning and interpretation in GAIA methodology. This edge makes GAIA accurate. As the TROPOS base on i* model that has clear notation and meaning of each symbol. Thus, it prevents users from misinterpretation. Same like GAIA, MaSE model, symbols, and notations have a clear meaning and interpretation. This advantage makes the MaSE accurate and prevent the developers from misinterpretation. All these methodologies are precise.

13. System View: The ability of a methodology to provide a microscopic system oriented model, to understand the whole working scenario of the developing system.

GAIA does not promote system view feature. Since, this methodology has a deficiency of hierarchical model and overall system model. In TROPOS the activity diagram demonstrates the general view of the system model. In MaSE, agent classes depict a vague concept of system view. Only TROPOS has somehow clear abstraction of system view.

14. Modularity: The ability of a methodology to develop a system in an iterative way. That allows adding new requirements without affecting the existing specifications.

GAIA is modular, due to its models (agent, service and, role). The changing in the role does not affect the whole system. This change only influences the internal structure of an agent or role. Within the TROPOS modularity is fully supported. Same like GAIA, MaSE support modularity because of models (task, agent, goal). TROPOS strongly support modularity as compared to GAIA and MaSE.

### 4.1.3 Development Process

A development process is a step by step guideline for developing a system from scratch. This process consists of a series of actions, functions, and models, that when performed, then the outcome is an operational computerized system. Basically, this section of evaluation framework deals with the different facets of system development process. The terminologies for checking the developing process are given in the below table:

| **[52]** | **[53]** | **[54]** | **[55]** | **[56]** | **[57]** | **[58]** | **[33]** | **[59]** | **[60]** | **GAIA V.2** | **TROPOS** | **MaSE** | **Development Process** |
|---|---|---|---|---|---|---|---|---|---|---|---|---|---|
| × | ✓ | × | × | × | × | × | × | × | × | ✓ | ✓ | ✓ | Architecture Design |
| × | × | × | × | × | × | × | × | × | × | × | ✓ | × | Requirement Norms |
| × | × | × | × | ✓ | ✓ | × | × | ✓ | × | ✓ | ✓ | ✓ | Legal Document |
| × | × | × | × | × | × | × | × | ✓ | × | × | ✓ | ✓ | System Design |
| ✓ | ✓ | × | × | × | × | × | ✓ | × | ✓ | ✓ | ✓ | ✓ | Development Life Cycle |
| × | ✓ | × | × | × | ✓ | × | × | × | ✓ | ✓ | ✓ | ✓ | Development Context |
| ✓ | ✓ | ✓ | ✓ | × | × | ✓ | ✓ | × | × | × | ✓ | ✓ | Implementation |
| × | × | × | ✓ | × | × | × | ✓ | × | ✓ | ✓ | ✓ | ✓ | SDLC coverage |
| × | ✓ | × | ✓ | × | × | × | × | × | × | ✓ | ✓ | ✓ | Development Approach |
| × | ✓ | × | × | × | × | × | ✓ | × | ✓ | ✓ | ✓ | ✓ | System Specification |
| × | × | × | × | ✓ | × | × | ✓ | × | ✓ | × | × | × | Project Management guideline |
| × | × | × | × | ✓ | × | × | ✓ | × | ✓ | ✓ | ✓ | ✓ | Verification and Validation |
| ✕ | ✓ | ✕ | ✕ | ✕ | ✕ | ✕ | ✓ | ✕ | ✓ | ✕ | ✕ | ✕ | Quality Assurance |

1. Architecture Design: The ability of a methodology to facilitate design by using patterns or modules.

GAIA provides almost all modules of a system such as interaction, role, environmental, service, agent models and also provides an architectural design of developing system. The architectural design phase of TROPOS explicates the overall system's global architecture in terms of sub-systems, interconnected through data and control flow. The MaSE have all architectural bricksin the orm of models such as agent model, goal model, role model, conversation model and task model. The agent architecture model represents the architectural design of the whole system. All these methodologies have architecture design property for system development.

2. Requirement Norm: The ability of a methodology to identify organization's norm during requirement analysis.

GAIA does not support the system requirement. This methodology performs development process after requirement specification. During early and late requirement analysis phases of TROPOS, this methodology identifies and formalize norms for the system to be developed. Same like GAIA, the MaSE methodology have analysis and design phase. It does not support system requirement analysis phase. Only TROPOS has a strong feature of building system norms.

3. System Design: The ability of a methodology to introduce the normative environment of the system as an integral part of the development phase.

GAIA does not support system design. TROPOS does not support system design. MaSE expresses system design through deployment diagrams. More work is required in the system design phase of all methodologies.

4. SDLS Coverage: The ability of an agent-oriented methodology to include elements of software development.

GAIA covers only two main phases of the development process such as analysis and design phase. This coverage is not sufficient for developing an outstanding system. The TROPOS methodology covers almost all phases of the development cycle. However, it does not deal with the testing stage. MaSE covers development cycle from analysis to implementation phase. However, its goal model, use cases and sequence diagrams in somehow manners support requirement phase. MaSE does not support only to testing phase. Only GAIA methodology has a deficiency of DLC coverage.

5. Development Context: The ability of a methodology to be remolded according to users need such as creating new software, reverse engineering, re-engineering systems by using reusability or creating new system property.

GAIA has the ability of creating new software, designing and re-engineering systems by using reusability propert or creating from scratch. It does not address implementation phase and does not support classical reverse engineering . The TROPOS methodology can be used for creating software systems from scratch and for prototyping, re-engineering and designing systems by using reusability property. However, TROPOS does not support reverse engineering. Because when going from one stage to the next stage, then several concepts undergo significant changes. The MaSE can be used in creating new systems, designing systems from scratch and re-engineering with reuseability property. However, MaSE does not support reverse engineering. All methodologies are incapable of reverse engineering.

6. Implementation Guidance: The capability of a methodology to deal with coding issues, quality, performance, libraries and debugging.

GAIA does not support implementation phase. TROPOS has implementation guidance property as it suggests to use JACK toolkit since it easily maps a BDI architecture. The MaSE methodology has a graphical agentTool, which is a fully human-interactive tool and guides each step of MaSE development. The agentTool have the ability for automatic verification of inter-agent communication, semi-automated design, and code generation for multiple MAS framework. These all methodologies are strong in implementation guidance property except GAIA.

7. Development Approach: The attribute of a methodology to guide system development in a specific way, such as top-down, bottom-up, mix.

GAIA follows a top-down development approach. TROPOS follows a top-down development approach. MaSE follows a top-down development approach. All these three methodologies follow a top-down approach.

8. System Specification: The ability of a methodology to accurately interpret the problem from specification document. And confirm that this is the right problem to be solved.

The analysis phase of GAIA, manifest the requirement specification of the system in the form of sub-organizations, role model, interaction model, and organizational rules. The actor model in the early requirement analysis phase of TROPOS embodies the requirement of the system. In the analysis phase the MaSE express system specification in the form of goal model, use cases, sequence diagram, roles and task model. All these methodologies identify and specify system specifications.

9. Project Management guideline: The ability of a methodology to effectively and efficiently guide all aspects of a project from conception through completion.

This issue is not dealt with in the GAIA. The TROPOS has lack to deal with project management issues. This issue is not dealt with in the MaSE. ALL these methodologies are unable to deal with the project management issues.

10. Verification and Validation: The ability of a methodology to provide a way for formal verification and validation.

GAIA performs verification and validation during the transformation of preliminary roles to role model. In TROPOS, there is no coverage checking with respect to the initial requirement. However, TROPOS extension "formal TROPOS" can be used for verification and validation. MaSE performs verification over its models by checking consistency, deadlocks and unused elements between the stages. ALL these methodologies have no specific verification and validation method. However, TROPOS overcome this lack by TROPOS formal method. GAIA and MaSE overcome this lack of transformation of models.

11. Quality Assurance: The ability of a methodology that ensures that the developed software meets and compiles with defined or standardize quality specifications.

This issue is not dealt with GAIA. This issue is not dealt with TROPOS. This issue is not dealt with MaSE. All methodologies have lack quality assurance property.

#### 4.1.4 Pragmatics

The pragmatics of a methodology, determine the industrial success of a methodology. Pragmatics determine that a methodology is applicable in the industry for developing complex and distributed systems. Moreover, this section determines that is the considered methodology have the ability of project management and determine that the considered methodology can be adapted within the organization according to the organizational budget and experience. The pragmatics checking terminologies are given below in the table.

| [52] | [53] | [54] | [55] | [56] | [57] | [58] | [33] | [59] | [60] | GAIA V.2 | TROPOS | MaSE | Pragmatics |
|---|---|---|---|---|---|---|---|---|---|---|---|---|---|
| × | ✓ | × | × | × | × | ✓ | × | × | × | × | ✓ | ✓ | Tools Available |
| × | ✓ | × | × | × | × | × | ✓ | × | ✓ | ✓ | ✗ | ✓ | Required Expetries |
| × | ✓ | × | × | × | ✓ | × | × | × | × | × | ✓ | × | Modeling suitability |
| × | ✓ | × | × | ✓ | × | ✓ | ✓ | × | ✓ | ✓ | ✓ | ✓ | Domain Applicability |
| × | ✓ | × | ✓ | × | × | × | ✓ | × | ✓ | ✓ | ✓ | ✓ | Scalability |
| × | × | × | ✓ | × | × | × | ✓ | × | ✓ | ✓ | ✓ | ✓ | Resources |
| × | × | × | × | × | × | × | ✓ | × | ✓ | ✓ | ✓ | ✓ | Language Suitability |

1. Tools available: The ability of a methodology to guide about available tools and tools ready to use.

The GAIA methodology does not provide any automated Tool. TROPOS provides several automated tools for animation, model checking, and reasoning. The famous one is JACK tool. The MaSE methodology has agentTool, which represents the behavior of internal components of agents and protocols. All these methodologies are somehow strong on this property except GAIA.

2. Required expertise: The ability of a methodology, be as simple as possible that users don't require any background knowledge.

The GAIA required a strong background knowledge of logic and temporal logic. These logics reduce its accessibility. Since many developers do not know or do not want to get familiar with logic. TROPOS does not require strong background knowledge. Since, its modeling notations are very simple. Same like GAIA, the MaSE methodology requires a strong background knowledge of logic and temporal logic. TROPOS is easy to develop at the first time as compared to MaSE and TROPOS.

3. Modeling Suitability: The ability of a methodology, consisting of a specific architecture.

The GAIA methodology has no specific modeling architecture. So, designers have no need of architectural information for development. TROPOS has BDI architecture, So designers have to explicitly describe beliefs, desires, and intentions. MaSE has no specific architecture, So designers have no need of architectural information for development. Only TROPOS methodology has architecture, and developers need to search if their system has BDI properties then they use this methodology.

4. Domain Applicability: The ability of a methodology to be suitable for a variety of domains.

GAIA is suitable for dynamic-open (where the agents are not known) and dynamic-close (where the agents are known) systems. However, this methodology is not suitable for developing applications with dynamic characteristics such as goals generation. Generally, TROPOS is suitable for developing componentized systems like e-business applications and BDI based systems. MaSE can be used for various types of agents and systems. Only TROPOS and MaSE support to dynamic characteristics in complex system development.

5. Scalability: The ability of a methodology to be adjusted to dealt with various application sizes.

GAIA has a simple structure, that's why it can be fitted in different sizes of applications. TROPOS does not provide information for subsets and supersets. So, developers have no idea of exact size must be to avoid complexity. MaSE also does not provide information for subsets and supersets. Thus, developers have no idea of exact size must be to avoid complexity. GAIA has strong scalability characteristics as compared to TROPOS and MaSE.

6. Resources: The ability of a methodology to be mature enough that, publish material, tools and training groups are easy to find.

The GAIA methodology does not provide any automated Tool. However, published papers describe the implementation of this methodology in detail. The TROPOS's Tools and published material are available. It fully supports system development from scratch. The MaSE Tools, web site, and various published studies are available. It fully supports system development from scratch. MaSE and TROPOS are well built in resources as compared to GAIA.

7. Language Suitability: The ability of a methodology to be coupled with a particular implementation language or a specific architecture.

As GAIA does not have any specific architecture and language. And it does not refer to the implementation issues. Thus, the specifications made using GAIA can be implemented in any language. As TROPOS based on the BDI concept, so its implementation will be biased towards BDI direction. MaSe does not target to any specific architecture, language, and a specific framework. MaSE and GAIA are more independent from any language suitability than the TROPOS.

**Exploratory Agent Based Modeling (EABM):**

Cognitive Agent-Based Computing (CABC) is a unified framework, proposed by Niazi and Hussain in the article [61] for modeling the complex systems of agents. This framework has four levels for developing complex systems. Exploratory Agent-Based Modeling (EABM) is the second level of CABC framework. The EABM guides the researchers to develop a proof of concept model of the developed system with the goal of performing multiple simulations for improving understanding about a particular real-world complex system [62].

<u>Breeds:</u>

There are two types of breed.

1. T-Cells of immune system
   a) Regulatory T-Cells (T-reg)
   b) Effectors T-Cells (T-eff)
   c) Active Regulatory T-Cells (A.T-reg)
   d) Active Effectory T-Cells (A.T-eff)
   e) External environmental virus (virus)
   f) Cytokines
2. Brain organs
   a. Blood Brain Barrier BBB
   b. Brain White Matter (Myelin)

a) T-reg

This agent search for the virus, if virus found, then it changes its state from inactive T-reg to active A.T-reg.

1. Move randomly
2. Search virus
3. Transform from T-reg to A.T-reg

b) T-eff

This agent search for the virus, if virus found, then it changes its state from inactive T-eff to Active A.T-eff.

1. Move randomly
2. Search virus
3. Transform from T-eff to A.T-eff

c) A.T-reg

When a virus attacks any inactive regulatory cell or inactive regulatory cell became successful to catch the virus , then T-reg regulatory cell changes its state from inactive to active A.T-reg and its color turns from black to blue. When A.T-reg blue agent finds any active Effector A.T-eff in radius, then it kills effector cell and does these jobs.

1. Move randomly
2. Kills active Effector T-cells
3. Gain energy
4. Duplicate A.T-reg

d) Effector T-Cells

When a virus attacks any Inactive Effector T-Cell T-eff or T-eff cell became successful to catch the virus, then effector cell changes its state from resting T-eff to active A.-eff and color turns from white to red. Active effector T-cell perform these tasks;

1. Move randomly
2. Search Myelin
3. Eat myelin
4. Gain energy
5. Duplicate A.T-eff
6. Produce Cytokines

e) Virus

The virus agent search for T-Cells T-reg and T-eff. If the virus finds any T-cell in its radius, then it makes them active. It performs the following tasks.

1. Move randomly
2. Search T-Cells
3. Make T-Cells active

f) Cytokines

Cytokines search for BBB. If they become successful to find the BBB then they attack BBB and damage it. After the attack, it died. It performs the following tasks.

1. Move randomly
2. Search BBB
3. Damage BBB
4. Die

2. Brain organs

a) Blood Brain Barrier BBB

BBB is a brain organ that stops all unwanted minerals, molecules, and cells to enter into the brain. If any cell tries to enter the brain, then it bounces back that cell. If the BBB is damaged, then it repairs BBB. It performs the following tasks:

1. Stop cells' entrance into the brain
2. Repair damaged BBB

b) Brain White Matter (Myelin)
Myelin covers the axonal part of the neuron cells and maintains communication between two neurons. If any damaged occur due to effectory cells attack, then it repairs the myelin to maintain communication. It performs the following tasks:

1. Maintain communication
2. Repair myelin

<u>Global Variables:</u>
Global variables are accessible anywhere in the simulation code and store particular value for agents during the one continued simulation experiment. There are four preferred global variables.
1. The recoverable brain patches (recoverable)
2. The unrecoverable brain patches (unrecoverable)
3. The total myelin in each brain patch (initmyelin)
4. The total energy of each agent (total)

Besides these variables, there are many other input variables in the user interface which can be adjusted by the user manually or else by using behavior space.

<u>Procedures:</u>
1. Setup
2. Setup-world
3. Reproduce-Tregs
4. reproduce-effectors
5. grow-myelin
6. eat-myelin
7. go
8. move
9. death
10. count-virus
11. count-Tegg
12. count-effectors
13. bounce
14. display-labels
15. catch-effectors
16. count unrecoverable
17. display label
18. do-ploting

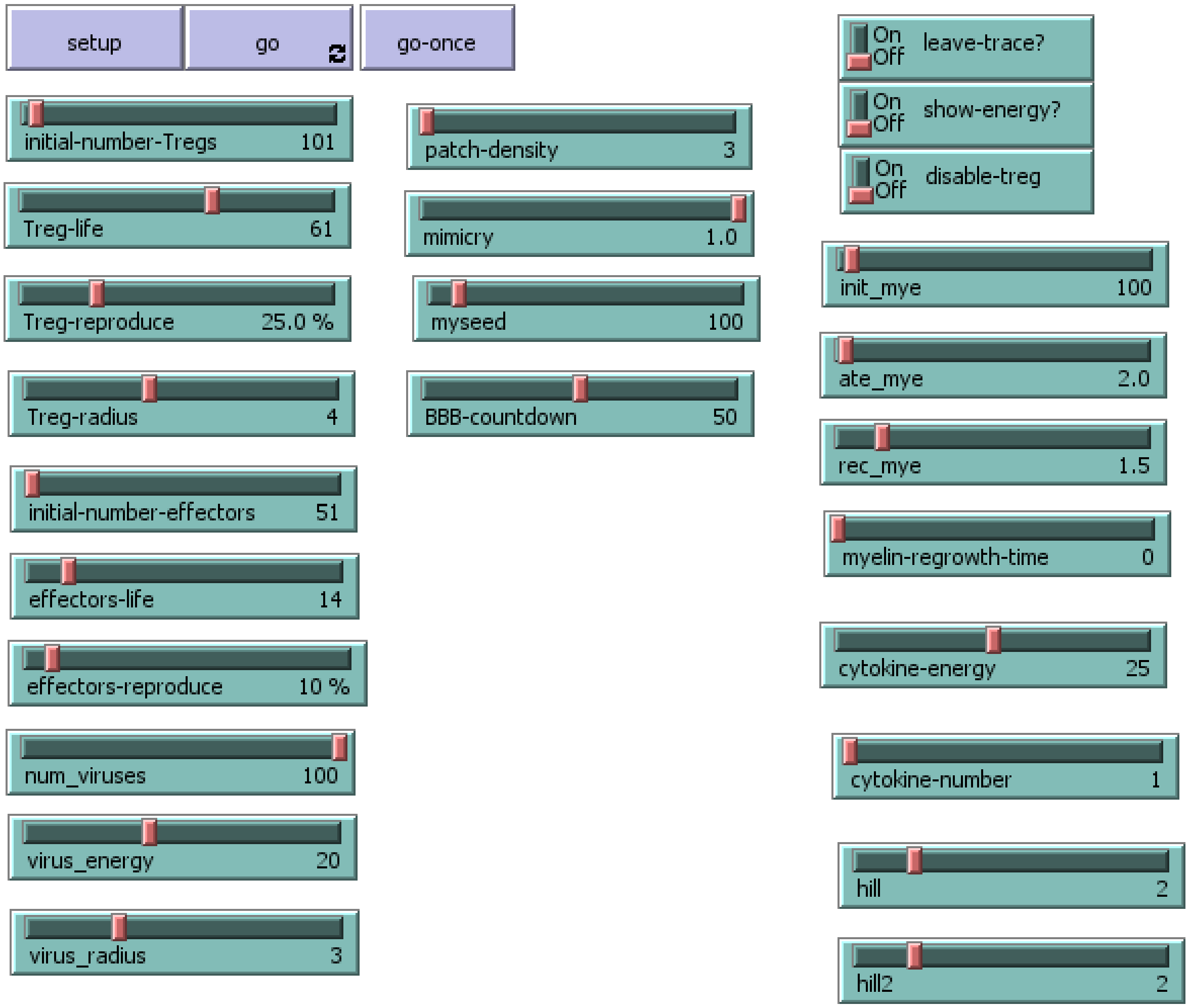

**Figure 33: These are various configurable values for the simulation**

1. Setup:

First of all setup functions clears all variable, agents and environment of the simulation. This is important because we need to run multiple simulations on a specific hypothesis. This procedure divides simulation environment into three types of patches 1) Blood / LYMPHATIC system, 2) Brain axonal area and 3) Blood Brain Barrier. After creating patches, the next step is to create three types of agents (regulatory, effectors, viruses).

Create-Tregs:

This function creates regulatory T-cells T-reg according to its input variable and sets its attributes. The input variable is adjustable from 1 to 2000 numbers. Finally, this procedure adjusts regulatory cells xy-coordinate in simulation environment.

Create-effectors:

This function creates effectors T-cells T-eff according to its input variable and sets its attributes. The input variable is adjustable from 1 to 2000 numbers. Finally, this setup procedure adjusts effector cells xy-coordinate in simulation environment.

Create-viruses:

This function creates an external environmental virus, according to its input variable and sets its attributes. The input variable is adjustable from 1 to 100 numbers. Finally, setup procedure adjusts virus cells xy-coordinate on simulation environment.

At the end, setup function calls the display-labels function. The detailed description of each function follows along with their function.

2. Setup-world:

This function divides 51*51 grid into patches of size 9. 0 to 13 y-coordinate patches are declared as blood portion and from 13 to 15 y-coordinate are declared as BBB portion. 15 to 35 y-coordinates represent brains white portion. 35 to 37 again represent BBB and 38 to 50 represent a blood portion of the brain.

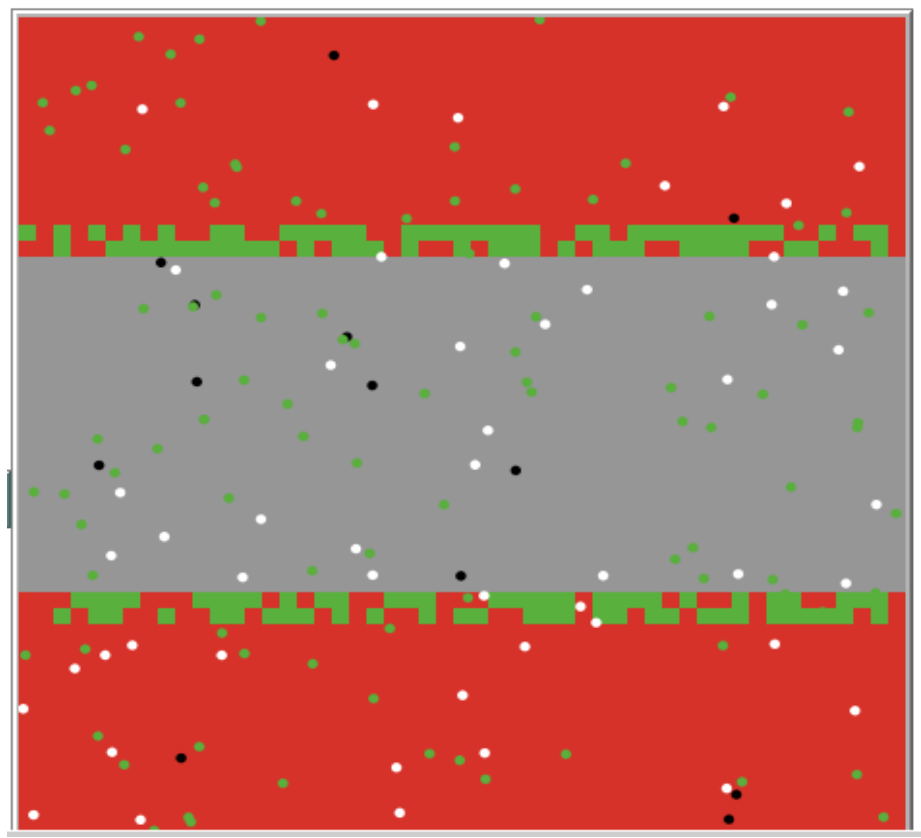

**Figure 34: model view after pressing the setup button**

3. Reproduce-Tregs:

This procedure reproduces regulatory T-cell according to probability. If random float of 100 is less than reg-reproduce the input variable, then it checks a condition. If the number of turtles on a single patch is less than the patch density, then it cuts down turtle energy by half and produces another regulatory T-cell. Otherwise, move forward according to move function.

4. Reproduce-effectors:

This procedure calculates the probability of reproduction of effector T-cells. If an effector successfully eats myelin and causes axonal damage then, it will be simulated to duplicate. And the duplication is modeled as a stochastic Bernoulli process. The duplication probability p is calculated for every duplication process according to the following law:

P = effector-dupl × myelin$^2$ / init-mye$^2$ × mean-Tregs / Treg-here + mean-Treg

Where myelin indicates the quantity of myelin in the current patch, effector-dupl is a duplication constant representing the maximum duplication rate of Teff, mean-Treg is a given threshold and Treg-here is the number of Tregs in a given radius Treg-radius. The term myelin$^2$ / init-mye$^2$ gives higher probabilities to duplicate if the patch has higher quantities of myelin, where the term mean-Tregs / Treg-here + mean-Treg is used to model the down-regulation of Teff duplication rates by Treg actions.

5. Grow-myelin:

Each patch with color gray shows myelin amount. Initially, this procedure sets each patch myelin to 100. For each time when an active effector eats myelin then, that's patched myelin amount decreased by -5 and then this procedure again check condition if the myelin amount reaches to zero then, that patch color turns to black. After each attack, the patch color turns from gray to black. After that, this black patch recovers myelin according to rec-mye input variable.

6. Eat-myelin:

This procedure sets a condition that if any active A.T-eff is on any patch with color gray and the patch myelin amount is greater than the eat-myelin input variable then minus -5 from that's patched myelin amount. If myelin amount reaches zero then turn patch color from gray to red.

7. Go:

First of all, this procedure calculates the total number of unrecoverable brain patches. Unrecoverable patches are those whose myelin amount reaches to 0 and their color turn from gray to black. After that, it calculates myelin of each patch and subtract it from the total myelin amount and assign the total myelin amount to each patch, which amount it has at this simulation time. This procedure also calculates the total number of recoverable patches. If any patch has some amount of myelin then it can recover from multiple sclerosis. Recoverable patches are calculated by subtracting unrecoverable patches from the total number of patches.

I. Create regulatory: This procedure creates regulatory T-cells according to a specific probability. The probability to create a regulatory T-Cells is, if random float of 1 is less than 10/365 then create a regulatory T-Cells according to the slider value.
II. Create effectors: This procedure calculates specific probability to create an effector T-cells. If random float of 1 is less than 10/365 then create an effector T-cells according to the slider value.
III. Create viruses: This procedure also creates external environmental viruses, according to a specific probability. The probability to create viruses is if random float of 1 is less than 10/365 then create regulatory T-cells according to the slider value.

After creation agents, this procedure asks effectors call bounce and move function. After that, this procedure asks active effectors eat myelin and produce 1 cytokine against one myelin attack. It also instructs inactive effectors that try for activation. If any virus is in radius, then they change their state to active else move forward and try to catch the virus.

After creation cytokines, this procedure instructs cytokines that if they are at patch with color green they change the patch color to red. Means if any cytokine collides with the blood brain barrier then barrier destroy and that patch start countdown. When the countdown reaches 0 then they again became a barrier.

Go function ask regulatory T-cells call to move and bounce function. They also instruct that if regulatory cells are active then catch effectors and produce one another regulatory cell or else try to catch virus to make himself active.

At the end, his function asks viruses call move, bounce and death function. And instruct patches call grow-myelin function.

8. Move:

This procedure decreases turtle's energy at every single step and sets turtle heading random 360. After that it forces turtle to take the next random step.

9. Death:

This procedure order to each turtle if its energy level is zero, then it must be die at any place in the simulation environment.

10. Count-virus:

This procedure reports total number of virus for every moment in the running simulation.

11. Count-Teg:

This procedure reports a total number of active regulatory T-cells with the color blue for every moment in the running simulation.

12. Count-effectors:

This procedure reports the total number of active effector T-cells with the color red for every moment in the running simulation.

13. Bounce:

This procedure ask all turtles if the next move patch's color is green, then bounce back and set heading random 360. This green patch is a blood Brain Barrier which blocks turtle's movement.

14. Display-labels:

This procedure asks turtles if the energy level switch is on then, they display their current energy level during the current simulation experiment.

15. Catch-effectors:

This procedure asks active regulatory T-cell if any active effector T-cell in its radius, then kill him and breed one regulatory T-cell. If there is no one effector in radius, then try to catch effectors with the color red.

16. Count unrecoverable

This procedure counts all patches with color black. These black color patches show to damaged brain that is unrecoverable from multiple sclerosis disease.

17. Display label

This procedure displays the current value of each patch and agent during running a simulation.

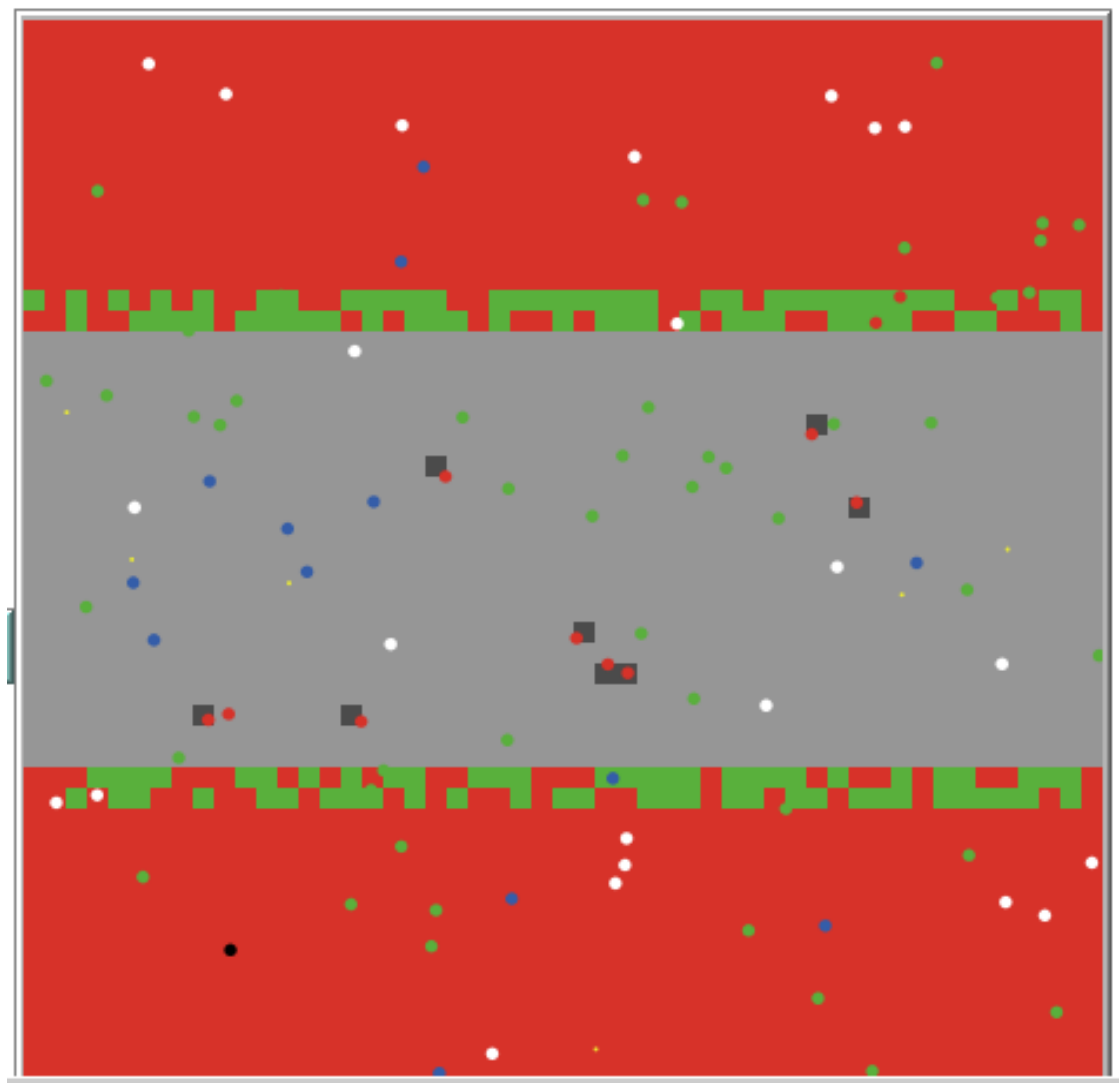

**Figure 35: model view during simulation**

| Parameters | Simulation 1 | Simulation 2 | Simulation 3 | Simulation 4 | Simulation 5 | Simulation 6 |
|---|---|---|---|---|---|---|
| Init-Treg-n | 100 | 50 | 100 | 100 | 50 | 100 |
| Treg-life | 60 | 30 | 60 | 60 | 30 | 60 |
| Treg-repro | 25% | 12 % | 25% | 25% | 12 | 25% |
| Treg-radius | 3 | 2 | 3 | 3 | 2 | 3 |
| Init-Teff-n | 100 | 100 | 50 | 100 | 100 | 100 |
| Teff-life | 60 | 60 | 30 | 60 | 60 | 60 |
| Teff-repro | 25% | 25% | 12% | 25% | 25% | 25% |
| Init-virus-n | 100 | 100 | 100 | 100 | 100 | 50 |
| V-energy | 20 | 20 | 20 | 20 | 20 | 10 |
| v-radius | 3 | 3 | 3 | 3 | 3 | 2 |
| Mimicry | 1.0 | 1.0 | 1.0 | 1.0 | 1.0 | 1.0 |
| Myseed | 100 | 100 | 100 | 100 | 100 | 100 |
| Show-enegy | off | off | off | off | off | off |
| Disable-Treg | off | off | off | off | off | off |
| Init-mye | 100 | 100 | 100 | 100 | 100 | 100 |
| Ate-mye | 2 | 2 | 2 | 5 | 5 | 2 |
| Rec-mye | 1.5 | 1.5 | 1.5 | 1.5 | 1.5 | 1.5 |
| Mye-rgrow-time | 2 | 2 | 2 | 0 | 0 | 2 |
| BBB-coun-down | 50 | 50 | 50 | 50 | 50 | 50 |
| Cytokine-energy | 25 | 25 | 25 | 25 | 25 | 12 |
| Cytokine-n | 1 | 1 | 1 | 1 | 1 | 1 |
| Hill1 | 2 | 2 | 2 | 2 | 2 | 2 |
| Hill2 | 1 | 1 | 1 | 1 | 1 | 1 |
| Patch-density | 3 | 3 | 3 | 3 | 3 | 1 |

**Table 20: values of all variables during simulation**

# First simulation:

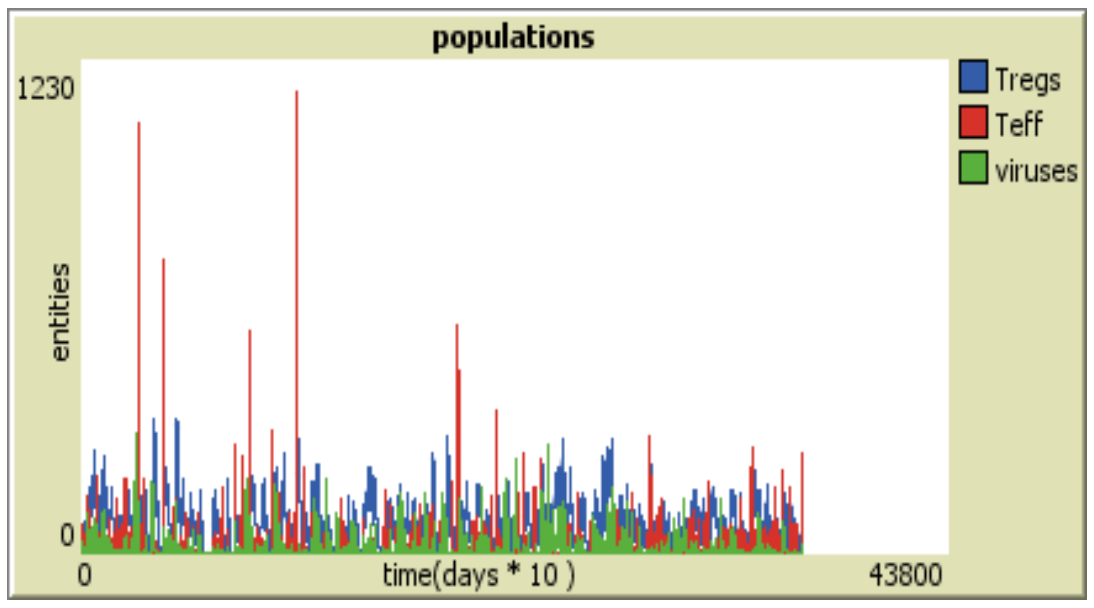

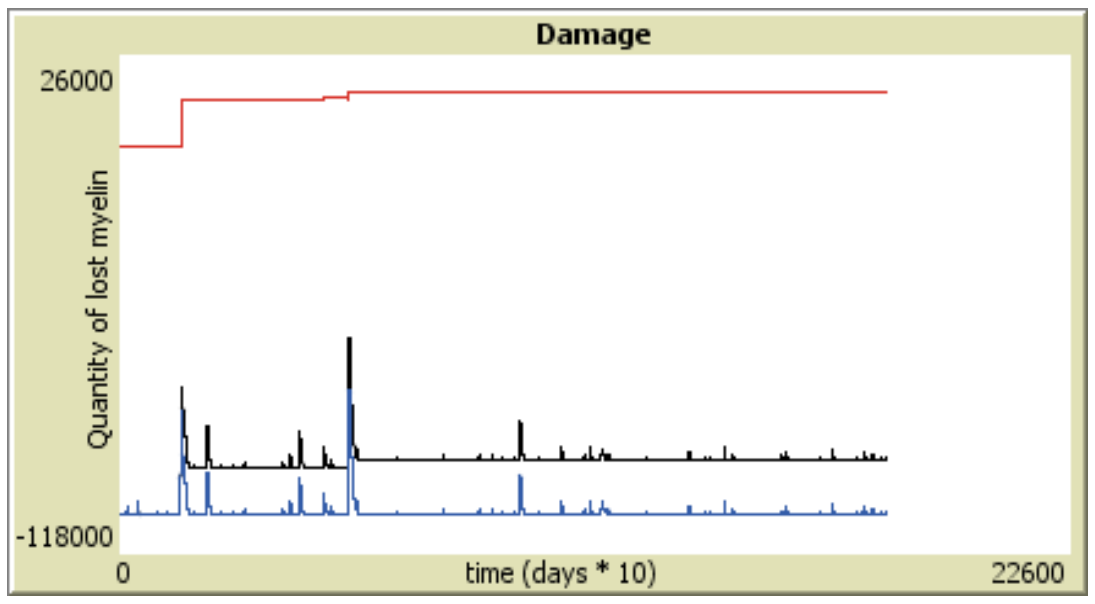

In this simulation the T-regulatory and T-effectors production and the initial birth ratio are same. In this case brain, axonal damage and brain recovery time are almost same. Sometimes T-effectors strongly attack brain axonal, at this time brains immune system gives quick response to damage and regulatory cells control T-effectors progression.

# 2$^{nd}$ simulation:

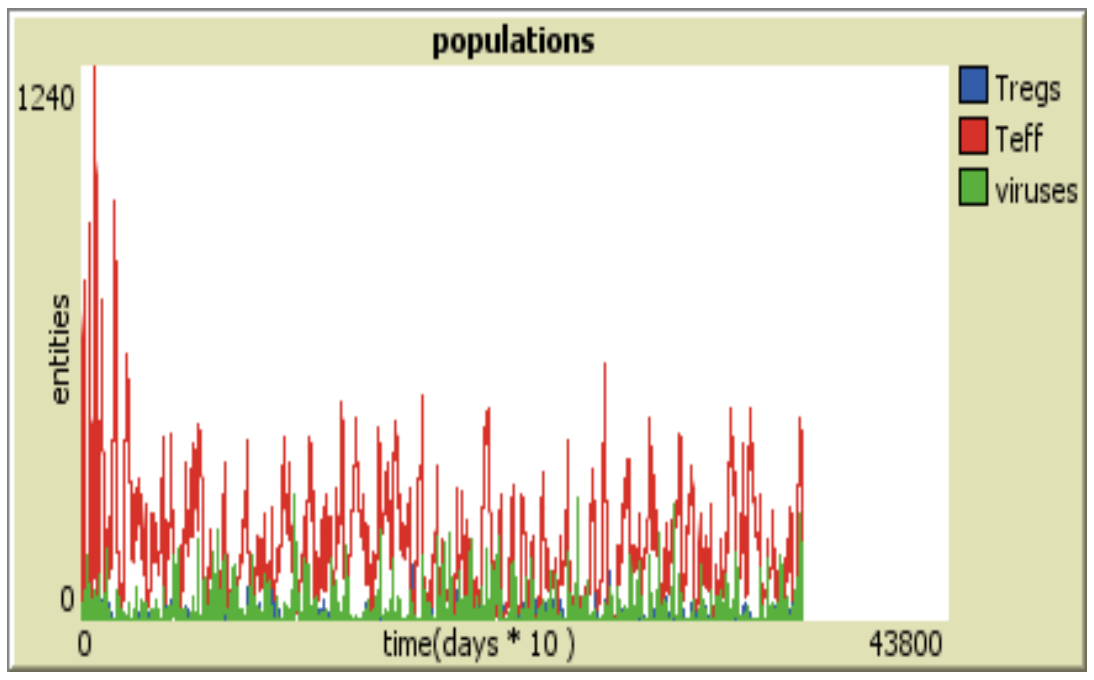

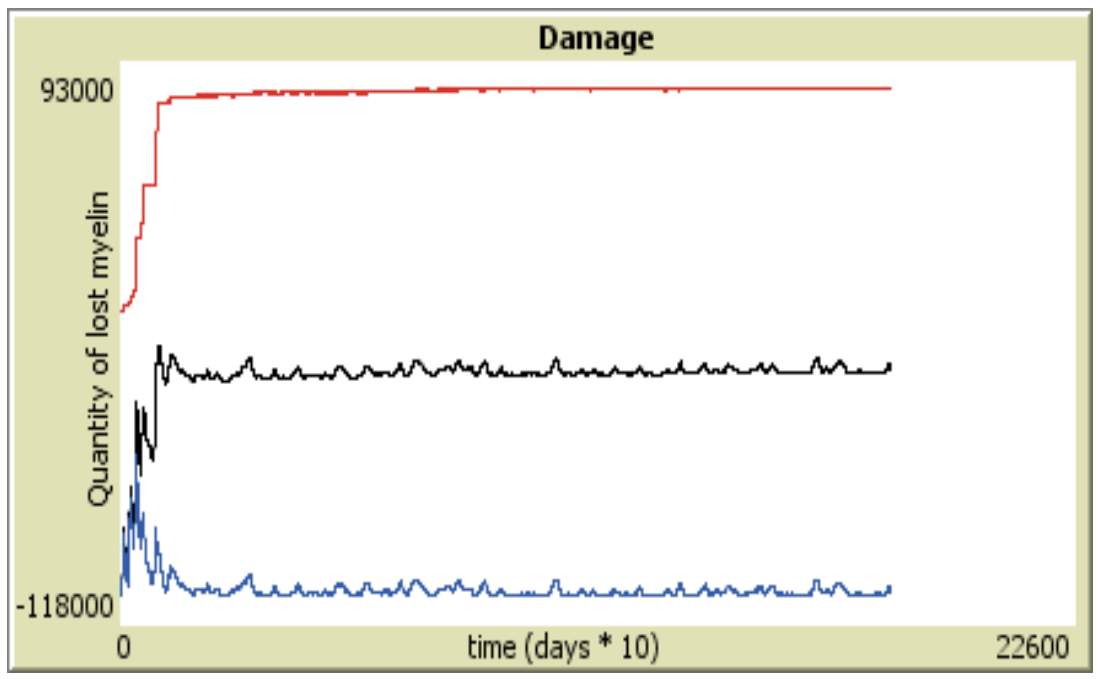

In this simulation, T-effectors cell production. Reproduction and initial number rate are strong. Less T-regulatory strength shows weak immune system. In which recovery cells and immune cells are weak or they are in very small amount. In this case, T-effectors strongly attack brain axonal area and cause severe brain damage. Most of the time this damage is unrecoverable and growing with time.

## 3rd simulation:

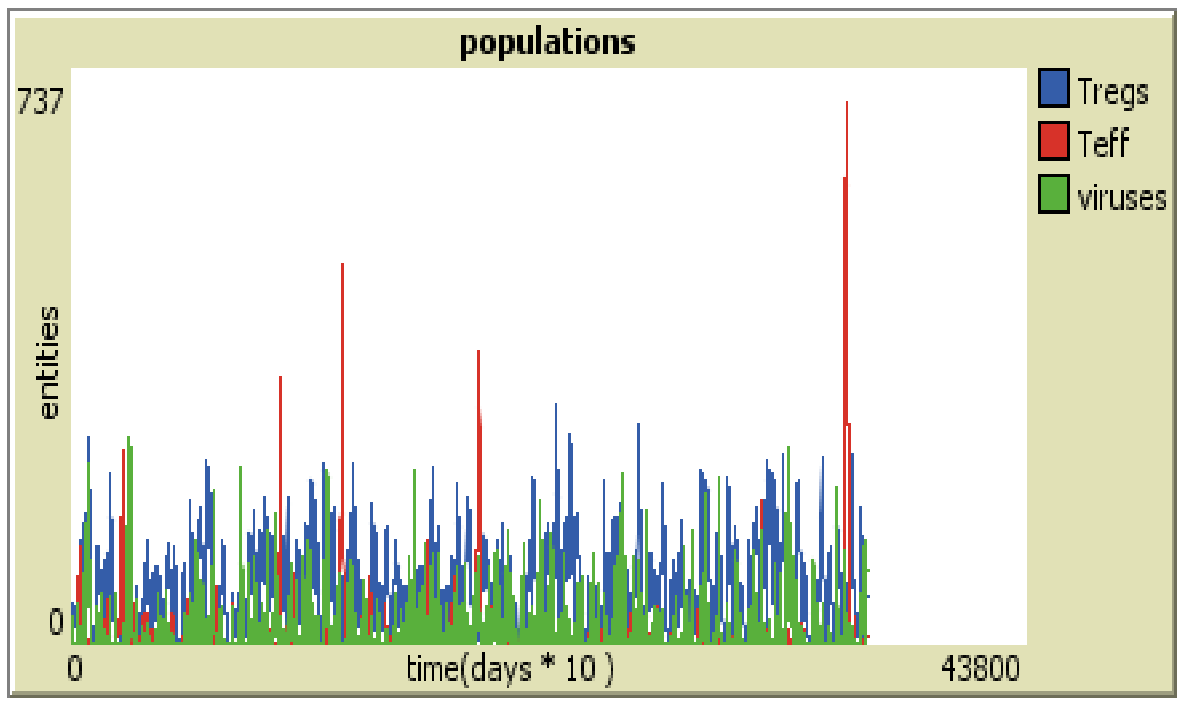

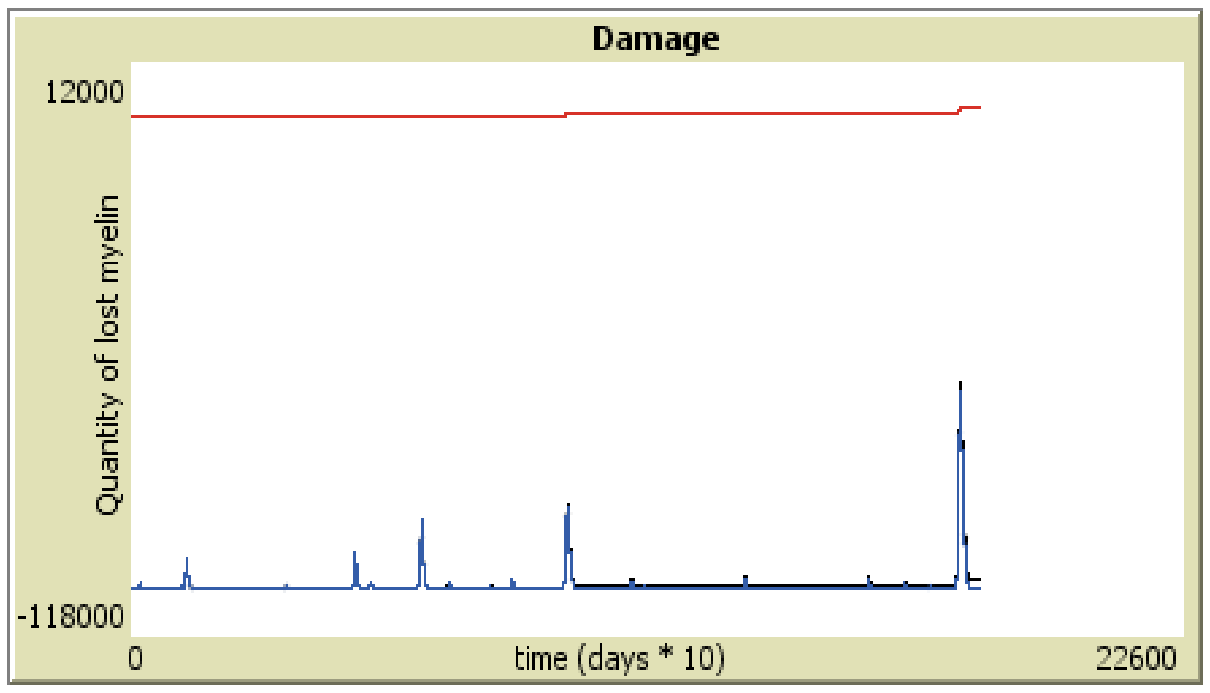

This simulation is adjusted according to the strong immune system. In which T-regulatory cells are more active than T-effector cells ant they give a quick response to external and internal disease attack. In this simulation, T-regulatory sells swiftly controls T-effectors activity without any significant brain damage.

## 4th simulation (slow recovery time)

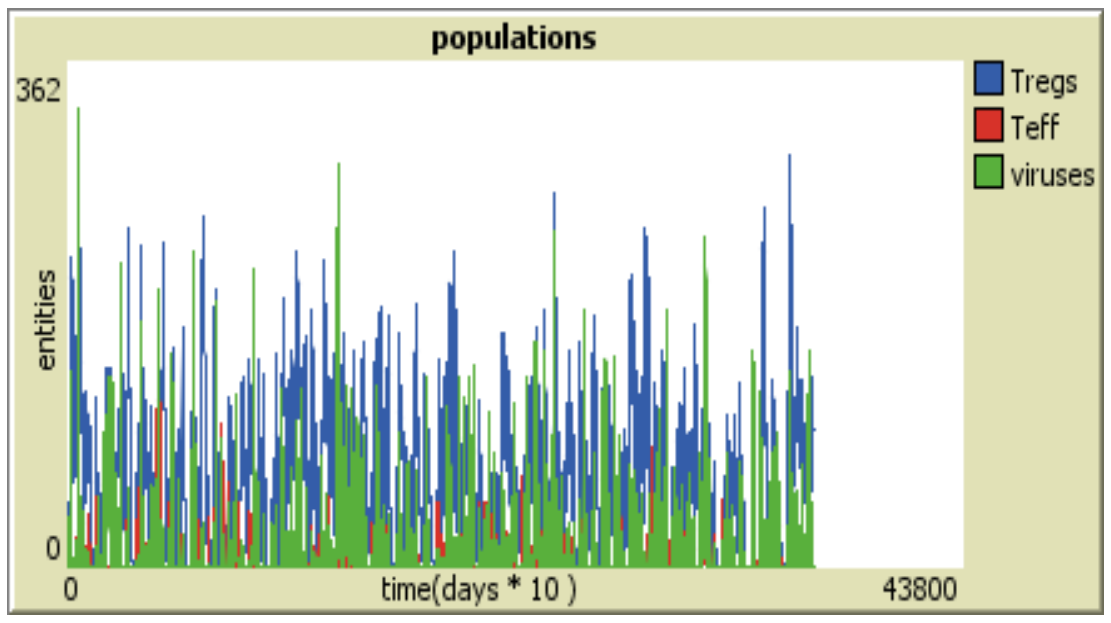

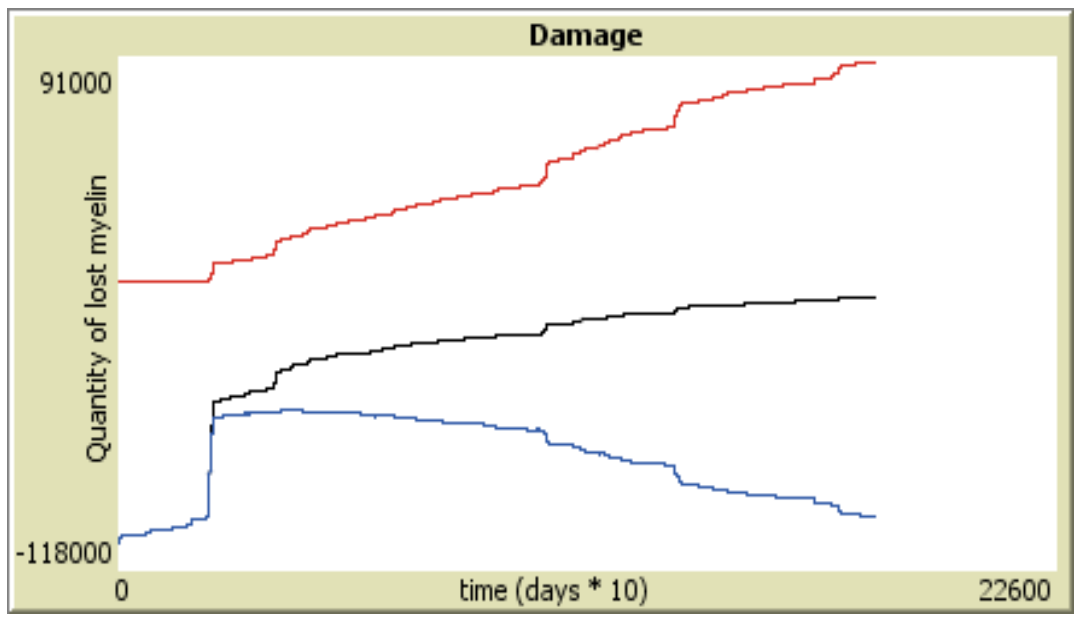

This simulation shows very slow healing power in patients. This simulation is adjusted as T-regulatory and T-effector cells production and reproduction ratio are same. When an effector attack on axonal then this damage goes to severe damage because of very slow recovery time. This type of brain is an easy target for disease. The brain damage rate increase, alternatively recovery process leads to slow.

## 5th simulation (weak immune system)

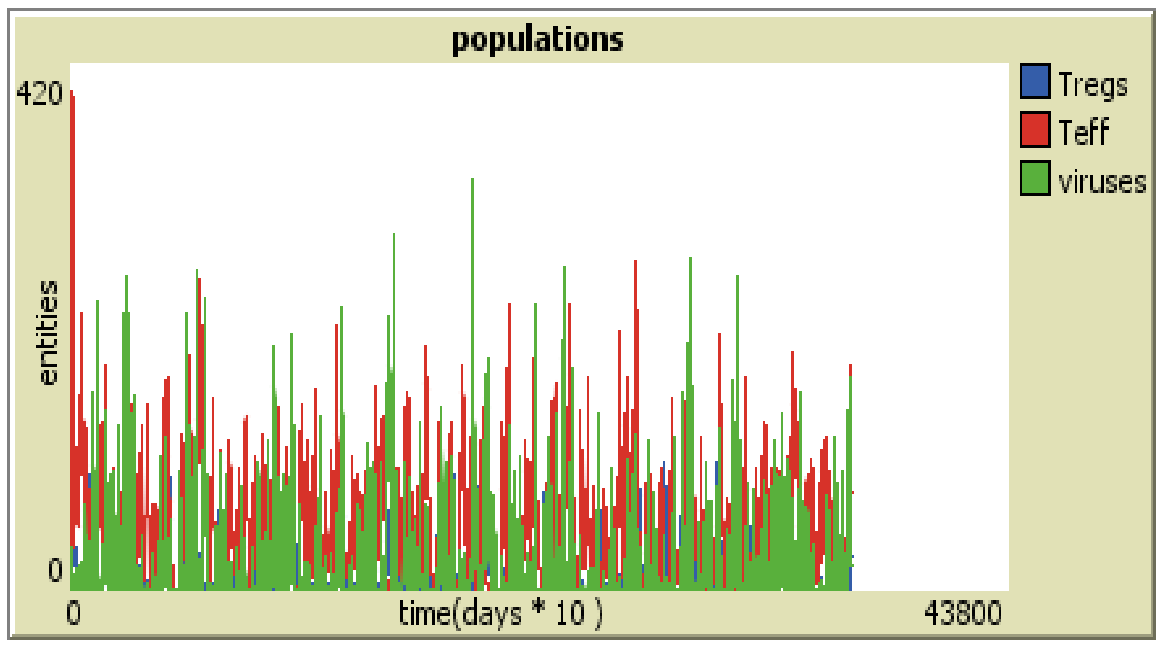

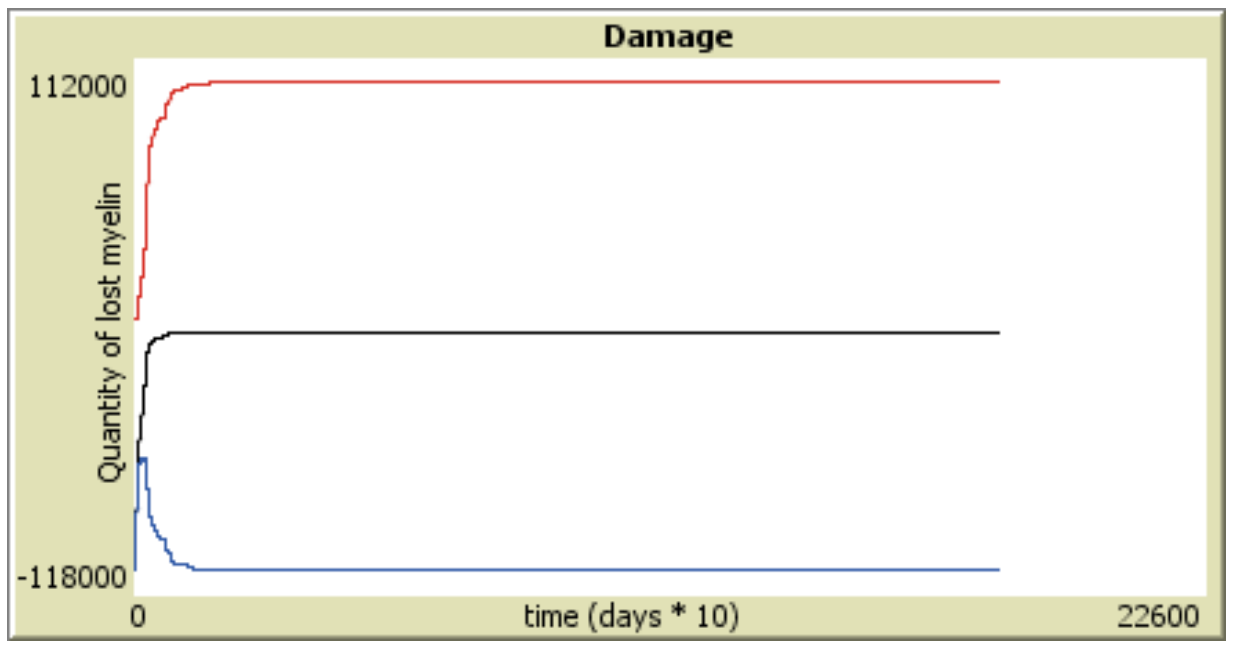

This simulation is adjusted as when an active T-effector attack on axonal it would be a severe damage and the production and reproduction of regulatory cells is half of the effector cells. This simulation shows a rapid damage rate and rapid decline in brain recovery time. At the end, most of the brain damage and became unrecoverable. This sign is a high severity of the disease.

## 6th simulation ()

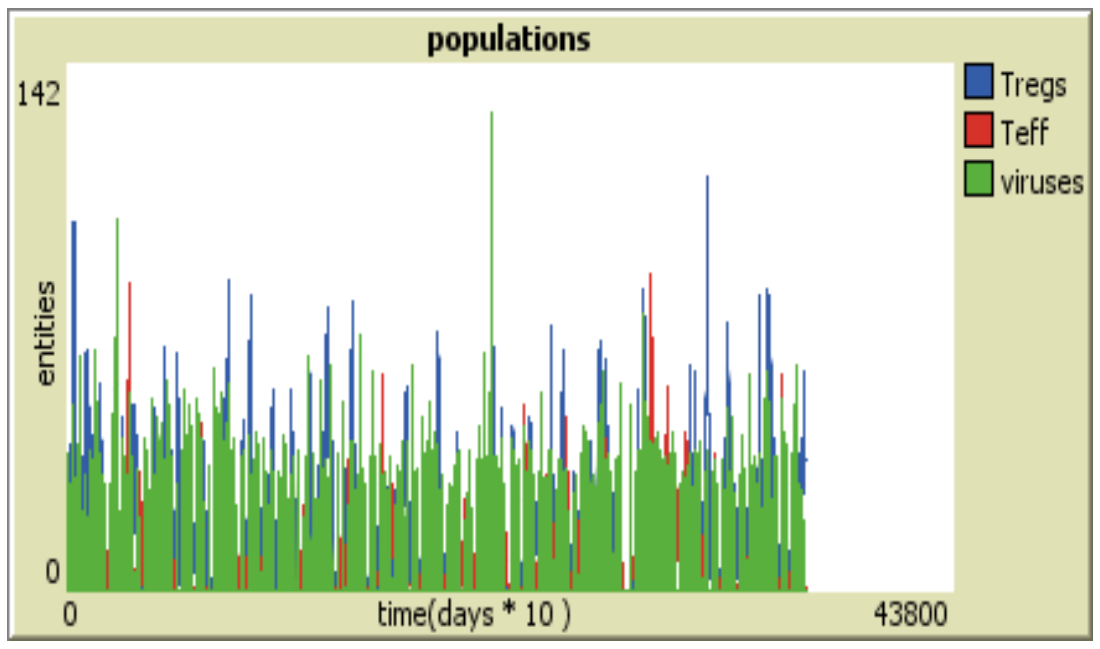

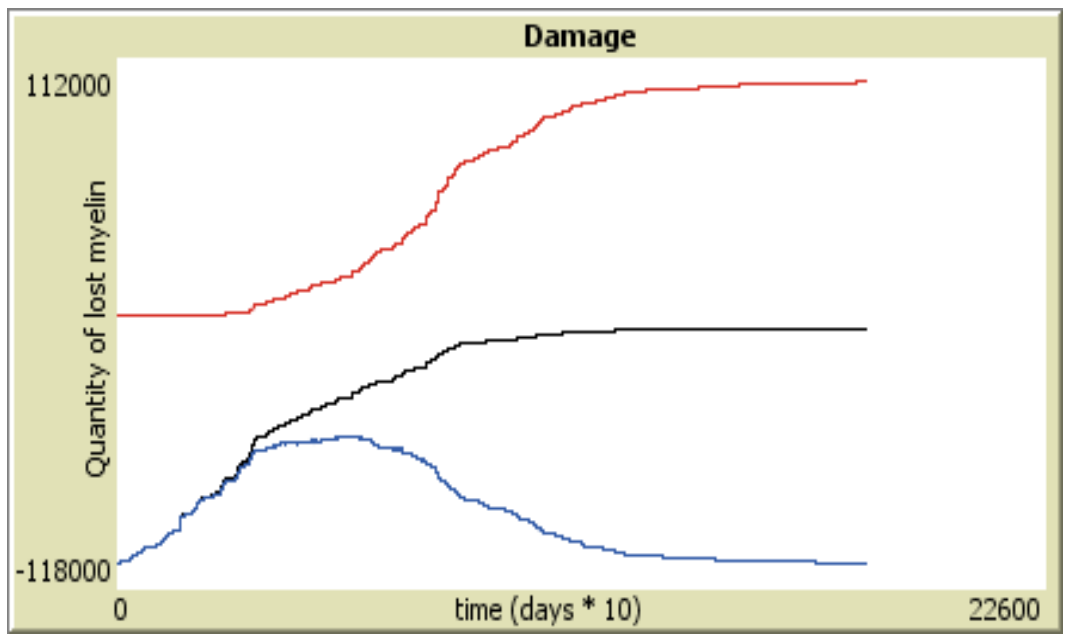

This simulation is adjusted as low virus attack on the immune system and the immune system have low cytokine energy. Cytokines support the immune system by decreasing activity of opposing cells. When T-effector cells attack axonal then T-effector cells give quick a response to effectors by controlling their activities. But alternatively, their produced cytokines have low energy. This case leads to increasing damage rate and decreasing recovery time.

### 6. Conclusion:

In this study, we developed a simulation model for MS disease by using AOSE methodologies. The novelty of our work is proposing an idea that, develop biological models with the help of AOSE methodologies, that provide developer support in the analysis, design, and implementation phase. After development, we have evaluated AOSE methodologies utilizing a framework that examines the various facets of a methodology. The proposed framework consists of four phases. The analysis of methodologies shows that although methodologies are mature, that can be used to develop biological or any other complex system, however, there are still open issues because methodologies do not provide a solid rule for agent identification from the requirement document and do not guide the transformation of roles to agents. In general, some software engineering issues such as quality assurance, cost estimation, and project management guidelines are not supported by any of the methodology. Besides all these limitations, AOSE methodologies have shown the potential for the development of the MS disease model. Moreover, this framework and comparison results can be utilized for selecting a methodology for developing an agent-based application.